\documentclass[a4paper,11pt]{article}
\pdfoutput=1 

\usepackage{jheppub}

\usepackage[utf8]{inputenc}

\usepackage[capitalise]{cleveref}
\usepackage{changepage}
\usepackage{tocloft}
\usepackage[T1]{fontenc}
\usepackage[utf8]{inputenc}

\usepackage{graphicx}
\usepackage{floatrow}
\usepackage{subfig}
\renewcommand{\figurename}{Fig.}
\usepackage{transparent}
\usepackage{bbold}
\usepackage{eso-pic}

\usepackage{tabu}
\usepackage{slashed}
\usepackage{multirow}
\usepackage{booktabs}
\usepackage{float}
\floatstyle{plaintop}
\restylefloat{table}

\usepackage{amsmath}
\usepackage{amssymb}
\usepackage{dsfont}
\usepackage{braket}
\usepackage{mathtools}
\usepackage{mhchem}
\usepackage{cancel}
\usepackage{dsfont}
\usepackage[normalem]{ulem}

\DeclarePairedDelimiter\abs{\lvert}{\rvert}
\DeclarePairedDelimiter\norm{\lVert}{\rVert}

\makeatletter
\let\oldabs\abs
\def\abs{\@ifstar{\oldabs}{\oldabs*}}
\let\oldnorm\norm
\def\norm{\@ifstar{\oldnorm}{\oldnorm*}}
\makeatother

\usepackage{fancyhdr}
\usepackage{afterpage}
\usepackage{titlesec}
\titlelabel{\thetitle.\quad}
\usepackage{blindtext}
\usepackage[bottom]{footmisc}

\usepackage{hyperref}
\hypersetup{
    allcolors=black
}

\titlespacing\section{0pt}{12pt plus 4pt minus 2pt}{-\parskip}\relax
\titlespacing\subsection{0pt}{12pt plus 4pt minus 2pt}{-\parskip}\relax

\usepackage{color, soul}
\usepackage{xcolor}
\usepackage{listings}

\usepackage{enumerate}
\usepackage{comment} 

\definecolor{mBlue}{rgb}{0,0,0.6}
\definecolor{mGray}{rgb}{0.2,0.2,0.2}
\definecolor{mRed}{rgb}{0.6,0,0}
\definecolor{backgroundColour}{rgb}{0.92,0.92,0.92}

\lstdefinestyle{CStyle}{
    backgroundcolor=\color{backgroundColour},   
    commentstyle=\color{mBlue},
    keywordstyle=\color{mRed},
    numberstyle=\tiny\color{mGray},
    stringstyle=\color{magenta},
    basicstyle=\footnotesize,
    breakatwhitespace=false,         
    breaklines=true,                 
    captionpos=b,                    
    keepspaces=true,                 
    numbers=left,                    
    numbersep=5pt,                  
    showspaces=false,                
    showstringspaces=false,
    showtabs=false,                  
    tabsize=2,
    language=C
}

\newcommand{\cevns}{{CE$\nu$NS\ }}

\title{EFT analysis of New Physics at COHERENT}

\author[a]{V\'{i}ctor Bres\'o-Pla}
\author[b]{, Adam Falkowski}
\author[a]{, Mart\'{i}n Gonz\'{a}lez-Alonso}
\author[a]{, Kevin Mons\'alvez-Pozo}

\affiliation[a]{Departament de F\'isica Te\`orica, IFIC, Universitat de Val\`encia - CSIC, Apt.  Correus 22085, E-46071 Val\`encia, Spain}
\affiliation[b]{Universit\'{e} Paris-Saclay, CNRS/IN2P3, IJCLab, 91405 Orsay, France}

\abstract{
Using an effective field theory approach, 
we study coherent neutrino scattering on nuclei, in the setup pertinent to the COHERENT experiment. 
We include non-standard effects both in neutrino production and detection,
with an arbitrary flavor structure, with all leading Wilson coefficients simultaneously present, and without assuming factorization in flux times cross section. 
A concise description of the COHERENT event rate is obtained by introducing   
three generalized weak charges, which can be associated (in a certain sense) to the production and 
scattering of $\nu_e$, $\nu_\mu$ and $\bar{\nu}_\mu$ on the nuclear target. 
Our results are presented in a convenient form that can be trivially applied to specific New Physics scenarios. 
In particular, we find that existing COHERENT measurements provide percent level constraints on two combinations of Wilson coefficients. 
These constraints have a visible impact on the global SMEFT fit, even in the constrained  flavor-blind setup. 
The improvement, which affects certain 4-fermion LLQQ operators, is significantly more important in a flavor-general SMEFT. 
Our work shows that COHERENT data should be included in electroweak precision studies from now on.
}

\usepackage{natbib}
\usepackage{graphicx}

\newcommand{\eref}[1]{Eq.~\eqref{eq:#1}}

\newcommand{\aref}[1]{Appendix~\ref{app:#1}}
\newcommand{\sref}[1]{Section~\ref{sec:#1}}
\newcommand{\tref}[1]{Table~\ref{tab:#1}}

\newcommand{\nn}{\nonumber \\}  
\newcommand{\nnl}{\nonumber \\}

\newcommand{\beq}{\begin{equation}} 
\newcommand{\eeq}{\end{equation}} 
\newcommand{\ba}{\begin{array}}  
\newcommand{\ea}{\end{array}} 
\newcommand{\bea}{\begin{eqnarray}}  
\newcommand{\eea}{\end{eqnarray} }  
\newcommand{\be}{\begin{eqnarray}}  
\newcommand{\ee}{\end{eqnarray} }  
\newcommand{\bal}{\begin{align}}
\newcommand{\eal}{\end{align}}   
\newcommand{\bi}{\begin{itemize}}  
\newcommand{\ei}{\end{itemize}}  
\newcommand{\ben}{\begin{enumerate}}  
\newcommand{\een}{\end{enumerate}}  
\newcommand{\bc}{\begin{center}}
\newcommand{\ec}{\end{center}} 
\newcommand{\bt}{\begin{table}}
\newcommand{\et}{\end{table}}  
\newcommand{\btb}{\begin{tabular}}
\newcommand{\etb}{\end{tabular}}

\newcommand{\cO}{{\mathcal O}}

\newcommand{\cM}{{\mathcal M}}

\newcommand{\re}{{\mathrm{Re}} \,}

\def\hc{{\rm h.c.}} 

\newcommand{\eps}{\epsilon}

\begin{document}

\maketitle

\section{Introduction }\vspace{0.4cm}
\label{sec:intro}

Precision measurements play an ever-increasing role in particle physics.  
A broad range of observables has been designed to test various aspects of the Standard Model (SM), such as accidental symmetries, CP violation, flavor structure, electroweak symmetry breaking, etc.   
Beyond the SM (BSM), precision measurements are sensitive to new particles and interactions, often well beyond the direct reach of high-energy colliders. 

Experiments where neutrinos are detected constitute a vital ingredient in the precision program.
Neutrino scattering on matter (electrons, nucleons, nuclei) probes the interaction strength and structure of the weak force. 
In the SM context it is a measure of the weak mixing angle, although currently it cannot compete with the much more precise determination based on the W mass and Z-pole physics. 
More generally, it is sensitive to new BSM particles interacting with neutrinos. 

Coherent elastic neutrino scattering on nuclei (CE$\nu$NS) is the newest member of the family of electroweak observables.
The process was theorized long ago~\cite{Freedman:1973yd,Drukier:1984vhf}, but the experimental challenges were overcome only recently by the COHERENT experiment~\cite{COHERENT:2017ipa}.  
The peculiarity of \cevns is that vector interactions between neutrino and nuclear constituents (nucleons or quarks) add up coherently. 
Consequently, possible small deviations between the actual neutrino interaction strength and that predicted by the SM are also amplified.  
This effect, roughly proportional to the number of neutrons in the nucleus, 
makes \cevns a powerful probe of physics beyond the SM. 

Assuming the BSM particles are heavy, the most convenient description of the non-standard effects in \cevns involves the effective field theory (EFT) framework~\cite{Weinberg:1980wa,Weinberg:2021exr}. 
There is in fact a ladder of relevant EFTs, with each consecutive rung involving higher energy scales, different degrees of freedom,  and more stringent assumptions about the scale of new physics~\cite{Pich:1998xt,Manohar:2020nzp}.
At the energy scales typical for CE$\nu$NS, nucleons (protons and neutrons) are convenient degrees of freedom. 
In this case, experiments probe the strength of effective  4-fermion contact interactions between neutrinos and nucleons, that is to say, neutron and proton weak charges in the standard parlance~\cite{Erler:2013xha}. 
Moreover, the observed \cevns rate is obviously also sensitive to BSM effects in neutrino production.
In the SM, the \cevns event rate is simply proportional to the square of  the weak charges weighted by the number of protons and neutrons in the target nucleus~\cite{Scholberg:2005qs}. 
At higher energies, above the QCD phase transition but below the electroweak scale, quarks become the relevant degrees of freedom. 
In this language \cevns experiments probe the Wilson coefficients of the effective 4-fermion neutral-current (NC) interactions between neutrinos and quarks~\cite{Barranco:2005yy}, as well as additional charge-current (CC) interactions involved in neutrino production (in pion and muon decay in the case of COHERENT experiment). 
We refer to this effective theory as the Weak EFT (WEFT)~\cite{Jenkins:2017jig}.\footnote{
For NC non-standard interactions (NSI)~\cite{Gago:2001xg} the NSI coefficients are in a simple correspondence with the Wilson coefficients of WEFT NC operators involving neutrinos~\cite{Campanelli:2002cc}.
For CC NSIs~\cite{Grossman:1995wx} the situation is much subtler, as discussed in Ref.~\cite{Falkowski:2019kfn}.
}

The Wilson coefficients of operators involving neutrinos can be translated into the language of non-standard interactions (NSI)~\cite{Grossman:1995wx} prevalent in the neutrino literature.

Finally, above the mass of the Z boson the relevant effective theory is the so-called SM Effective Field Theory (SMEFT)~\cite{Buchmuller:1985jz,Grzadkowski:2010es}, where the degrees of freedom are those of the SM and the full $SU(3)_C\times SU(2)_L\times U(1)_Y$ gauge symmetry is realized linearly. 
This is the most used EFT in the particle physics community because it can be directly (and now automatically~\cite{Carmona:2021xtq}) matched to a host of popular BSM models with supersymmetric particles, leptoquarks, W'/Z' bosons, etc.

In this paper we interpret the results of the COHERENT experiment in the language of the EFTs. We analyze the entirety of the available COHERENT data to extract the nuclear and nucleon weak charges (which we properly generalize so that production effects are taken into account). 
We translate these into constraints on the WEFT Wilson coefficients. 
Our results are provided  in a completely general form that is straightforward to apply to any BSM model or EFT analysis.
In fact, we analyze the impact of current \cevns data in a global SMEFT fit to electroweak precision observables (EWPO), which include such emblematic data as the $W$ boson mass measured in hadron colliders,  $Z$ boson decays measured in LEP-1, or atomic parity violation in cesium. 
We will demonstrate that this impact is non-trivial, significantly reducing the allowed parameter space for the SMEFT Wilson coefficients. 
Moreover, from the point of view of EWPO and SMEFT fits, we will show that COHERENT is the most relevant neutrino-detection experiment to date!

There are several ways in which this paper improves on the previous literature regarding the EFT approach to COHERENT data~\cite{Barranco:2005yy,Scholberg:2005qs,Coloma:2017ncl,Papoulias:2017qdn,Shoemaker:2017lzs,Liao:2017uzy,AristizabalSierra:2018eqm,Denton:2018xmq,Esteban:2018ppq,Khan:2019cvi,Giunti:2019xpr,Arcadi:2019uif,Coloma:2019mbs,Denton:2020hop,Miranda:2020tif,Coloma:2022avw,AtzoriCorona:2022qrf,DeRomeri:2022twg}:
\vspace{-0.15cm}
\begin{itemize}
\item 
Our work, together with the recent Refs.~\cite{AtzoriCorona:2022qrf,DeRomeri:2022twg}, are the first ones  to use the entire COHERENT dataset presently available (2D energy and time distributions with argon and cesium-iodine targets)~\cite{COHERENT:2020iec,COHERENT:2021xmm} to probe fundamental interactions, and thus represents the current state of the art.
\item 
We perform for the first time a complete analysis in the framework of WEFT, including the nonlinear effects of the various Wilson coefficients  in detection and production. 
The latter, as well as the non-trivial interplay between the two, have not been correctly discussed before. 
We remark that correlated effects in production and detection are generic in new physics models, since the $SU(2)_L$ gauge symmetry relates CC and NC interactions. 
\item
We identify the linear combinations of EFT Wilson coefficients that are strongly constrained by COHERENT.
\item The model-independent SMEFT analysis of COHERENT data and the combination with other EWPO is performed for the first time. 
\end{itemize}

The rest of this article is organized as follows. In~\sref{TH} we describe the EFT framework, and in~\sref{rate} we obtain the EFT prediction for the event rate measured at COHERENT. In~\sref{exp} we discuss the various COHERENT measurements and how they will be implemented in our numerical analysis, which is presented in~\sref{numerical} using the WEFT setup. The implications for the SMEFT and the combination with Electroweak Precision Observables is discussed in~\sref{EWPO}. Finally,~\sref{conclusions} contains our conclusions.

\section{Theory framework: EFT ladder} \vspace{0.3cm}
\label{sec:TH}
%
\subsection{EFT below the electroweak scale} \vspace{0.4cm}

Given the low energies involved in CE$\nu$NS, our starting point will be the most general effective Lagrangian below the electroweak scale with the SM particle content, the WEFT Lagrangian~\cite{Jenkins:2017jig}. In this effective theory, electroweak gauge bosons, the Higgs boson, and the top quark have been integrated out and the electroweak symmetry is explicitly broken. Let us stress that right-handed neutrinos are not part of the WEFT particle content.

Here we focus on the lepton-number-conserving parts of the WEFT Lagrangian that are relevant for COHERENT physics, namely the NC interactions mediating \cevns and the CC interactions involved in pion and muon decay. Let us first present the NC interactions between neutrinos and quarks~\cite{Campanelli:2002cc,Jenkins:2017jig}:
\begin{align}
\label{eq:LagrangianNC}
    \mathcal{L}_{\rm WEFT} \supset &   - \frac{1}{v^2} \sum_{q=u,d}\left\{~~
    [g_V^{qq} \,\mathbb{1} + \epsilon_V^{qq}]_{\alpha \beta} \left( \bar{q}\gamma^{\mu}q\right)\left(\bar{\nu}_{\alpha}\gamma_{\mu}P_L \nu_{\beta} \right) \right. \nnl & 
     ~~~~~~~~~~~~~~ \left. +~ [g_A^{qq} \,\mathbb{1} + \epsilon_A^{qq}]_{\alpha \beta} \left( \bar{q}\gamma^{\mu}\gamma^5 q\right)\left(\bar{\nu}_{\alpha}\gamma_{\mu}P_L \nu_{\beta} \right) \right\},
\end{align}
where $P_{L,R}=\left(1 \mp \gamma^5 \right)/2$ are the chirality projection operators and $\mathbb{1}$ is the unit matrix. The dimensionful normalization factor is related to the measured value of the Fermi constant  $G_F\approx 1.166\times 10^{-5}~{\rm GeV}^{-2}$~\cite{ParticleDataGroup:2022pth} via $v \equiv (\sqrt{2} G_F)^{-1/2}\approx 246.22$~GeV.
In the coefficients of the operators we separated the SM values $g_{V,A}^{qq}$ from the new physics contributions $\epsilon_{V,A}^{qq}$. 
The former are given at tree-level by
\begin{equation}
 g_V^{qq} = T^3_q - 2 Q_q \sin^2 \theta_W,  
\qquad 
g_A^{qq}   = - T^3_q ~, 
\end{equation}
where $T_q^3$ and $Q_q$ are the weak isospin and charges of the quark $q$ and $\theta_W$ is the weak mixing angle. 
See Ref.~\cite{Erler:2013xha} for the discussion of radiative corrections and numerical values of these SM couplings. 
The parameters $\epsilon_{V,A}^{qq}$ are zero in the SM limit and more generally they are Hermitian matrices (in the neutrino indices). 
As is well known~\cite{Freedman:1973yd}, the contribution of the vector Wilson coefficients to the coherent neutrino scattering  rate is enhanced by $(A-Z)^2$, where $A-Z$ is the number of neutrons in the nucleus.
On the other hand, the contribution of the axial Wilson coefficients does not benefit from such an enhancement, and moreover it vanishes completely for scattering on spin-zero nuclei. 
In view of the current experimental uncertainties, we can thus
neglect the axial contributions in the following, leaving us only with the vector Wilson coefficients.\footnote{See Ref.~\cite{Hoferichter:2020osn} for a detailed discussion of the axial contributions to coherent neutrino scattering.} 
To lighten the notation the vector subindex will be omitted and we will work instead with the more compact notation $\epsilon^{qq}_{\alpha \beta}\equiv [\epsilon_V^{qq}]_{\alpha \beta}$.

Let us now introduce the parts of the WEFT Lagrangian that describe the interactions relevant for neutrino production at COHERENT. Pion decay is described by~\cite{Cirigliano:2009wk,Jenkins:2017jig}
\begin{align}
\label{eq:Lagrangianpion}
    \mathcal{L}_{\rm WEFT} \supset - \frac{2\, V_{ud}}{v^2} \Big\{
    & [1+\epsilon^{ud}_L]_{\alpha \beta} \left( \bar{u}\gamma^{\mu}P_L d\right)\left(\bar{\ell}_{\alpha}\gamma_{\mu}P_L \nu_{\beta} \right) 
    + [\epsilon^{ud}_R]_{\alpha \beta} \left( \bar{u}\gamma^{\mu}P_R d\right)\left(\bar{\ell}_{\alpha}\gamma_{\mu}P_L \nu_{\beta} \right) 
    \nn
    & \left. +\frac{1}{2}\left[ \epsilon_S^{ud}\right]_{\alpha \beta} \left(\bar{u}d \right)\left(\bar{\ell}_{\alpha}P_L \nu_{\beta} \right)
    -\frac{1}{2}\left[ \epsilon_P^{ud}\right]_{\alpha \beta} \left(\bar{u} \gamma^5 d \right)\left(\bar{\ell}_{\alpha}P_L \nu_{\beta} \right)\right. 
    \nn
    & +  \frac{1}{4}\left[ \epsilon_T^{ud}\right]_{\alpha \beta} \left(\bar{u} \sigma^{\mu \nu} P_L d \right)\left(\bar{\ell}_{\alpha} \sigma_{\mu \nu}P_L \nu_{\beta} \right) \Big\} ~+\rm{h.c.}~,
\end{align}
where $\ell_\alpha = e,\mu,\tau$ are charged lepton fields,  $V_{ud}$ is the (1,1) element of the Cabibbo--Kobayashi--Maskawa~(CKM) matrix. 
Here, new physics is parametrized by $\epsilon^{ud}_{L,R,S,P,T}$, which are general complex matrices in the lepton indices.
Finally, the WEFT interactions describing muon decay are~\cite{Cirigliano:2009wk,Jenkins:2017jig}: 
\begin{align}
\label{eq:Lagrangianmuon}
    \mathcal{L}_{\rm WEFT} \supset 
    - \frac{2}{v_0^2} \Big\{ 
    &
    \left( \delta_{\alpha a} \delta_{\beta b} + \left[\rho_L \right]_{a \alpha \beta b}\right) \left( \bar{\ell}_{a}\gamma^{\mu}P_L \nu_{\alpha}\right)\left(\bar{\nu}_{\beta}\gamma_{\mu}P_L \ell_{b} \right) 
    \nn
    &-2 \left[\rho_R \right]_{a \alpha \beta b} \left( \bar{\ell}_{a}P_L \nu_{\alpha}\right)\left(\bar{\nu}_{\beta}P_R \ell_{b} \right) \Big\}.
\end{align}
Here new physics is parametrized by the $[\rho_{L,R}]_{a \alpha \beta b}$ tensors.
Hermiticity of the Lagrangian implies 
$\left[\rho_{L,R} \right]^*_{a \alpha \beta b} = \left[\rho_{L,R} \right]_{b \beta \alpha a}$. 
Contrary to~\cref{eq:LagrangianNC,eq:Lagrangianpion}, in the case of~\cref{eq:Lagrangianmuon} we cannot normalize the interactions using $v$. 
This is because that parameter is defined via $G_F$, which is extracted from the experimental measurement of the muon decay rate, to which the interactions in~\cref{eq:Lagrangianmuon}  contribute. 
Instead, we use the tree-level value of the Higgs vacuum expectation value, $v_0$. Using \cref{eq:Lagrangianmuon} it is trivial to calculate the muon decay width and relate $v$ and $v_0$, finding $v_0^2 = v^2\big (1 + 2\,\re [\rho_L]_{\mu\mu ee} + \sum_{\alpha \beta} \sum_{X=L,R} \big |[\rho_X]_{\mu \alpha \beta e}\big |^2 \big )^{1/2}_{\mu\simeq m_\mu}$.  In other words, we are working with an input scheme such that the $\mu \to e \bar{\nu} \nu$ decay rate is controlled, exactly as in the SM, by $v$ (equivalently, by $G_F$), without any tree-level dependence on the $\rho_{L,R}$ Wilson coefficients.\footnote{The discussion is much simpler if one truncates the WEFT predictions at the linear level, as it is commonly done. In that case  it is sufficient to replace $v_0 \to v$ in~\cref{eq:Lagrangianmuon}, with the restriction that  $\re [\rho_L]_{\mu\mu ee}$ vanishes at the scale $\mu \simeq m_\mu$~\cite{Falkowski:2017pss}. In this paper, however, we will analyze the COHERENT constraints beyond the linear level, and for this reason we introduce the more general input scheme.}

Likewise,  one should also take into account that new physics effects in~\eref{Lagrangianpion} affect the extraction of the CKM factor $V_{ud}$.
See Refs.~\cite{Gonzalez-Alonso:2018omy,Falkowski:2020pma} for the discussion of this issue in the context of  extracting $V_{ud}$ from beta decays.
For the present purpose, however, $V_{ud}$ appears in COHERENT observables in combination with the pion decay constant $f_{\pi^\pm}$, and the product of the two is usually extracted from the experimental measurements of $\pi \to \mu \nu$ decay width.

In the WEFT Lagrangian above, the quarks $u$, $d$ and charged leptons $\ell_{\alpha}$ are in the basis where their kinetic and mass terms are diagonal, whereas the neutrino fields $\nu_\alpha$ are taken in the flavor  basis. The latter are connected to neutrino mass eigenstates through the Pontecorvo-Maki-Nakagawa-Sakata (PMNS) mixing matrix: $\nu_{\alpha} = \sum_{n=1}^3 U_{\alpha n} \nu_n$. 
As usual, flavor indices are denoted with Greek letters, whereas mass eigenstate indices are denoted with Roman letters.  

The Wilson coefficients $\epsilon_X^{ff'}$ and $\rho_X$ parametrize the effect of new interactions mediated by non-standard heavy particles (heavy compared with the COHERENT physics scale), such as,  e.g. leptoquarks or $Z'$ and $W'$ vector bosons. 
It is customary to define the Wilson coefficients in \cref{eq:Lagrangianpion} 
at the renormalization scale $\mu = 2\,{\rm GeV}$ in the $\overline{\rm MS}$ scheme, since lattice QCD provides hadronic decay constants at that scale and scheme. 
We make the analogous scale choice for the Wilson coefficients $\epsilon^{qq}_{\alpha\beta}$ in \cref{eq:LagrangianNC}. 
Below 2~GeV, the quark-level Lagrangians will be matched in the  next subsection to the nucleon-level EFT. 
On the other hand, for the Wilson coefficients in the leptonic  interactions in \cref{eq:Lagrangianmuon} it is convenient to choose a lower renormalization scale, $\mu \simeq m_\mu$. 
In any case, in this paper we will take into account only the one-loop QCD running of the Wilson coefficients, which can lead to substantial effects, but only concerns the parameters $\epsilon_{S,P,T}$ in \cref{eq:Lagrangianpion}~\cite{Gonzalez-Alonso:2017iyc}. 
Although the Wilson coefficients are complex in general we will treat them as real throughout this paper for the sake of simplicity.

\subsection{Nucleon-level EFT} \vspace{0.4cm}

The EFT Lagrangian in the preceding subsection contains quarks fields.
However,  quarks are not useful degrees of freedom at the energies relevant for \cevns experiments. 
In particular, in the COHERENT experiment  some of the neutrinos are produced in pion decay, and they subsequently scatter on heavy nuclei. 
Thus, we need to connect the quark-level formalism to hadronic- and nuclear-level observables. 
Regarding pions, it is customary to connect the WEFT Lagrangian in \cref{eq:Lagrangianpion} directly to the decay amplitude, 
using the matrix element of the quark bilinears between a pion state and the vacuum, see \cref{sec:amplitudes}. 
On the other hand, regarding scattering on nuclei, it is convenient to introduce an intermediate step in the form of an effective Lagrangian where the degrees of freedom are nucleons rather than quarks. To match the nucleon- and quark-level Lagrangians we need the matrix elements of the quark operators between the nucleon states. Let us denote the incoming nucleon momentum $k$, and the outgoing nucleon momentum $k'$. In the near-zero recoil limit, $k -  k' \approx 0$, we have 
\begin{eqnarray}
\bra{N(k',s')} \bar q \gamma^\mu q \ket {N(k,s)}  & = & F_1^{q,N} \bar u_N(k',s') \gamma^\mu u_N(k,s),
\end{eqnarray} 
where $N = p,n$, and $u_N(k,s)$ is the usual Dirac spinor wave function of the nucleon $N$. 
Isospin symmetry implies $F_1^{u,p} = F_1^{d,n} $ and $F_1^{u,n} = F_1^{d,p}$. 
For the vector part, conservation of the electromagnetic current implies 
\begin{eqnarray}
\bra{p(k',s')} (2/3) \bar u \gamma^\mu u -  (1/3) \bar d \gamma^\mu d  \ket {p(k,s)}  & = &  (+1) \bar u_p(k',s') \gamma^\mu u_p(k,s), 
\nnl 
\bra{n(k',s')} (2/3) \bar u \gamma^\mu u -  (1/3) \bar d \gamma^\mu d  \ket {n(k,s)}  & = &  0 . 
\end{eqnarray}  
Thus, in the limit where the strange content of the nucleon is ignored\footnote{See Ref.~\cite{Hoferichter:2020osn} for the corrections due to the strange content.
Taking this into account in our analysis would introduce a weak dependence of the COHERENT observables on the WEFT Wilson coefficients $\epsilon_V^{ss}$. However, currently COHERENT is sensitive only to  $|\epsilon^{ss}_{\alpha\beta}|\gtrsim 1$, and for this reason we ignore the strange content of the nucleon. }
one finds that 
$(2/3) F_1^{u,p} - (1/3) F_1^{d,p} = 1$ and
$2 F_1^{u,n} = F_1^{d,n}$ (thus $2 F_1^{d,p} = F_1^{u,p}$),  
so finally 
\begin{equation}
 F_1^{d,p}  =  F_1^{u,n}  = 1, \qquad 
 F_1^{u,p}  =  F_1^{d,n}  = 2 .   
\end{equation} 
Using these matrix elements, we can write down the effective Lagrangian for vector NC interactions between neutrinos and nucleons:
\begin{equation}
\label{eq:FORM_Lnucleon}
{\cal L}_{\rm nucleon} \supset 
- {1 \over 2 v^2} 
\sum_{N = p,n} \sum_{\alpha,\beta = e,\mu,\tau} 
(\bar \nu_\alpha \gamma_\mu P_L \nu_\beta )
g^{\nu N}_{\alpha \beta} (\bar N \gamma^\mu N)~,
\end{equation} 
where the leading order matching of the Wilson coefficients of the two EFTs reads 
\begin{align}
\label{eq:FRAME_nucleonToWEFTmap}
g^{\nu p}_{\alpha \beta}  &=  2\, \left[ (2\,g_V^{uu} + g_V^{dd}) \, \mathbb{1} + (2\,\epsilon^{uu} + \epsilon^{dd}) \right]_{\alpha \beta}~.\nn
g^{\nu n}_{\alpha \beta}  &=  2\, \left[ (g_V^{uu}+2g_V^{dd})\, \mathbb{1} + (\epsilon^{uu} + 2\epsilon^{dd})\right]_{\alpha \beta}~.
\end{align}
In the SM limit at tree level we have 
\begin{align}
 \, [g^{\nu p}_{\alpha \beta}]_{\rm SM} = & 
 \big ( 1   - 4   \sin^2 \theta_W \big) \delta_{\alpha \beta} 
 \nnl  
 \, [g^{\nu n}_{\alpha \beta}]_{\rm SM} = & 
  -  \delta_{\alpha \beta} . 
\end{align}
One can see that the NC interactions between neutrinos and vector nucleon currents are approximately protophobic, due to the accidental fact that $\sin^2 \theta_W \approx 1/4$.
For this reason, low-energy neutrinos scatter mostly on neutrons in nuclei.

At leading order, the nucleon weak charge $Q_w^N$ can be defined simply as the value of the effective couplings in \cref{eq:FORM_Lnucleon} at some fixed renormalization scale, $[Q_w^N]_{\alpha \beta} \equiv g^{\nu N}_{\alpha \beta}(\mu)$. 
One can generalize this definition so that it includes radiative corrections, which cancels the renormalization scale dependence, but it becomes process dependent~\cite{Tomalak:2020zfh}. In that approach, the SM values for the weak charges are the following
\begin{align}
\label{eq:FORM_weakChargesSM}
\, [Q_w^p]_{ee}^{\rm SM} = &   0.0747(34), \qquad [Q_w^p]_{\mu\mu}^{\rm SM} =  0.0582(34), 
\nnl 
\, [Q_w^n]_{\alpha\alpha}^{\rm SM} = & 
-1.02352(25). 
\end{align}
Note that with this definition the proton weak charge depends slightly on the neutrino flavor. The difference between the proton weak charge experienced by electron- and muon-neutrinos is however numerically irrelevant at present, given the accuracy of the COHERENT experiment. Accordingly, we will just neglect such differences and work with the muonic weak charge also in relation with electron neutrinos (see Section~\ref{sec:amplitudes}).

 In order to connect the nucleon-level EFT
to the nuclear scattering observables, it is convenient to take the non-relativistic limit of the Lagrangian in \cref{eq:FORM_Lnucleon}, since that will allow us to calculate observables for nuclei of arbitrary spin.
At the zero-recoil level it takes the form 
\begin{equation}
 {\cal L}_{\rm NR} \supset  
- {1 \over 2 v^2} \sum_{N = p,n} \sum_{\alpha,\beta = e,\mu,\tau} 
g^{\nu N}_{\alpha \beta}  (\psi_N^\dagger \psi_N)  ( \bar \nu_\alpha \gamma^0 P_L \nu_\beta )
+ \cO(\nabla/m_N).  
 \end{equation} 
Above, we traded the relativistic nucleon Dirac fields for the non-relativistic  ones denoted as $\psi_N$, which satisfy the Schr\"{o}dinger equations of motion.  
We also dropped all terms containing spatial derivatives $\nabla$, which correspond to recoil effects.\footnote{This formalism can be generalized to include recoil effects, see Ref.~\cite{Falkowski:2021vdg} for a study along these lines in the context of  beta decay.} 
Now, coherent neutrino scattering amplitudes will involve matrix elements of $\psi_N^\dagger \psi_N$ between nuclear states. 
For a nucleus ${\cal N}$ with momentum $k$, energy $E_{\cal N}$, charge $Z$, mass number $A$, spin $J$, and  spin projection along the z-axis $J_z$,  the rotational and isospin symmetry  requires the matrix elements to take the form 
\begin{align}
\bra{ {\cal N}(k', J_z') } \psi_p^\dagger \psi_p  \ket{ {\cal N}(k, J_z) }   =&  
2 \sqrt{E_{\cal N} E_{\cal N}'} \,  \mathcal{F}_{p} (q^2)  Z \delta_{J_z' J_z}  , 
\nnl 
\bra{ {\cal N}(k', J_z') } \psi_n^\dagger \psi_n  \ket{ {\cal N}(k, J_z) }   =&  
2 \sqrt{E_{\cal N} E_{\cal N}'} \,  \mathcal{F}_{n} (q^2)  (A-Z)\delta_{J_z' J_z}  , 
\end{align} 
where the primed variables refer the final nuclear state ($E_{\cal N} = m_{\cal N} \approx E_{\cal N}'$ in the target rest frame). 
The nucleon form factors $\mathcal{F}_{p,n} (q^2)$ are equal to 1 at $q^2\equiv (k -  k')^2=0$ due to isospin symmetry, although we will not take that limit in our studies. The factor $A-Z$ in the neutron  matrix element is at the origin of the coherent enhancement of the neutrino scattering on heavy nuclei.

\section{COHERENT event rate} \vspace{0.4cm}
\label{sec:rate}

Let us consider neutrinos produced by a source $S$ through the process $S\to X_{\alpha} \nu_{k}$, where $X_{\alpha}$ is a one- or more-body final state that contains a charged lepton $\ell_{\alpha}=e,\mu, \tau$, and $\nu_{k}$ is a neutrino-mass eigenstate ($k=1,2,3$). These neutrinos propagate a distance $L$ --- conserving its mass index $k$ --- and are detected via the process $\nu_{k}{\cal N}\to \nu_{j}{\cal N}$, where $j=1,2,3$ is again a mass index and ${\cal N}$ denotes the target nucleus. 
Let us consider the differential number of detected events per time $t$, incident neutrino energy $E_{\nu}$, nuclear recoil energy $T$ and target particle
\begin{equation}
\label{eq:DifferentialNumberEventsFunctionR}
R^S_{\alpha} (t,T,E_\nu)\equiv \frac{1}{N_T}\frac{dN^S_{\alpha}}{dt \, dE_{\nu}\, dT}~,
\end{equation}
where $N_{T}$ stands for the number of target particles. 
Previous calculations of the \cevns rate~\cite{Freedman:1973yd,Barranco:2005yy,Scholberg:2005qs,Hoferichter:2020osn} have been carried out within the SM or under the assumption that the neutrino production is unaffected by New Physics (NP) and that one can thus simply calculate the rate as a flux times cross-section, i.e., $R_\alpha^S = \sum_\beta d\Phi_{SM} (S\!\to\!X_\alpha\nu_\alpha) / dE_\nu \times  d\sigma(\nu_\alpha\,{\cal N} \!\to\! \nu_\beta\,{\cal N}) / dT$, where $\nu_{\alpha,\beta}$ are neutrino flavor eigenstates. In this work we present a derivation that is more general in the treatment of new physics contributions.  
We calculate the event rate in terms of the WEFT Wilson coefficients introduced in~\sref{TH}. 
The interactions of neutrinos relevant for their production and detection are assumed to be most general at the leading order in the WEFT expansion, i.e., we allow for the simultaneous presence of all interactions described in~\cref{eq:LagrangianNC,eq:Lagrangianpion,eq:Lagrangianmuon}. We allow for arbitrary flavor mixing, both via the PMNS matrix, as well as via the various WEFT Wilson coefficients of 4-fermion interactions. Finally, we take into account that NP affecting neutrino production also affects the muon and pion decay widths, which are used in the COHERENT analysis and to determine some of the input observables (such as $G_F$). 

For this derivation, we need to connect the observable event rate in \cref{eq:DifferentialNumberEventsFunctionR} with the production and detection QFT amplitudes, denoted by ${\cal M}_{\alpha k}^{\text{P}}\equiv {\cal M}\left( S \to X_{\alpha} \nu_{k} \right)$ and ${\cal M}_{jk}^{\text{D}}\equiv {\cal M}\left(  \nu_{k} {\cal N} \to \nu_{j} {\cal N}  \right)$, which encode the fundamental physics taking place at production and detection. This connection was obtained in Ref.~\cite{Falkowski:2019kfn} using a QFT approach for the case of charged-current interactions both at production and detection. The main idea behind such derivation was, instead of considering the neutrino production and detection separately, to treat both interactions as a single process~\cite{Giunti:1993se}. In our CC-NC configuration, that translates into the following process
\begin{equation}
S {\cal N} \to X_{\alpha} {\cal N} \nu_{j} \, ,
\end{equation} 
where the $\nu_k$ neutrino is considered just as an intermediate particle in the amplitude. 
We can adapt the result found in Ref.~\cite{Falkowski:2019kfn} to describe the CE$\nu$NS rate observed at COHERENT taking into account the various differences. 
In NC neutrino scattering we have no information about the neutrino final mass eigenstate (or flavor) and hence, we should sum over the corresponding mass index $j$. 
Secondly, the time variation of the number of source particles $N_S$ cannot be neglected in this case (in fact, it produces a time-dependent signal that is measured). 
Furthermore, at COHERENT one measures the differential number of events per recoil energy $T$, and thus we will not integrate over that detection kinematic variable. 
Finally, the incident neutrino energy is not observed in COHERENT, which can be taken into account trivially integrating $R_\alpha^S$ over that variable. We note that the assumption of neutrinos emitted isotropically from a source at rest applies to the case of COHERENT.

All in all, the event rate for a source $S$ is given by
\begin{equation}
\label{eq:DifferentialRateEvents}
    R_{\alpha}^S = \frac{N_S(t)}{32 \pi L^{2} m_S m_{\cal N} E_{\nu}} 
    \sum_{j,k,l}
    e^{-i\frac{L \Delta m_{kl}^{2}}{2E_{\nu}}} \int d \Pi_{P^{\prime}} \mathcal{M}^{P}_{\alpha k} \bar{\mathcal{M}}^{P}_{\alpha l} \int d \Pi_{D^{\prime}} \mathcal{M}^{D}_{j k} \bar{\mathcal{M}}^{D}_{j l} ~,
\end{equation}
where complex conjugation is denoted with a bar, $m_{S,{\cal N}}$ are the masses of the source and target particles respectively and $\Delta m_{kl}^{2}\equiv m_{k}^{2}-m_{l}^{2}$ is the mass squared difference between neutrino (mass) eigenstates, which appears in the formula through the usual $e^{-i \,L \Delta m_{kl}^2 / (2E_\nu)}$ oscillatory factor, which we can simply approximate as one for the case of COHERENT given its very short baseline.  
The phase space elements for the production and detection processes, $d \Pi_{P}$ and $d \Pi_{D}$, are defined as usual: 
{$d \Pi \equiv {d^3 k_1 \over (2 \pi)^3 2 E_1} \dots  {d^3 k_n \over (2 \pi)^3 2 E_n} (2\pi)^4 \delta^4(\sum p_n - \sum k_i )$}, 
where $k_i$ and $E_i$ are the 4-momenta and energies of the final states and $\sum p_n$ is the total 4-momentum of the initial state. 
However, in order to obtain the observable of interest at COHERENT, we use a slight modification of the standard production and detection phase spaces denoted by primed subindices and defined by $d \Pi_{P}\equiv d \Pi_{P^{\prime}} dE_{\nu} $ and $d \Pi_{D}\equiv d \Pi_{D^{\prime}} dT $, such that $R_\alpha^S$ then provides the differential number of events per incident neutrino energy $E_{\nu}$ and recoil energy $T$ via Eq.~\eqref{eq:DifferentialNumberEventsFunctionR}. The integral sign involves both integration as well as sum and averaging over all unobserved  degrees of freedom such as spin. 
Finally $N_S (t)=n_{\rm POT}\,r_{S/p}\,\tau_S \,g_S(t)$ is the time-dependent number of source particles $S$, where $n_{\rm POT}$ is the total number of protons on target delivered at the Spallation Neutron Source (SNS), $r_{S/p}$ is the number of $S$ particles produced per proton, $\tau_S$ is the $S$ lifetime and $g_S(t)$ encodes the $N_S$ time dependence (normalized to one over each bunch cycle
characteristic of the proton beam at the SNS). The precise form of $g_S(t)$ depends on several factors such as the $S$ lifetime, time-dependent efficiency and the time-structure of the $S$ pulse. 

For antineutrinos the event rate is defined equivalently.

\subsection{Production and detection amplitudes} \vspace{0.4cm}
\label{sec:amplitudes}
Let us first discuss the detection process. The amplitude for neutrino scattering  on nuclei as a function of the nucleon-level EFT parameters in \cref{eq:FORM_Lnucleon}
reads
\begin{eqnarray}
{\cal M}^{D}_{jk} 
&\equiv& 
{\cal M}(\nu_k {\cal N}  \to \nu_j {\cal N}  )
\nn
&=&  
- {1 \over v^2}  \sqrt{E_{\cal N} E_{\cal N}'}  
\bigg ( Z [U^\dagger g^{\nu p} U]_{jk} {\cal F}_p(q^2)  + (A-Z) [U^\dagger  g^{\nu n} U]_{jk} {\cal F}_n(q^2)  \bigg ) \,
(\bar{u}_{\nu_j} \gamma^0\,P_L u_{\nu_k})\,\delta_{J_z J_z'}, ~~~~~
\end{eqnarray}
where $u_{\nu_{k,j}}$ are the Dirac wave functions of the incoming and outgoing neutrinos. 
Taking into account the current uncertainties affecting the COHERENT event rate, it is convenient to approximate neutron and proton form factors to be equal, ${\cal F}_p(q^2) = {\cal F}_n(q^2) \equiv {\cal F}(q^2)$, which allows us to write the detection amplitude in the more compact form:
\begin{eqnarray}
{\cal M}^{D}_{jk} 
&=&    -  {1 \over v^2}
\sqrt{E_{\cal N} E_{\cal N}'}\,
[ U^\dagger{\mathcal Q} \, U]_{jk}\,
(\bar{u}_{\nu_j} \gamma^0\,P_L u_{\nu_k})\,{\cal F}(q^2)\,\delta_{J_z J_z'}~, 
\end{eqnarray} 
where the (dimensionless and Hermitian) nuclear weak charge is defined as
\begin{equation}
\label{eq:RATE_calQ}
 [ {\mathcal Q}]_{\alpha \beta}  =  
  Z g^{\nu p}_{\alpha \beta}  + (A-Z) g^{\nu n}_{\alpha \beta} ~.
\end{equation} 
To work instead with different neutron and proton form factors would entail working with $q^2$-dependent weak charges (or, equivalently, with the $g^{\nu N}$ coefficients instead of the weak charges), which would make our subsequent discussion and intermediate results more cumbersome. Additionally, the final results for the quark-level WEFT Wilson coefficients are not expected to be affected by this approximation given current uncertainties. On the other hand, the $q^2$ dependence in the form factor can not be neglected in the studied recoil energy range. To describe it, we will make use of the the Helm parametrization~\cite{Helm:1956zz} (see~\aref{simdetails} for further details). Other recoil effects are numerically 
unimportant given the current experimental precision.

On the production side we have the 2-body leptonic pion decay $\pi^+ \to \ell_\alpha^+ \nu_k$ ($\alpha = e,\mu$) and the 3-body muon decay $\mu^+ \to \bar{\nu}_m e^+ \nu_k$.
Their amplitudes are given by\footnote{In the muon decay amplitude ${\cal M}^{P,\mu}_{mk}$ we omit the charged lepton subindex ($\alpha=e$ in this case) and we include both the neutrino and antineutrino mass eigenstate indices. This allows us to use the same notation for the neutrino and antineutrino rates.}
\begin{eqnarray}
{\cal M}^{P,\pi}_{\alpha k} 
&\equiv& 
\cM (\pi^+ \to \ell_\alpha^+ \nu_k)  
= 
 - i \, m_{\ell_\alpha}   f_{\pi^\pm}  \frac{V_{ud}}{v^2} [ {\cal P} U]_{\alpha k}^*(\bar{u}_{\nu_k} P_L v_{\ell_\alpha})~,
 \nn
{\cal M}^{P,\mu}_{mk} 
&\equiv &
{\cal M}(\mu^{+} \to \bar{\nu}_m e^+ \nu_{k} ) = -\frac{2}{v_0^2} [U^\dagger {\cal P}^T_L U]_{km} (\bar{v}_\mu \gamma^{\mu}P_{L} v_{\bar{\nu}_m})(\bar{u}_{\nu_{k}}\gamma_{\mu}P_L v_e)
\nn 
&&\hspace{3.6cm}
+\frac{4}{v_0^2} [U^\dagger {\cal P}^T_R U]_{km} (\bar{v}_\mu P_{L} v_{\bar{\nu}_m})(\bar{u}_{\nu_{k}}P_R v_e) ~,
\label{eq:muondecayamplitude}
\end{eqnarray}
where $v_\alpha$, $u_{\nu_k}$, 
$ v_{\bar{\nu}_m}$  are the Dirac spinor wave functions of the charged lepton, the neutrino, and the antineutrino,  respectively, 
and the pion decay constant $f_{\pi^\pm}$ is defined by $\langle 0 | \bar d \gamma_\mu \gamma_5 u(0) |\pi^+(p) \rangle   =   i p_\mu f_{\pi^\pm}$. Above we introduced the shorthand notation
\begin{eqnarray}
\label{eq:RATE_calP}
 [ {\cal P}]_{\alpha \beta} 
 &\equiv& 
\delta_{\alpha \beta}  +   [\epsilon^{ud}_L]_{\alpha \beta} -  [\epsilon^{ud}_R]_{\alpha \beta} 
-    [\epsilon^{ud}_P]_{\alpha \beta}   { m_{\pi^\pm}^2  \over   m_{\ell_\alpha}  (m_u + m_d) }     ~,
\nn
\left[{\cal P}_{L}\right]_{\alpha \beta} 
&\equiv& \delta_{\alpha \mu } \delta_{\beta e} + [\rho_{L}]_{\mu \alpha \beta e}~,
\nn
\left[{\cal P}_R\right]_{\alpha \beta}  
&\equiv& [\rho_R]_{\mu \alpha \beta e} \, .
\label{eq:Pdefs}
\end{eqnarray}
The transposition in the ${\cal P}_{L,R}$ matrices in \eref{muondecayamplitude} is defined such that it only affects the two neutrino indices when applied to the $\rho_{L,R}$ Wilson coefficients.

\subsection{Amplitudes product and phase space integrations} \vspace{0.4cm}
\label{sec:amplitudesproduct}
On the detection side, for neutrinos we find
\begin{eqnarray}
\label{eq:COH_detection}
\sum_j \int  d \Pi_{D'} {\cal M}_{jk}^D \bar{{\cal M}}_{jl}^D 
= [ U^\dagger {\mathcal Q}^2 U]_{l k} 
{
 {\cal F}(q^2)^2 
\,m_{\cal N} E_\nu (  T + m_{\cal N} )  \over 2 \pi v^4}    \bigg ( 
1  - { (m_{\cal N}   + 2 E_{\nu})  \,T \over  2 E_\nu^2 } 
 \bigg )~,~~  
\end{eqnarray}
where $T=E_{\cal N}' - m_{\cal N}$ is the (kinematic) nuclear recoil energy, and thus $q^2=-2M_{\cal N}\,T$. The same result holds for antineutrinos except for the ordering of the $(l,k)$ indices on the right-hand side.

On the production side, for pion-decay neutrinos we obtain
\begin{eqnarray}
\label{eq:COH_productionPion} 
\int  d \Pi_{P'}\cM_{\mu k}^{P,\pi} \bar{\cM}_{\mu l}^{P,\pi} 
&=& 
  [ {\cal P} U ]_{\mu l}   [U^\dagger  {\cal P}^\dagger]_{k \mu} 
  \frac{V_{ud}^2   f_{\pi^\pm}^2 m_\mu^2 ( m_{\pi^\pm}^2  - m_\mu^2 )^2  }{8 \pi m_{\pi^\pm}^2 v^4}  \delta \bigg ( E_\nu - E_{\nu,\pi} \bigg )~,
\end{eqnarray}
where $E_{\nu,\pi} = (m^2_{\pi^{\pm}}-m_{\mu}^2)/(2m_{\pi^{\pm}})$ is the energy of the neutrino emitted in the 2-body pion decay. On the other hand, for muon-decay neutrinos and muon-decay antineutrinos we find, respectively, the following results\footnote{For the sake of clarity we have written explicitly the sum over the (anti)neutrino of mass $m_k$ instead of considering it implicit inside the integral sign.}
\begin{eqnarray}
\sum_m  \int  d \Pi_{P'}{\cal M}^{P,\mu}_{mk}  \bar{{\cal M}}^{P,\mu}_{ml} 
&=& 
\frac{m_{\mu}^{5}}{\pi^{3} v_{0}^{4}}  \left(\frac{E_{\nu}}{m_{\mu}}\right)^{2} \left( \frac{1}{2} -\frac{E_{\nu}}{m_{\mu}} \right) \left[
U^{\dagger} P_{L}^{ T} {\cal P}_{L}^{ *}U
+  p_{RR}~U^{\dagger} P_R^T {\cal P}_R^{ *} U \right]_{k l}\,, 
\nn
\sum_m   \int  d \Pi_{P'}{\cal M}^{P,\mu}_{km} \bar{{\cal M}}^{P,\mu}_{lm} 
&=&
\frac{m_{\mu}^{5}}{3 \pi^{3}v_{0}^{4}} \left(\frac{E_{\nu}}{m_{\mu}}\right)^{2}\left( \frac{3}{4} -\frac{E_{\nu}}{m_{\mu}} \right) \left[
U^{\dagger} P_{L}^{*} {\cal P}_{L}^{ T}U
+ \bar{p}_{RR}~ U^{\dagger} P_R^{*} {\cal P}_R^{T}U \right]_{l k}   ~, \nn 
\label{eq:COH_productionMuon}
\end{eqnarray}
where $p_{RR} = 1/\bar{p}_{RR} = (3 m_{\mu}-4 E_{\nu})/(6 m_{\mu}-12 E_{\nu})$. We have neglected $\cO(m_e/m_\mu)$ corrections, which include the crossed $RL$ and $LR$ terms.

We would like to write these results in terms of the observable pion and muon decay widths,  since COHERENT uses those quantities in their flux predictions. They can be obtained from the expressions above integrating over the (anti)neutrino energy and working with equal neutrino indices in ${\cal M}^P$ and ${\bar{\cal M}}^P$, which are summed over. This gives
\begin{align}
\Gamma_{\pi\to\mu\nu} =& 
 {1 \over 2 m_{\pi^\pm}} \sum_{k} \int  dE_\nu \int  d \Pi_{P'}\cM^{P,\pi}_{\mu k}   {\bar \cM}^{P,\pi}_{\mu k}  
= \frac{V_{ud}^2   f_{\pi^\pm}^2 m_\mu^2 ( m_{\pi^\pm}^2  - m_\mu^2 )^2  }{16 \pi m_{\pi^\pm}^3  v^4}
  [ {\cal P}  {\cal P}^\dagger]_{\mu \mu}~,
\nnl 
    \Gamma_{\mu\to e\bar{\nu}\nu} 
    =& \frac{1}{2m_\mu} \sum_{k,m} \int dE_\nu 
    \int  d \Pi_{P'}{\cal M}^{P,\mu}_{mk}  \bar{{\cal M}}^{P,\mu}_{mk} 
    = 
    \frac{m_\mu^5}{384\pi^3 v_0^4}
    \, {\rm Tr} \left( {\cal P}_{L}{\cal P}_{L}^{\dagger}+ {\cal P}_{R}{\cal P}_{R}^{\dagger}\right ) 
    ~,
\label{eq:widths}
\end{align}
up to radiative and $m_e/m_\mu$ corrections. The muon decay width can be expressed as $\Gamma_{\mu\to e\bar{\nu}\nu} = m_\mu^5 / (384\pi^3 v^4)$, which represents the definition of $v$ (equivalently, of $G_F \equiv (\sqrt 2 v^2)^{-1}$), and remains valid in the presence of new physics.\footnote{
It is straightforward to see this implies $v_0^2 = v^2 \left(\text{Tr}\left( {\cal P}_L {\cal P}_L^\dagger + {\cal P}_R {\cal P}_R^\dagger \right)\right)^{1/2}_{\mu\simeq m_\mu}$, which is consistent with the discussion after~\cref{eq:Lagrangianmuon}.} 
In other words, in our input scheme the possible new physics contamination in the determination of the Fermi constant is absorbed into the parameter $v$, whose value is fixed by experiment. 
These NP effects are also absorbed in the Wilson coefficients $\epsilon_{V,A}^{qq}$ and $\epsilon^{ud}_X$, since $v$ is used in their definitions,~{\it cf.}~\cref{eq:LagrangianNC,eq:Lagrangianpion}. This can be seen explicitly matching to the Warsaw-basis SMEFT,~{\it cf.}~\cref{sec:EWPO,app:SMEFT}.

Using these results we can rewrite~\eref{COH_productionPion} and~\eref{COH_productionMuon} as
\small{\begin{eqnarray}
\int  d \Pi_{P'}\cM_{\mu k}^{P,\pi} \bar{\cM}_{\mu l}^{P,\pi} 
&=& 
  2\,m_{\pi^\pm} \Gamma_{\pi\to\mu\nu} \frac{[ {\cal P} U ]_{\mu l}   [U^\dagger  {\cal P}^\dagger]_{k \mu}} {[{\cal P}{\cal P}^\dagger]_{\mu\mu}}
  \,\delta \bigg ( E_\nu - E_{\nu,\pi} \bigg )~,
\label{eq:COH_productionMuon2}
\\
\sum_m  \int  d \Pi_{P'}{\cal M}^{P,\mu}_{mk}  \bar{{\cal M}}^{P,\mu}_{ml} 
&=& 
384\, \Gamma_{\mu\to e\bar{\nu}\nu} \left(\frac{E_{\nu}}{m_{\mu}}\right)^{2} \left( \frac{1}{2} -\frac{E_{\nu}}{m_{\mu}} \right) 
\frac{\left[ U^{\dagger} P_{L}^{ T} {\cal P}_{L}^{ *}U
+  p_{RR}~U^{\dagger} P_R^T {\cal P}_R^{ *} U \right]_{k l}}{{\rm Tr} \left( {\cal P}_{L}{\cal P}_{L}^{\dagger}+ {\cal P}_{R}{\cal P}_{R}^{\dagger}\right)}\,, 
\nn
\sum_m   \int  d \Pi_{P'}{\cal M}^{P,\mu}_{km} \bar{{\cal M}}^{P,\mu}_{lm} 
&=&
128\, \Gamma_{\mu\to e\bar{\nu}\nu} \left(\frac{E_{\nu}}{m_{\mu}}\right)^{2}\left( \frac{3}{4} -\frac{E_{\nu}}{m_{\mu}} \right) 
\frac{\left[ U^{\dagger} P_{L}^{*} {\cal P}_{L}^{ T}U
+ \bar{p}_{RR}~ U^{\dagger} P_R^{*} {\cal P}_R^{T}U \right]_{l k}}{{\rm Tr} \left( {\cal P}_{L}{\cal P}_{L}^{\dagger}+ {\cal P}_{R}{\cal P}_{R}^{\dagger}\right)}\,.
\nonumber
\end{eqnarray}}
\normalsize 
\hspace{-0.18cm}The NP effects that appear in the numerators (involving the PMNS matrix) are those contributing directly to the production amplitudes in~\eref{COH_productionPion} and~\eref{COH_productionMuon}, whereas those in the denominators enter indirectly because they affect the pion and muon decay widths. 
We will refer to these contributions as direct and indirect NP effects respectively.\footnote{Equivalently, indirect effects account for the NP contamination introduced through the extraction of $v$ and $V_{ud}f_{\pi^\pm}$ from the experimental pion and muon decay widths.} We stress that both contributions appear at the same order and are generated by the same EFT operators, so it is not consistent to include only the direct piece.
\subsection{Event rate} \vspace{0.4cm}
\label{sec:ratesubsection}
The total rate per recoil energy $T$ and time $t$ detected at COHERENT is given by: 
\begin{equation}
\label{eq:NumberofEventsPerRecoilEnergy}
 \frac{d N}{dt\,dT} = N_T \int d E_{\nu} \left( R_\mu^\pi + R_e^\mu + \bar{R}_e^\mu \right) \, ,
\end{equation}
where $R_\mu^\pi$ and $R_e^\mu$ ($\bar{R}_e^\mu$) denote the event rates mediated by neutrinos produced in pion decay and by (anti)neutrinos produced in muon decays respectively. These three event rates are obtained plugging the results of~\eref{COH_productionMuon2} in~\cref{eq:DifferentialRateEvents}. 
The integral over the neutrino energy is trivial in the pion-decay case since the neutrino energy is fixed, whereas for muon decay the lower integration limit is $E_\nu^{min}(T)=\frac{T}{2}\left(1+\sqrt{1+2\frac{m_{\cal N}}{T}}\right)$ (i.e. the minimum energy required to produce \cevns with a recoil energy $T$) and the upper one is simply $m_\mu/2$. 
Working in the $L=0$ limit and separating the events in prompt (i.e. produced in pion decays) and delayed (i.e. produced in muon decays), we  can rewrite Eq.~\eqref{eq:NumberofEventsPerRecoilEnergy} as 
\begin{eqnarray}
      \frac{dN}{dt\, dT} 
      = 
      g_\pi(t)\,\frac{dN^{\rm prompt}}{dT}
      + g_\mu(t)\,  \frac{dN^{\rm delayed}}{dT}~,
\label{eq:dNdTdt}
\end{eqnarray}
where $g_{\pi,\mu}(t)$ encode the $N_{\pi,\mu}$ time dependences (normalized to one over each bunch cycle), as discussed at the beginning of~\sref{rate}. The prompt and delayed components are given by
\begin{eqnarray}
      \frac{dN^{\rm prompt}}{dT} 
      &=& 
      n_{\rm POT}f^\pi_{\nu/p} \frac{ N_T\, {\cal F}(q^2)^2\, \left(m_{\cal N} +T \right)}{32\,\pi^2 v^4 L^2} 
      f_{\mu}(T) \tilde{Q}_\mu^2~,
\nn
     \frac{dN^{\rm delayed}}{dT}
     &=& 
     n_{\rm POT}f^\mu_{\nu/p} \frac{N_T\, {\cal F}(q^2)^2\, \left(m_{\cal N} +T \right)}{32\,\pi^2 v^4 L^2} 
     \left( 
     f_e(T) \,\tilde{Q}_e^2 
     + f_{\bar{\mu}}(T) \,\tilde{Q}_{\bar{\mu}}^2 
      \right) ~.
\label{eq:THprediction}
\end{eqnarray}
The generalized squared charges $\tilde{Q}_f^2$ are defined as the following positive and target-dependent quantities\footnote{Let us note that the RR term in $\tilde{Q}^2_{\bar{\mu}}$ ($\tilde{Q}^2_e$) is generated by non-standard effects in $\nu$-mediated ($\bar{\nu}$-mediated) events. Thus one should only identify the $f_{\bar{\mu}} \,\tilde{Q}_{\bar{\mu}}^2$ ($f_e \,\tilde{Q}_e^2$) terms in the delayed event rate with $\bar{\nu}$-mediated ($\nu$-mediated) events if the RR contributions are zero. The same caveat holds for the flavor indices $\bar{\mu}$ and $e$ used in those two terms, which only refer to the flavor of the mediating (anti)neutrino in the case of flavor-diagonal interactions in production (and no RR terms). \label{RRpieces}}
\begin{eqnarray}
\label{eq:Qdefs}
\tilde{Q}_\mu^2 
    &\equiv& \frac{\left[\mathcal{P} \mathcal{Q}^2 \mathcal{P}^{\dagger} \right]_{\mu\mu}}{\left(\mathcal{P}\mathcal{P}^{\dagger}\right)_{\mu\mu}}~,
    \nn
    \tilde{Q}_e^2
    &=&
    \frac{
    {\rm Tr} \left( {\cal P}_L^{*}\mathcal{Q}^2 P_{L}^{T}
    + {\cal P}_R^{T} \mathcal{Q}^2 P_R^{*} \right)
    }{
    {\rm Tr} \left( {\cal P}_{L}{\cal P}_{L}^{\dagger} +  {\cal P}_{R}{\cal P}_{R}^{\dagger}\right) }~,
\qquad 
\tilde{Q}_{\bar{\mu}}^2 
    \equiv 
    \frac{
{\rm Tr} \left( {\cal P}_L^T \mathcal{Q}^2 P_L^{*} + {\cal P}_R^{*}\mathcal{Q}^2 P_R^{T} \right)
}{
{\rm Tr} \left( {\cal P}_{L}{\cal P}_{L}^{\dagger} +  {\cal P}_{R}{\cal P}_{R}^{\dagger}\right)
}~.
\end{eqnarray}
The explicit form of the $f_f(T)$ functions, which is not very enlightening, can be found in~\eref{ffunctions}.   
Finally the number of (anti)neutrinos produced per proton via pion (muon) decay, $f^{\pi(\mu)}_{\nu/p}$, are given by
\begin{eqnarray}
f^{\pi}_{\nu/p} 
&\equiv& r_{\pi/p} \,\tau_\pi \,\Gamma_{\pi\to\mu\nu}
= r_{\pi/p} \,BR(\pi\to\mu\nu)_{\rm exp}
\approx r_{\pi/p}~,
\nn
f^{\mu}_{\nu/p} 
&\equiv& r_{\mu/p} \,\tau_\mu \,\Gamma_{\mu\to e\bar{\nu}\nu}
= r_{\mu/p} \,BR(\mu\to e\bar{\nu}\nu)_{\rm exp}~ \approx r_{\mu/p}~.
\label{eq:ffactors}
\end{eqnarray}
Finally, the allowed values for the recoil energy are $T\in \{0,T^a_{max}\}$ where $T_{max}^\text{prompt}=
2E_{\nu,\pi}^2/(m_{\cal N}+2E_{\nu,\pi})$ for prompt neutrinos and  $T_{max}^\text{delayed} = m_{\mu}^2/(2(m_{\mu}+m_{\cal N}))$ for the delayed ones.

The prompt and delayed event rates in~\eref{THprediction} represent one of the main results of this work, which is thus worth analyzing in some detail. First, let us note that the PMNS factors are not present anymore (they were removed using the unitarity condition $UU^\dagger=U^\dagger U=1$), which means that COHERENT is not sensitive to the PMNS mixing angles and phases, as expected in an $L\approx 0$ experiment. Secondly, let us note that expressions for the rates in~\eref{THprediction} are equal to the SM expressions except for the fact that the nuclear weak charge has been replaced by a generalized weak charge that (i) is different for muon neutrino, electron neutrino and muon antineutrino; and (ii) contains non-standard effects affecting detection {\it and} production. 
To put this in more explicit terms, we can re-write the prompt and delayed event rates as follows:
\begin{eqnarray}
      \frac{dN^{\rm prompt}}{dT} 
      &=& 
      N_T\, \int dE_\nu\,\frac{d\Phi_{\nu_\mu}}{dE_\nu}
      \,\frac{d\tilde{\sigma}_{\nu_\mu}}{dT}~, 
\nn
    \frac{dN^{\rm delayed}}{dT}
     &=& 
     N_T\, \int dE_\nu\,\Bigg( 
     \frac{d\Phi_{\nu_e}}{dE_\nu} \frac{d\tilde{\sigma}_{\nu_e}}{dT} 
     + \frac{d\Phi_{\bar{\nu}_\mu}}{dE_\nu} \frac{d\tilde{\sigma}_{\bar{\nu}_\mu}}{dT}  
     \Bigg)~,
\label{eq:THprediction2}
\end{eqnarray}
where the fluxes are the usual ones and the cross sections $\tilde{\sigma}_f$ are defined in the usual form but using the generalized charges $\tilde{Q}_f$, {\it cf.}~\aref{fluxesAndXsections}.

Even though we started from very general premises, the final result is similar to the SM formulas and to the usual NSI expressions (with NP only in the detection side), except for the introduction of the generalized weak charges.  
This unexpected result makes the phenomenological analysis very simple, since it represents a simple modification with respect to the standard approach in the previous literature. 
The conceptual change is however much deeper and one should keep in mind that in our general analysis the generalized weak charges contain non-standard lepton flavor violating effects affecting neutrino production. For instance, our general expression includes possible contributions through the process $\pi\to\mu\,\nu_\tau$,  despite not having introduced a $\nu_\tau$ flux in~\eref{THprediction2}. 
Thus one should keep in mind that the event rate in~\eref{THprediction2} is just a practical parametrization, but the factorization in fluxes and cross section, as well as the subindices $\nu_\mu$, $\nu_e$ and $\bar{\nu}_\mu$, do not have physical meaning except in the SM case and some simple BSM scenarios. 

In general, it is not possible to carry out a naive factorization of the event rate in~\eref{THprediction} in fluxes and cross sections, since there is a matrix multiplication between production and detection quantities in the generalized squared charges ${\tilde Q}_f^2$ in~\eref{Qdefs}. For simplicity let us consider the case of pion decay production and \cevns detection, where one can easily see that 
$R_\mu^\pi \neq \sum_{\alpha\beta} d\Phi (\pi\!\to\!\mu\nu_\alpha) / (dE_\nu dt) \times  d\sigma(\nu_\alpha\,{\cal N} \!\to\! \nu_\beta\,{\cal N}) / dT$ 
simply because 
$\left[\mathcal{P} \mathcal{Q}^2 \mathcal{P}^{\dagger} \right]_{\mu\mu} \neq \sum_{\alpha\beta} |\mathcal{P}_{\mu\alpha}|^2 \, |\mathcal{Q}_{\alpha\beta}|^2$.  

Before discussing some interesting specific cases, let us mention briefly how the analysis is modified if we consider different $q^2$-dependent form factors for neutron and proton, i.e., ${\cal F}_n (q^2) \neq {\cal F}_p (q^2) \neq {\rm const}$. In that case, the ${\tilde Q}^2_f$ parameters are not convenient objects to summarize experimental results, because they become $q^2$ dependent. 
In the COHERENT rate expression, the product of the weak charge matrix squared and the form factor squared, ${\cal Q}^2\,({\cal F}(q^2))^2$, would be replaced by the matrices $(g^{\nu p})^2$, $(g^{\nu n})^2$, $g^{\nu p}  g^{\nu n}$, and $g^{\nu n} g^{\nu p}$, accompanied by the appropriate powers of the $Z{\cal F}_p(q^2)$ and $(A-Z){\cal F}_n(q^2)$ functions. 
Thus, we would go from 3 parameters per target (${\tilde Q}^2_{f= e, \mu, \bar \mu}$) to 12 target-independent parameters. They are reduced to 9 parameters if the (production and detection) NP parameters are real, because the generalized squared charges obtained ``replacing'' ${\mathcal Q}^2$ with $g^{\nu p}g^{\nu n}$ and $g^{\nu n} g^{\nu p}$ are equal.

\subsection{Interesting limits} \vspace{0.4cm}
\label{sec:limits}

\noindent {\bf SM limit}.  
If all NP effects are switched off we recover the SM prediction, with a single nuclear weak charge
\begin{equation}
\label{eq:RATE_Qsm}
   Q_{SM}^2 \equiv   \tilde{Q}_\mu^2  = \tilde{Q}_{\bar{\mu}}^2 
    \simeq 
    \tilde{Q}_e^2 , 
    \end{equation}
where 
\begin{equation}
\label{eq:RATE_QSM}
Q_{SM}  = 
Z [{\mathcal Q_w^p}]_{\mu\mu}^{\rm SM}
+ (A-Z) [{\mathcal Q_w^n}]_{\alpha\alpha}^{\rm SM}
    \end{equation}
and the nucleon weak charges are given in \cref{eq:FORM_weakChargesSM}. 
Note that in principle the muon and electron weak charges have slightly different values, 
however the difference is irrelevant given the COHERENT accuracy, {\it cf.} Eq.~\eqref{eq:FORM_weakChargesSM}. 
In our analysis we will take the muon weak charge as the reference value.  
The SM scenario has of course been thoroughly studied before~\cite{Erler:2013xha,Tomalak:2020zfh}. 
For each target nucleus there is a single quantity, $Q_{SM}$, to be extracted from experiment, which is predicted in the SM in function of the weak mixing angle. 
Thus, in the SM limit coherent elastic neutrino-nucleus scattering can be regarded as a probe of the weak mixing angle, see e.g.~\cite{Lindner:2016wff, Papoulias:2017qdn, Canas:2018rng,Huang:2019ene,AtzoriCorona:2022qrf, DeRomeri:2022twg}. 
One should remark however that other probes, such as $Z$-pole physics, atomic parity violation, or parity-violating electron scattering have currently a much better sensitivity to the weak mixing angle.

\noindent {\bf New physics in production}. 
We move to the case where new physics affects the COHERENT observables via neutrino production in pion and muon decay.  
The matrices  ${\cal P}$ and ${\cal P}_{L,R}$, which encode these effects, are allowed to be completely generic.  
On the other hand,  we assume here that we can ignore new physics in detection. 
This implies that the weak charge ${\cal Q}$ defined in \cref{eq:RATE_calQ} is proportional to the unit matrix and thus it commutes with ${\cal P}$ and ${\cal P}_{L,R}$. 
It then follows from \eref{Qdefs} that the new physics production effects completely cancel out in the generalized weak charges, and we recover the SM limit in \cref{eq:RATE_Qsm} with a single nuclear weak charge. 
All in all, COHERENT data are completely insensitive to new physics affecting only the  CC semileptonic and leptonic  WEFT operators  due to cancellations between direct and indirect new physics effects.  This observation invalidates the bounds found in Ref.~\cite{Khan:2021wzy}, where the indirect NP effect was not taken into account. 

A more intuitive way of understanding this null sensitivity is the following. 
The CC operators in \cref{eq:Lagrangianpion} certainly affect the pion decay rate to muon and neutrino, but they do not distort the kinematics. Their effect has been fully absorbed into the experimental value of BR$(\pi\to\mu\nu)$, which is used to calculate the neutrino flux. Moreover, in the particular  case at hand BR$\,\approx \!1$, that is, $\sim$100\% of the pions will decay to muon and neutrino for any reasonable values of $[\epsilon_{L,R,P}]_{\alpha\beta}$. 
Similarly, new leptonic operators in  \cref{eq:Lagrangianmuon} affect the muon decay rate, but their effect has been fully absorbed into the experimental value of the Fermi constant.

Our work is the first one that takes into account the direct and indirect effects in production, as well as the possible cancellations. 
Let us stress however that new physics in production cannot be ignored completely:  its effects do not cancel out if there are accompanying new physics effects in detection.

\noindent {\bf New physics in detection}.  
If we neglect NP in production our expressions reduce to those found previously in the NSI literature~\cite{Barranco:2005yy,Scholberg:2005qs,Lindner:2016wff}, where two free parameters are present (instead of three). Namely
\begin{align}
\label{eq:onlydetection}
    \tilde{Q}_\mu^2 
    =& \tilde{Q}_{\bar{\mu}}^2 = \left[{\cal Q}^2\right]_{\mu\mu}
    = \sum_{\alpha} | [{\cal Q}]_{\alpha \mu}|^2 
    =
    \sum_{\alpha} \bigg| Z g^{\nu p}_{\alpha \mu}  + (A-Z) g^{\nu n}_{\alpha \mu}   \bigg|^2
    \nnl =& 
    4 \sum_{\alpha} \bigg[ (A+Z) (g_V^{uu} \,\mathbb{1} + \epsilon^{uu})  + (2A-Z) (g_V^{dd}\,\mathbb{1} + \epsilon^{dd})  \bigg]_{\alpha \mu}^2
    ~,\nnl
    \tilde{Q}_e^2 
    =& 
    \left[{\cal Q}^2\right]_{ee}
    = \sum_{\alpha} | [{\cal Q}]_{\alpha e}|^2
    =
    \sum_{\alpha} \bigg| \bigg ( Z g^{\nu p}_{\alpha e}  + (A-Z) g^{\nu n}_{\alpha e}  \bigg ) \bigg|^2  
    \nnl =& 
    4 \sum_{\alpha} \bigg[ (A+Z) (g_V^{uu}\,\mathbb{1} + \epsilon^{uu})  + (2A-Z) (g_V^{dd}\,\mathbb{1} + \epsilon^{dd})  \bigg]_{\alpha e}^2~. 
\end{align}

\noindent {\bf Linear new physics terms}.   
Finally, let us consider the case where only corrections linear in non-standard Wilson coefficients are kept. 
At this order, direct and indirect BSM effects in production cancel (even if there are NP in detection) and we arrive at a linearized version of~\eref{onlydetection}:
\begin{eqnarray}
    \tilde{Q}_\mu^2 
    &=& \tilde{Q}_{\bar{\mu}}^2 = \left[{\cal Q}^2\right]_{\mu\mu}
    = Q_{SM}^2 + 2\,Q_{SM}\,[\delta{\cal Q}]_{\mu \mu}
    ~,\nn
    \tilde{Q}_e^2 
    &=& \left[{\cal Q}^2\right]_{ee}
    = Q_{SM}^2 + 2\,Q_{SM} \,[\delta{\cal Q}]_{ee}
    \label{eq:linear1}
\end{eqnarray}
where $Q_{SM}$ is given in \cref{eq:RATE_QSM}, and 
\begin{equation}
\label{eq:deltaQ} 
 \delta {\cal Q} 
= Z \,\delta g^{\nu p}  + (A-Z) \,\delta g^{\nu n}
    = 2 \bigg ( (A+Z) \epsilon^{uu}  + (2A-Z) \epsilon^{dd} \bigg ) 
    ~, 
\end{equation}
where we defined $\delta g^{\nu N} \equiv g^{\nu N} - [g^{\nu N}]_{SM}$.
That is, for a given target, COHERENT is linearly sensitive only to two linear combinations of the four WEFT Wilson coefficients: $\epsilon^{uu}_{\mu\mu}$, $\epsilon^{dd}_{\mu\mu}$, $\epsilon^{uu}_{ee}$, and $\epsilon^{dd}_{ee}$, 
which describe flavor-diagonal NC interactions between neutrinos and quarks.\footnote{
This statement depends on the definition of the WEFT coefficients and on the input scheme. In particular, if one uses $v_0$ (instead of $v$) in~\cref{eq:LagrangianNC}, then we would find that COHERENT is also linearly sensitive to the CC interaction $[\rho_L]_{e\mu\mu e}$ because of its effect on the muon decay, and hence on $G_F$ (or $v$), which is used to calculate the \cevns cross section. 
In our approach such effects have been absorbed inside the NC coefficients $\epsilon^{qq}_{ll}$.
Both approaches are of course equivalent, as can be seen explicitly when they are matched to the Warsaw-basis SMEFT, ~{\it cf.}~\cref{sec:EWPO,app:SMEFT}. Last, we note that this caveat also applies to the previous discussion about NP in production.}

\section{Experimental input} \vspace{0.4cm}
\label{sec:exp}

The COHERENT collaboration uses a series of detectors to detect neutrinos produced by the Spallation Neutron Source (Oak Ridge National Laboratory) through CE$\nu$NS. At this facility, high-energy protons with $E\sim 1$ GeV hit a mercury target to produce $\pi^+$ and $\pi^-$. The latter are absorbed, whereas the positive pions decay at rest into the \emph{prompt} neutrinos and positive muons. The latter correspond to the source particles of the \emph{delayed} (anti)neutrinos. 

We will analyze the two available measurements of the CE$\nu$NS interaction performed by this experiment: one performed on a liquid argon target (LAr)~\cite{COHERENT:2020iec} and another one using a target consisting of a mixture of cesium and iodine (CsI)~\cite{COHERENT:2021xmm}. The input needed to calculate the number of prompt and delayed events for these two measurements through~\eref{THprediction} is summarized in~\tref{dNdTinput}. The number of target particles $N_T$ is obtained as the ratio of the active mass of the detector $m_{\text{det}}$ and the mass of the interacting nuclei. 
For the CsI measurement, we treat cesium and iodine as a single nucleus with $(Z, A)=(54, 130)$. This will allow us to analyze CsI data in terms of only 3 charges (instead of 6) and we do not expect it to have any impact in the final bounds on the WEFT Wilson coefficients, since the atomic numbers for Cs and I are very similar (namely $Z_{Cs}=55,Z_I=53$). 
\begin{table}
\begin{center}
\begin{tabular}{ccc}
\hline
{Parameter} &  {CsI}~\cite{COHERENT:2021xmm} &   {LAr}~\cite{COHERENT:2020iec}  \\  \hline 
$(Z,N)$     &   $(54,130)$    &   $(18,40)$ \\
$n_{\rm POT}$   &   $31.98 \times 10^{22}$   &   $13.77 \times 10^{22}$ \\
$f_{\nu/p}$   &   0.0848    &   0.09 \\
$m_{\text{det}}$ (kg)  &   14.6   &   24.4 \\
$L$ (m)        &   19.3   &   27.5 \\ \hline
\end{tabular}
\end{center}
\caption{Input used to calculate the differential number of prompt and delayed events $dN/dT$ for each setup, taken directly from the corresponding experimental article. Following those references we approximate the number of pions and muons per proton to be equal $f_{\nu/p}\equiv f^\pi_{\nu/p}=f^\mu_{\nu/p}$.}
\label{tab:dNdTinput}
\end{table}

Naively, one only has to integrate the expression in~\eref{THprediction} over the recoil energy $T$ in each bin to obtain the expected number of prompt/delayed events (for a given value of the generalized weak charges $\tilde{Q}_f$), that is
\begin{eqnarray}
\label{eq:naive}
N^a_i = \int_{\rm i-th\,bin} \frac{dN^a}{dT} \,dT~,~\qquad(a={\rm prompt,delayed})~.
\end{eqnarray}
However, this simple step needs to be modified to take into account various experimental effects.
The first thing to consider is that COHERENT does not measure its events in nuclear recoil energy ($T$), but in electron-equivalent recoil energy ($T_{ee}$). These two magnitudes are related as follows
\begin{equation}
\label{eq:QFElectronEquivalentRecoilEnergy}
T_{ee}=\text{QF}(T)\times  T\, ,
\end{equation}
where $\text{QF}(T)$ is the so-called quenching factor, which depends on the detector and the recoil energy $T$, as we indicated explicitly. 
Moreover, one has to introduce an energy resolution function ${\cal R}(T_{ee}^{\text{rec}},T_{ee})$, which relates the true value of the electron-equivalent recoil energy, $T_{ee}$, with the reconstructed one, $T_{ee}^{\text{rec}}$, that is registered at the detector. 
Finally, the efficiency of the detector, $\epsilon(T_{ee}^{\text{rec}})$, should also be taken into account. These considerations are collected in the following modified expression for the number of prompt and delayed events in the $i$-th $T_{ee}^{\text{rec}}$ bin~\cite{COHERENT:2020ybo, COHERENT:2021xmm}
\begin{equation}
\label{eq:NumberofEventsWithExperimentalEffects}
N^a_{i}=\int_{T_{ee}^{\text{rec}, i}}^{T_{ee}^{\text{rec}, i+1}} dT_{ee}^{\text{rec}} \,\epsilon(T_{ee}^{\text{rec}}) \int_{T_{\text{min}}}^{T^a_{\text{max}}} dT \,{\cal R}(T_{ee}^{\text{rec}},T_{ee}(T)) \frac{dN^a}{dT}\left( T \right)\, ,
\end{equation}
where $a={\rm prompt/delayed}$. The theoretical prediction for the (differential) number of events, $dN^a/dT$, is given in terms of the nuclear recoil energy $T$, as provided in~\eref{THprediction}. Note that we have expressed the energy resolution ${\cal R}(T_{ee}^{\text{rec}},T_{ee})$ in terms of $T$ instead of $T_{ee}$ by means of the QF. That way, the integral over $T$ allows us to go from the nuclear recoil energy $T$ to the reconstructed electron-equivalent recoil energy $T_{ee}^{\text{rec}}$. The integration limits, the quenching factor and the efficiency and energy resolution functions that we use for each dataset are discussed in~\aref{simdetails}.

Additionally, the CsI analysis presents  the data in terms of the number of photoelectrons (PE) that are recorded for each event instead of using the electron-equivalent recoil energy.  These two magnitudes are simply connected by $\text{PE}=\text{LY}\times T_{ee}^{\text rec}$, where the light yield LY is the number of PE produced by an electron recoil of one keV. 
Thus, the general expression in~\eref{NumberofEventsWithExperimentalEffects} holds also for CsI with the replacement $T_{ee}^{\text rec}\to$ PE. 

Above we presented the expression for the prompt and delayed events, which have to be summed to provide the observed number of \cevns events. 
Equivalently this result can be obtained integrating $t$ in~\eref{dNdTdt} over the entire bunch cycle and taking into account that the $g_a(t)$ functions are normalized to one. If instead one is interested in the double distribution in nuclear recoil energy $T$ and time $t$, then we will have
\begin{eqnarray}
N_{ij}^{\rm signal} &=& g_j^{\rm prompt} N_i^{\rm prompt} + g_j^{\rm delayed} N_i^{\rm delayed}~,
\label{eq:NijTH}
\end{eqnarray}
where we introduced a second index $j$ that refers to the $j$-th time bin. The $g_j^a$ factors are the prompt/delayed probability distributions for the timing of the events (calculated integrating the $g_a(t)$ functions over the $j$-th time bin), which can be extracted from the COHERENT publications, {\it cf.}~\aref{simdetails}.

Finally, one also has to include background events and nuisance parameters to parametrize the most relevant uncertainty sources. Thus, the predicted number of events has the following generic form
\begin{eqnarray}
N_{ij}^{\rm th}\left( \vec{Q}_{\cal N}^2;\vec{x}\right)=N_{ij}^{\rm signal}\left( \vec{Q}_{\cal N}^2\right) \left(1 +h_{ij}^{\rm signal}(\vec{x})\right) + \sum_a N_{ij}^{\text{bkg},a}\left(1 +h_{ij}^{\text{bkg},a}(\vec{x})\right)~,
\label{eq:NiTH}
\end{eqnarray}
where we have indicated explicitly the dependence of the expected number of \cevns events on the generalized squared weak charges for the nucleus ${\cal N}$, denoted by $\vec{Q}_{\cal N}^2\equiv \{ \tilde{Q}_\mu^2,\tilde{Q}_{\bar{\mu}}^2, \tilde{Q}_e^2\}$. 
The expected number of background events of type $a$, denoted by $N_{ij}^{\text{bkg},a}$, is obtained following COHERENT prescriptions, as described in detail in~\aref{simdetails}. The typical background sources are the steady state (SS) background, the neutrino-induced neutron (NIN) background and the beam-related neutron (BRN) background, although the way each of them is characterized differs slightly in every measurement. Finally, the $h_{ij}^{\text{signal/bkg},a}(\vec{x})$ functions are linear in the nuisance parameters $\vec{x}$ and vanish at their central values $\vec{x}=\vec{0}$. The specific form of these functions for each experimental analysis is given in~\aref{simdetails} following once again the COHERENT prescription. 

In our numerical analysis we use the 2D distributions (in time and recoil) measured in the CsI and Ar works~\cite{COHERENT:2020iec,COHERENT:2021xmm}. For each of these two datasets we work with a Poissonian chi-squared function with the following generic form
\begin{eqnarray}
\label{eq:chi2}
\chi^2 = 
    \sum_{i, j} 2 &&\!\!\!\!\left( -N_{ij}^{\rm exp}+N_{ij}^{\rm th}\left( \vec{Q}_{\cal N}^2;\vec{x}\right) + N_{ij}^{\rm exp}~\text{ln}\left( \frac{N_{ij}^{\rm exp}}{N_{ij}^{\rm th}\left( \vec{Q}_{\cal N}^2;\vec{x}\right)}\right)\right) + \sum_n \left(\frac{x_n}{\sigma_n}\right)^2,\,~~~~~~
\end{eqnarray}
where $\sigma_n$ is the uncertainty of the $x_n$ nuisance parameter. 
All in all we have 52x12 bins in CsI and 4x10 bins in LAr, {\it cf.}~\aref{simdetails} for further details. 

\section{Numerical analysis} 
\label{sec:numerical}
\vspace{0.3cm}

\subsection{Generalized nuclear weak charges} \vspace{0.4cm}
In this section we present the results of the analysis of LAr and CsI data in terms of the generalized nucleus-dependent weak charges ${\tilde Q}_f$. Since the event rate depends quadratically on these charges, it is convenient to work with their squared values ${\tilde Q}_f^2$. 

\subsubsection{Argon charges} 
We carry out a 2D fit to the nuclear recoil energy and time distributions, as described in the previous section. This fit has 40 experimental inputs(with their associated uncertainties and backgrounds), 9 nuisance parameters (with their uncertainties) and the three $\tilde{Q}_f^2$ charges that contain the UV information. We find that the distribution of these three charges is approximately Gaussian, with the following results:\footnote{The squared charges have to be positive. Our results are approximately Gaussian (before imposing this prior) so we will present them in the usual form, i.e., central values, diagonal errors, and correlation matrix. It is straightforward to impose the $\tilde{Q}_f^2 \geq 0$ constraint a posteriori. This will induce a large non-Gaussianity if and only if the (Gaussian) errors are large.}
\begin{equation}
\label{eq:3qAr}
\left( \begin{matrix}\tilde{Q}_{\mu}^2 \\ \tilde{Q}_{\bar{\mu}}^2\\ \tilde{Q}_e^2
\end{matrix} \right)_{\!\!\!\!\rm Ar}
\frac{1}{Q^2_{\rm SM, Ar}}
= 
\left( \begin{matrix}
1.00 \pm 0.82 \\ 
0.4 \pm 6.2 \\ 
1.9 \pm 8.2 \end{matrix} \right)~ \quad \rho 
= 
\left( \begin{matrix}
1 &  0.29 & -0.31 \\ 
0.29 & 1 & -0.99 \\ 
-0.31 & -0.99 & 1 
\end{matrix} \right)~,
\end{equation}
along with the nine nuisance parameters that we do not display. 
We have normalized the results using the SM value, $\mathcal{Q}^2_{\rm SM, Ar} = 461$, with an associated small error that can be neglected for the purpose of this work.
The results in \cref{eq:3qAr} are in perfect agreement with the SM prediction $\tilde{Q}_f^2/Q^2_{SM} = 1$. 
We can disentangle the first charge, ${\tilde Q}_\mu^2$, from the other two thanks to its different time dependence: the former enters the rate via (prompt) pion decay, while the latter do it via (delayed) muon decay. On the other hand the recoil energy distribution only allows for a mild separation of ${\tilde Q}_e^2$ and ${\tilde Q}_{\bar\mu}^2$. 
To get rid of the large correlations, which obscure the strength of the results, let us rewrite them as the following uncorrelated bounds
\begin{eqnarray}
\label{eq:ArBoundsModCharges}
\left( 
\begin{matrix}
-0.14 & -3.48 & 4.62  \\ 
-0.69 & 0.98 & 0.71 \\
0.55 & 0.25 & 0.20 
\end{matrix} \right)
\left( \begin{matrix}\tilde{Q}_{\mu}^2 \\ \tilde{Q}_{\bar{\mu}}^2\\ \tilde{Q}_e^2
\end{matrix} \right)_{\!\!\!\!\rm Ar} \frac{1}{Q^2_{\rm SM, Ar}}
= \left( \begin{matrix}
6 \pm 59 \\
\bf{1.0 \pm 1.2} \\ 
\bf{1.03 \pm 0.48}
\end{matrix} \right)~,
\end{eqnarray}
where we have highlighted the most stringent constraints (note that the SM prediction is one by construction). 
A particularly interesting case is the SM supplemented by the following contributions: $ \tilde{Q}_{\mu}^2 = \tilde{Q}_{\bar{\mu}}^2 =\left[\mathcal{Q}^2 \right]_{\mu\mu}$ and $\tilde{Q}_e^2 = \left[\mathcal{Q}^2 \right]_{ee}$, which is the most general setup that we can have when considering NP only at detection or in a linear analysis (see~\cref{sec:limits}). In this case we find:
\begin{equation}
\label{eq:2qAr}
\left( \begin{matrix}
 \left[\mathcal{Q}^2 \right]_{\mu\mu}  \\  \left[\mathcal{Q}^2 \right]_{ee}  \\ 
\end{matrix} \right)_{\!\!\!\rm Ar} 
\frac{1}{Q^2_{\rm SM, Ar}}
= \left( \begin{matrix}1.02 \pm 0.81 \\ 1.1 \pm 1.9  
\end{matrix} \right)~\quad \rho = \left( \begin{matrix}1 & -0.68 \\ -0.68 & 1   \end{matrix} \right)~,
\end{equation}
which can be rewritten as the following uncorrelated bounds:
\begin{align}
\label{eq:Ar2Dfit}
-0.48\,  \left(\left[\mathcal{Q}^2 \right]_{\mu\mu}/Q_{SM}^2\right)_{\rm Ar} + 1.48\,  \left(\left[\mathcal{Q}^2 \right]_{ee}/Q_{SM}^2\right)_{\rm Ar} =&\, 1.1 \pm 3.0~,
\nnl 
0.75\,  \left(\left[\mathcal{Q}^2 \right]_{\mu\mu}/Q_{SM}^2\right)_{\rm Ar} + 0.25\,  \left(\left[\mathcal{Q}^2 \right]_{ee}/Q_{SM}^2\right)_{\rm Ar} =& \,\bf{1.03 \pm 0.45}.
\end{align}
The results of this 2D fit are shown in the left panel of~\cref{fig:2chargesfit}, where we also present the allowed regions if one only uses the total number of events, the energy distribution or the time distribution.

Finally in the SM scenario there is only one weak charge ($\tilde{Q}_{\mu}^2 = \tilde{Q}_{\bar{\mu}}^2 = \tilde{Q}_e^2 \equiv Q^2$), and we find $\left( Q^2/Q_{SM}^2\right)_{\rm Ar} = 1.03 \pm 0.45$.
%

\subsubsection{CsI charges} 
We have carried out a fit to the 2D distributions (nuclear recoil energy and time) provided in the CsI analysis. 
This fit has 624 experimental inputs (with their associated uncertainties and backgrounds), 4 nuisance parameters (with their uncertainties) and the three CsI generalized weak charges.
Once again we find that the distribution of the  $\tilde{Q}_f^2$ charges is approximately Gaussian, with the following results:
\begin{equation}
\label{eq:3qCsI}
\left( \begin{matrix}\tilde{Q}_{\mu}^2 \\ \tilde{Q}_{\bar{\mu}}^2\\ \tilde{Q}_e^2
\end{matrix} \right)_{\!\!\!\!\rm CsI} \frac{1}{Q^2_{\rm SM, CsI}}
= \left( \begin{matrix}1.33 \pm 0.35 \\ -1.4 \pm 1.5 \\ 4.4 \pm 2.3
\end{matrix} \right)
~ \quad \rho = \left( \begin{matrix}1 &  0.12 & -0.09  \\ 0.12 & 1 & -0.98 \\ -0.09 & -0.98 & 1   \end{matrix} \right)~,
\end{equation}
along with the nuisance parameters. Here, $Q^2_{\rm SM, CsI} = 5572$. As in the LAr case, we can separate much better $\tilde{Q}_\mu^2$ from the other two charges thanks to the use of the time information. 
The results can be rewritten as the following uncorrelated bounds:
\begin{eqnarray}
\left( 
\begin{matrix}
-0.04 & -1.80 & 2.85 \\
0.80 & 0.12 & 0.09   \\
-0.15 & 0.71 & 0.45   
\end{matrix} \right)
\left( \begin{matrix}\tilde{Q}_{\mu}^2 \\ \tilde{Q}_{\bar{\mu}}^2\\ \tilde{Q}_e^2
\end{matrix} \right)_{\!\!\!\!\rm CsI} \frac{1}{Q^2_{\rm SM, CsI}}
= \left( \begin{matrix}
15.1 \pm 9.1 \\ 
{\bf 1.28 \pm 0.28} \\
 {\bf 0.81 \pm 0.19 }
\end{matrix} \right)~,
\end{eqnarray}
where we find again good agreement with the SM predictions (one). 

\begin{figure}
\centering
\includegraphics[scale=0.401]{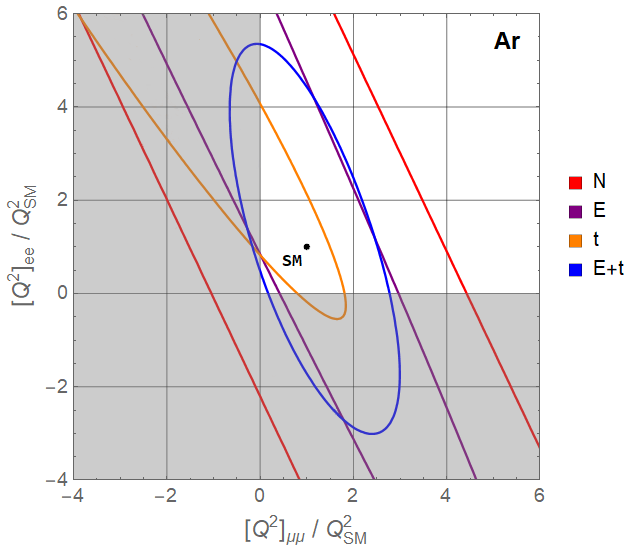}
\hspace{0.35cm}
\includegraphics[scale=0.4]{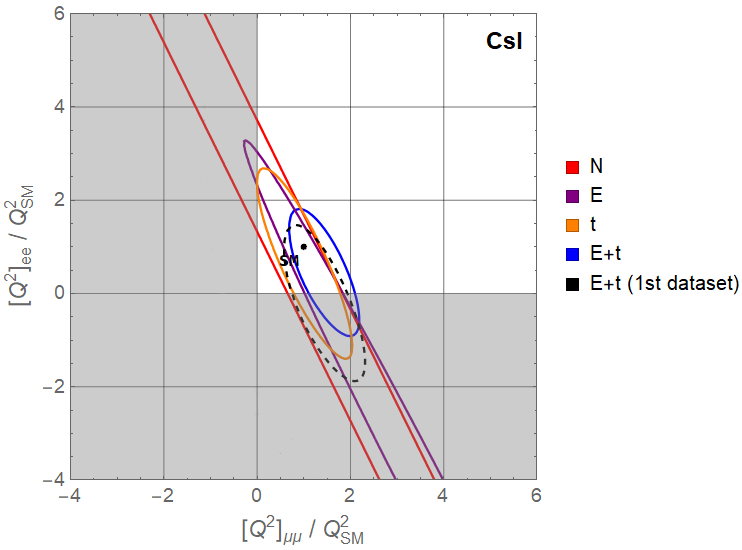}
\caption{90\% CL allowed regions ($\Delta\chi^2=4.61$) for the squared weak charges  (normalized to their SM values) using Ar (left) and CsI (right) data and assuming only NP at detection. For each dataset we include information from the total number of events (N), from the recoil energy distribution (E), from the timing distribution (t) and from the 2D distribution (E+t). In the right panel we also show the contour obtained using only the first CsI dataset~\cite{COHERENT:2017ipa}. The shaded area indicates the unphysical region (negative squared charges).}
\label{fig:2chargesfit}
\end{figure}

Considering only NP at detection we find
\begin{equation}
\label{eq:2qCsI}
\left( \begin{matrix}
\left[\mathcal{Q}^2 \right]_{\mu\mu} \\ 
\left[\mathcal{Q}^2 \right]_{ee}   
\end{matrix} \right)_{\!\!\!\rm CsI} \frac{1}{Q^2_{\rm SM, CsI}} = 
\left( \begin{matrix} 1.32\pm 0.34 \\ 0.44 \pm 0.61  \end{matrix} \right)
~\quad \rho = \left( \begin{matrix}1 & -0.73 \\ -0.73 & 1   \end{matrix} \right)~,
\end{equation}
which can be re-written as the following uncorrelated bounds: 
\begin{eqnarray}
-0.84\, \left(\left[\mathcal{Q}^2 \right]_{\mu\mu}/Q_{SM}^2\right)_{\rm CsI} + 1.84  \left(\left[\mathcal{Q}^2 \right]_{ee}/Q_{SM}^2\right)_{\rm CsI} &=& ({\bf -0.3 \pm 1.4})~,
\nn
0.69 \left(\left[\mathcal{Q}^2 \right]_{\mu\mu}/Q_{SM}^2\right)_{\rm CsI} +0.31  \left(\left[\mathcal{Q}^2 \right]_{ee}/Q_{SM}^2\right)_{\rm CsI} &=& ({\bf1.04 \pm 0.16})~.
\end{eqnarray}
The results of this 2D fit are shown in the right panel of~\cref{fig:2chargesfit}. We present as well the allowed regions if one uses only the total number of events, the energy distribution or the time distribution. Finally, we also show the result obtained using the full 2D distribution of the first CsI dataset~\cite{COHERENT:2017ipa}, which is in good agreement with Fig.~6 in Ref.~\cite{Coloma:2019mbs}. One observes a clear improvement when the entire CsI dataset is used.

Finally in the SM scenario there is only one weak charge ($\tilde{Q}_{\mu}^2 = \tilde{Q}_{\bar{\mu}}^2 = \tilde{Q}_e^2 \equiv Q^2$), and we find $\left( Q^2/Q_{SM}^2\right)_{\rm CsI} = 1.04 \pm 0.16$.

\subsection{WEFT coefficients}
\vspace{0.3cm}
In this section we move to consider the constraints on the nucleon- and quark-level EFT Wilson coefficients, stemming from our analysis of LAr and CsI CE$\nu$NS data.

\subsubsection{Linear BSM expansion} 
As shown in~\eref{linear1}, at linear order in New Physics there are only 2 free parameters per target: the flavor diagonal muon and electron weak charges, $[{\cal Q}^2]^{\cal N}_{\mu\mu,ee}$. Using~\eref{deltaQ} we can express our bounds on the four weak charges (with ${\cal N}=$ Ar, CsI, see \cref{eq:2qAr,eq:2qCsI}) as bounds on the four nucleon-level EFT Wilson coefficients $\delta g^{\nu p}_{ee,\mu\mu}$ and $\delta g^{\nu n}_{ee,\mu\mu}$. 
We find that we can constrain the following orthogonal and uncorrelated linear combinations of couplings:
\begin{eqnarray}
\left( \begin{matrix}
0.55 &  -0.19 & -0.77 & 0.26 \\ 
-0.18 & -0.56 & 0.26 & 0.76 \\ 
0.74 & -0.32 & 0.53 & -0.24 \\ 
0.32 & 0.74 & 0.23 & 0.54 
\end{matrix} \right) \left( \begin{matrix}
\delta g^{\nu p}_{ee} \\
\delta g^{\nu p}_{\mu\mu} \\
\delta g^{\nu n}_{ee} \\
\delta g^{\nu n}_{\mu\mu} 
\end{matrix} \right)  = \left( \begin{matrix}
4 \pm 12 \\
0.5 \pm 3.4 \\
0.22 \pm 0.25 \\
-0.021 \pm 0.079 
\end{matrix} \right)~.
\end{eqnarray}
At the quark level, using the map in \cref{eq:FRAME_nucleonToWEFTmap}, we can  translate these results into constraints on the following orthogonal and uncorrelated linear combinations of WEFT Wilson coefficients:
\begin{align}
\label{eq:epsilonbounds}
\left( \begin{matrix}
0.63 & -0.70 & -0.22 & 0.24 \\ 
0.21 & -0.24 & 0.63 & -0.70 \\ 
-0.68 & -0.61 & 0.30 & 0.27 \\ 
0.30 & 0.27 & 0.68 & 0.61
\end{matrix} \right) \left( \begin{matrix}
\epsilon^{dd}_{ee}  \\
\epsilon^{uu}_{ee}  \\
\epsilon^{dd}_{\mu\mu} \\
\epsilon^{uu}_{\mu\mu}
\end{matrix} \right) = \left( \begin{matrix}
2.0 \pm 5.7 \\
-0.2 \pm 1.7 \\
-0.037 \pm 0.042 \\
-0.004 \pm 0.013
\end{matrix} \right)~.
\end{align}
The last two constraints in~\eref{epsilonbounds} are not expected to change with the inclusion of quadratic corrections\footnote{This is true in the vicinity of the SM value. For large $\epsilon_{ll}^{qq}$ values one can find new allowed regions, the so-called dark solutions.} and they represent another central result of our work. Once again, their errors are Gaussian to a good approximation and the application of these EFT constraints to more specific setups is straightforward. 

We stress that we did not neglect NP affecting production, as usually done in the past literature. Instead, we showed that, with our Lagrangian input choice, they are absent at this order in the EFT expansion. 

 It is interesting to discuss the results in a more constrained WEFT scenario when lepton-flavor universality of the relevant Wilson coefficients is assumed:  
$\epsilon^{uu}_{\mu\mu} = \epsilon^{uu}_{ee} \equiv \epsilon_u$
and 
$\epsilon^{dd}_{\mu\mu} = \epsilon^{dd}_{ee} \equiv \epsilon_d$. 
Then the 4-parameter fit in \cref{eq:epsilonbounds} reduces to the two-parameter one: 
\begin{align}
\label{eq:NUM_epsilonboundsLFU}
\epsilon_u = -0.01 \pm 0.98, 
\qquad 
\epsilon_d = 0.01 \pm 0.88, 
\end{align}
with the highly degenerate correlation coefficient $\rho = -0.99988$. 
Disentangling the correlation one finds one strong constraint:
\begin{align}
\label{eq:NUM_epsilonboundsLFUstrong}
 0.67 \epsilon_u + 0.74 \epsilon_d =  & 
-0.002 \pm 0.010, 
\end{align}
while the orthogonal combination is very loosely constrained: 
$0.74 \epsilon_u - 0.67 \epsilon_d  = 0.0 \pm 1.3$.

The results above lead to weak and effectively meaningless marginalized constraints on $\delta g^{\nu N}_{\ell\ell}$ and $\epsilon^{qq}_{\ell\ell}$ because only two of the constraints have uncertainties $\lesssim{\cal O}(1)$, where the linear expansion makes sense.
The situation changes when only one operator is present, which leads to stringent and reliable individual limits:
\begin{equation}
\left( \begin{matrix}
\epsilon^{dd}_{ee} \\
\epsilon^{uu}_{ee}  \\
\epsilon^{dd}_{\mu\mu} \\
\epsilon^{uu}_{\mu\mu} \\
\end{matrix} \right) = \left( \begin{matrix}
0.011 \pm 0.036 \\
0.012 \pm 0.040\\
-0.008 \pm 0.019\\
-0.008 \pm 0.022\\
\end{matrix} \right)~.
\label{eq:NCindividual}
\end{equation} 
\begin{figure}
\centering
\includegraphics[scale=0.57]{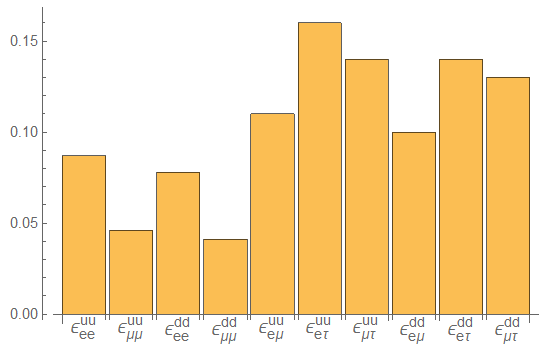}
\caption{90\% CL one-at-a-time limits for the absolute value of the NSIs probed by COHERENT.}
\label{fig:oneatatimefit}
\end{figure}

\subsubsection{Non-linear effects} 
In this section we take into account the full expression for the COHERENT event rate, including non-linear terms in the new physics Wilson coefficients, and discuss how this changes the results compared to the linear analysis.
For simplicity, we start by discussing the impact of non-linear effects in analyses where only one operator at a time is present.

As expected, the linear bounds presented in~\eref{NCindividual} for the neutral-current Wilson coefficients, $\epsilon^{qq}_{\alpha\alpha}$, are barely affected by non-linear terms,~{\it cf.}~\tref{WC1atatime}. 
The only qualitative difference is the presence of dark solutions placed far away from the SM values (which are not shown in~\tref{WC1atatime}).  
They appear because COHERENT is only sensitive to the squared charges $\Tilde{Q}^2_\alpha$, and so there are allowed regions near the $\pm Q_{SM}$ values. 

On the other hand, nonlinear terms give us access to NC flavor-changing operators, $\epsilon_{\alpha\beta}^{qq}~(\alpha\neq \beta)$, which only enter the event rate at quadratic order in NP. The corresponding results, obtained putting one operator at a time, can be found in Table~\ref{tab:WC1atatime}.\footnote{In this work we have assumed that all Wilson coefficients are real. However, our one-at-a-time bounds on lepton-flavor off-diagonal coefficients are trivially generalized to bounds on their modulus squared if they are complex, since they do not interfere with the SM contributions.} They are also compared with the flavor-diagonal ones in Fig.~\ref{fig:oneatatimefit}. Let us stress that, with our definition of the WEFT coefficients and input parameters, the effect of charged-current operators cancels in the rate (at all orders) if only one operator at a time is considered, as discussed in~\sref{limits}.

\begin{figure}
\centering
\includegraphics[scale=0.36]{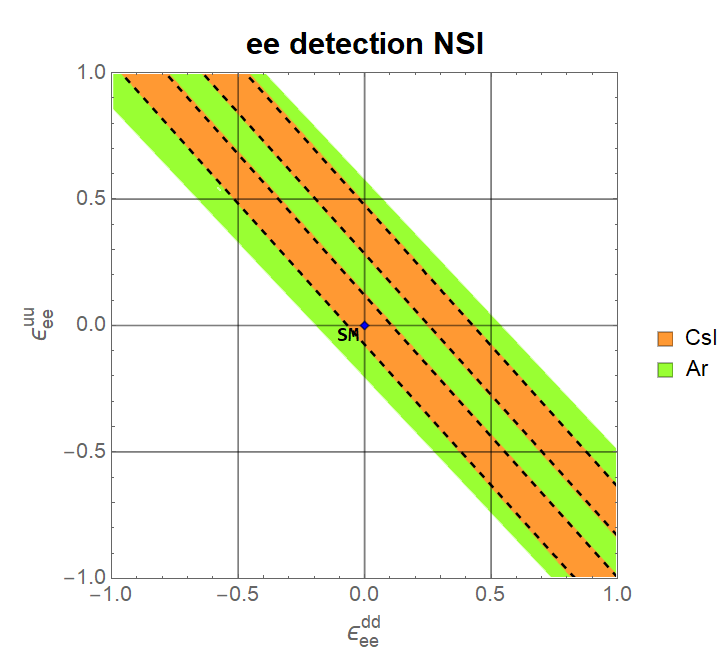} 
\hspace{0.5cm}
\includegraphics[scale=0.36]{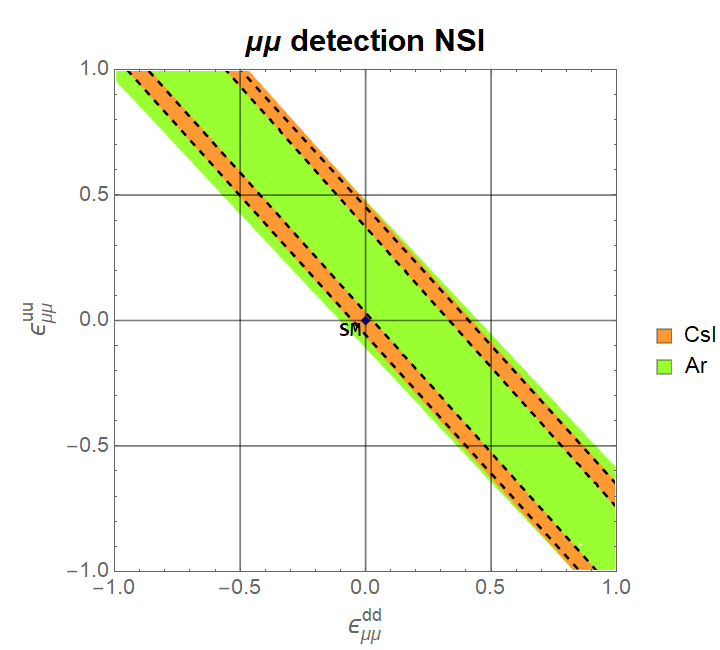}
\\
\vspace{0.5cm}
\includegraphics[scale=0.36]{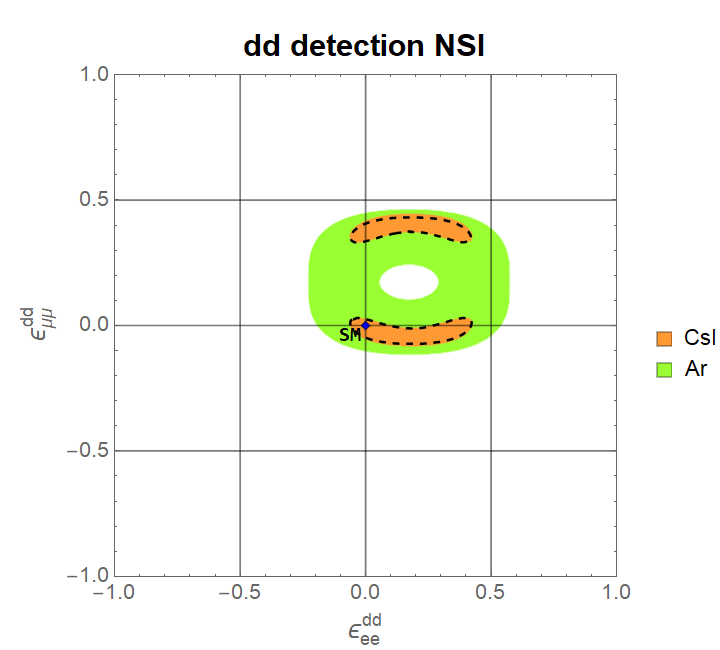}
\hspace{0.5cm}
\includegraphics[scale=0.36]{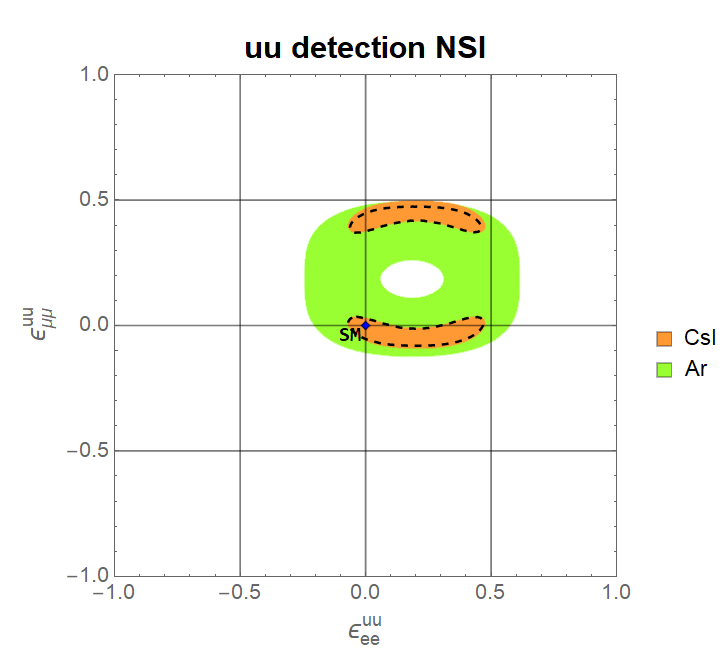}
\caption{$\Delta\chi^2=4.61$ regions for flavor-preserving Wilson coefficients as probed by COHERENT data. In each of the upper four panels the two Wilson coefficients displayed are the only non-zero NSI parameters. We display the bounds that the CsI and Ar datasets yield separately and the combination of the two through dashed lines.}
\label{fig:2WCfit}
\end{figure}

\begin{figure}
\centering
\includegraphics[scale=0.55]{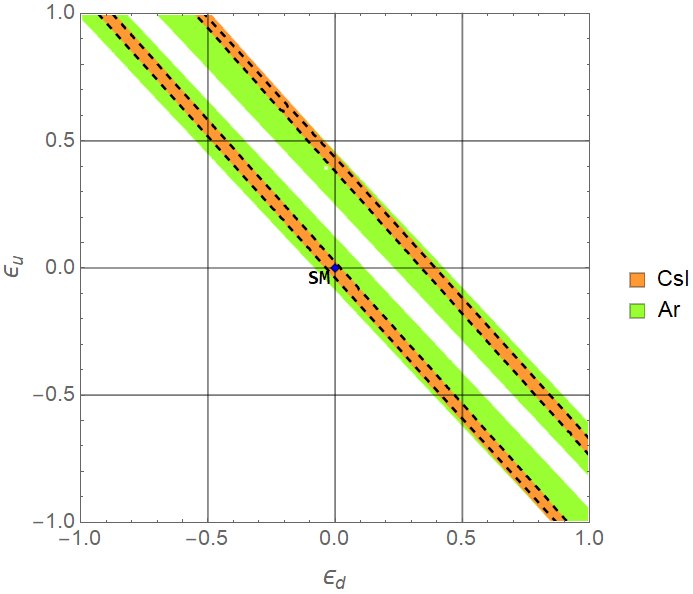}
\caption{$\Delta\chi^2=4.61$ regions for flavor-preserving Wilson coefficients as probed by COHERENT data assuming lepton universality: $\epsilon^{uu}_{\mu\mu} = \epsilon^{uu}_{ee} \equiv \epsilon_u$
and $\epsilon^{dd}_{\mu\mu} = \epsilon^{dd}_{ee} \equiv \epsilon_d$. We display the bounds that the CsI and Ar datasets yield separately and the combination of the two through dashed lines.}
\label{fig:2WCfit_universal}
\end{figure}

\begin{table}
\parbox{.45\linewidth}{
\centering
{\setlength{\extrarowheight}{5pt}
\begin{tabular}{c|c|c}
\cline{1-3}
\multicolumn{3}{c}{Flavour diagonal NSI}  \\ [0.8ex]
\cline{1-3}
WC & CV $\pm$ 1$\sigma$ & 90\% C.L. \\ [0.8ex] 
 \cline{1-3}
 $\epsilon_{ee}^{dd}$ & $~~0.012^{+0.045}_{-0.035}$ & $ 0.097$ \\ [0.8ex]
 \cline{1-3}
  $\epsilon_{ee}^{uu}$ & $~~0.013^{+0.050}_{-0.039}$ & $ 0.11$\\ [0.8ex]
\cline{1-3}
\cline{1-3}
\cline{1-3}
 $\epsilon_{\mu\mu}^{dd}$ & $-0.007^{+0.020}_{-0.018}$ & $ 0.036$ \\ [0.8ex]
 \cline{1-3}
   $\epsilon_{\mu \mu}^{uu}$ & $-0.008^{+0.022}_{-0.020}$ & $ 0.040$ \\ [0.8ex]
 \cline{1-3} 
\end{tabular}}
}
\hfill
\parbox{.45\linewidth}{
\centering
{\setlength{\extrarowheight}{5pt}
\begin{tabular}{c|c|c}
\cline{1-3}
\multicolumn{3}{c}{Flavour off-diagonal NSI}  \\ [0.8ex]
\cline{1-3}
WC & CV $\pm$  1$\sigma$ &  90\% C.L.   \\ [0.8ex]
 \cline{1-3}
 $\epsilon_{e\mu}^{uu}$ & 
 $\pm \left(0.031^{+0.053}_{-0.114} \right)$ & 
 $ 0.10$  \\ [0.8ex]
 \cline{1-3}
 $\epsilon_{e\tau}^{uu}$ &  
 $0 \pm 0.11$ 
 &$0.15$   \\ [0.8ex]
  \cline{1-3}
$\epsilon_{\mu \tau}^{uu}$ & 
$\pm \left( 0.057^{+0.052}_{-0.167} \right)$ & $0.13$  \\ [0.8ex]
\cline{1-3}
\cline{1-3}
\cline{1-3}
$\epsilon_{e\mu}^{dd}$ & 
$\pm \left( 0.027^{+0.048}_{-0.103}\right)$ & $0.094$  \\ [0.8ex]
 \cline{1-3}
$\epsilon_{e\tau}^{dd}$ & 
$0 \pm 0.098$ & $ 0.13$   \\ [0.8ex]
 \cline{1-3}
$\epsilon_{\mu \tau}^{dd}$ & 
$\pm \left( 0.052^{+0.047}_{-0.150} \right)$ & $0.12$   \\ [0.8ex]
  \cline{1-3}
\end{tabular}}
\caption{\label{tab:WC1atatime}
One-at-a-time bounds on neutrino NSIs. The second column shows the central value (CV) and their associated 1-sigma errors. The last column shows the 90\% C.L. bound for the absolute value of the associated Wilson coefficient (WC),  $|\epsilon_{\alpha \beta}^{qq}|$.}
}
\end{table}
Let us now discuss some cases where more than one operator is present at the same time. In Figs.~\ref{fig:2WCfit} and~\ref{fig:2WCfit_universal} we consider scenarios with two free NP parameters, with the remaining ones set to zero. A quick glance reveals that CsI data drive the constraining power of the fit in all cases. 

Comparing the upper two panels, we can see that the fit with only electron couplings yields noticeably weaker constraints than the one with muon parameters. This is to be expected since the contribution to the event rate coming from the muonic neutrinos is larger than the one coming from the electronic ones. We see that the CsI data is precise enough to separate the allowed region in two bands: one compatible with the SM and a second one corresponding to the above-mentioned dark solution. 
The well-known blind directions that these panels display is due to the potential cancellations among the linear combinations of up- and down-quark couplings in the weak charges, namely $(A+Z) \epsilon_V^{uu}  + (2A-Z) \epsilon_V^{dd} = \rm{constant}$, {\it cf.}~Eq.~\eqref{eq:onlydetection}. Since the blind directions are almost parallel for CsI and Ar, the combined dataset also shows this feature.

The second row in Fig.~\ref{fig:2WCfit} studies the cases where neutrinos are coupled only to down quarks and only to up quarks. In these fits we have NP in both the electron and muon charges and thus there are four solutions, corresponding to $([{\cal Q}]_{ee}=\pm Q_{SM},[{\cal Q}]_{\mu\mu}=\pm Q_{SM})$. Current COHERENT data is only able to separate the upper two dark solutions, whereas a third dark solution remains connected with the SM one. Adding \cevns measurements at reactors isolates the SM solution~\cite{Coloma:2022avw}.

Finally, Fig.~\ref{fig:2WCfit_universal} shows the lepton-universal case, where all four couplings are present, but electron and muon couplings are equal. The result is similar to the upper plots in~Fig.~\ref{fig:2WCfit}, but with a larger sensitivity.

The scenarios that we have discussed above
have been thoroughly discussed in previous works using different COHERENT datasets and projections~\cite{Coloma:2017ncl,Papoulias:2017qdn,AristizabalSierra:2018eqm,Khan:2019cvi,Giunti:2019xpr,Coloma:2019mbs,Denton:2020hop,Miranda:2020tif,Coloma:2022avw,AtzoriCorona:2022qrf}. Our work and the recent work of Ref.~\cite{DeRomeri:2022twg} are the first ones to use the entire COHERENT dataset presently available (2D distribution in LAr and CsI)~\cite{COHERENT:2020iec,COHERENT:2021xmm} to constrain NSI coefficients, and thus represent the current state of the art. Since the CsI detector has been decommissioned, they represent the final results with that target. Previous works have studied these NP scenarios using the SM fluxes and modified cross sections. Our more general approach reduces to this ``factorized'' NSI description if one neglects NP effects in production, as discussed in~\sref{limits}. Our numerical results for the separated LAr and CsI analyses agree well with previous works, including COHERENT analyses~\cite{COHERENT:2020iec, COHERENT:2021xmm}. Our combined bounds (LAr+CsI) are also in good agreement with the recent results in Ref.~\cite{DeRomeri:2022twg}.

Let us now compare our COHERENT results with other NSI probes, which were reviewed and compiled in Ref.~\cite{Farzan:2017xzy}. For simplicity we focus on bounds obtained switching on one operator at a time. 
For the muonic couplings, $\epsilon_{\mu\mu}^{dd}$ and $\epsilon_{\mu\mu}^{uu}$, the bounds obtained from COHERENT data match the best existing constraints, which come from atmospheric and accelerator neutrino data~\cite{Escrihuela:2011cf}. For the electronic couplings, $\epsilon_{ee}^{dd}$ and $\epsilon_{ee}^{uu}$, our COHERENT results are much stronger than the limits extracted from CHARM data~\cite{Davidson:2003ha} and comparable to those obtained from Dresden-II reactor data~\cite{Colaresi:2021kus, Coloma:2022avw}.   
For the flavor violating NSIs, our results for $\epsilon_{\mu\tau}^{qq}$ are roughly $20\times$ weaker than those obtained from IceCube~\cite{Salvado:2016uqu}. 
Finally, our one-at-a-time bounds on $\epsilon_{e\mu}^{qq}$ coefficients from COHERENT data are roughly two times weaker than those obtained in a global fit to oscillation data~\cite{Coloma:2019mbs}, whereas for the $\epsilon_{e\tau}^{qq}$ coefficients they are similar.  
The relatively weak sensitivity from our analysis to off-diagonal NSIs is to be expected since oscillation observables are linearly sensitive to them, whereas \cevns is only quadratically sensitive. On the other hand, \cevns is best suited to study flavor-conserving NSIs, with interesting synergies observed in combined analyses with oscillation data~\cite{Coloma:2019mbs}.

\subsubsection{Production and detection effects together}
Our general approach allows us to go beyond the well-known cases discussed above, and study situations where NP effects are present both in production and detection. 

\begin{figure}
\centering
\includegraphics[scale=0.391]{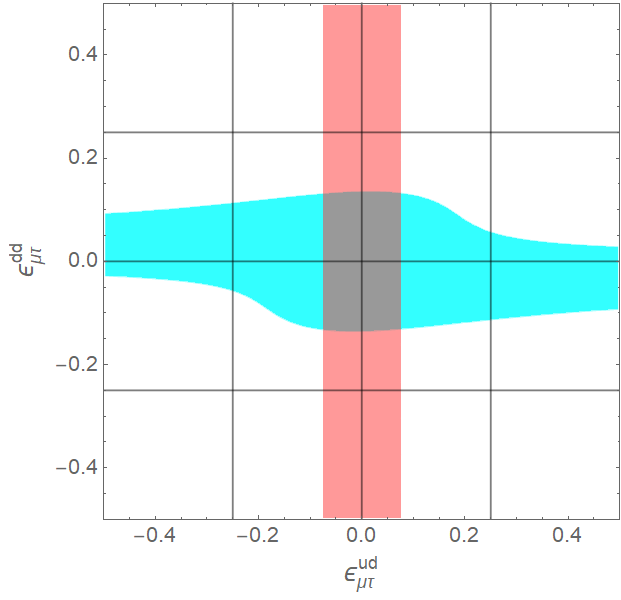} \hspace{0.4cm}
\includegraphics[scale=0.39]{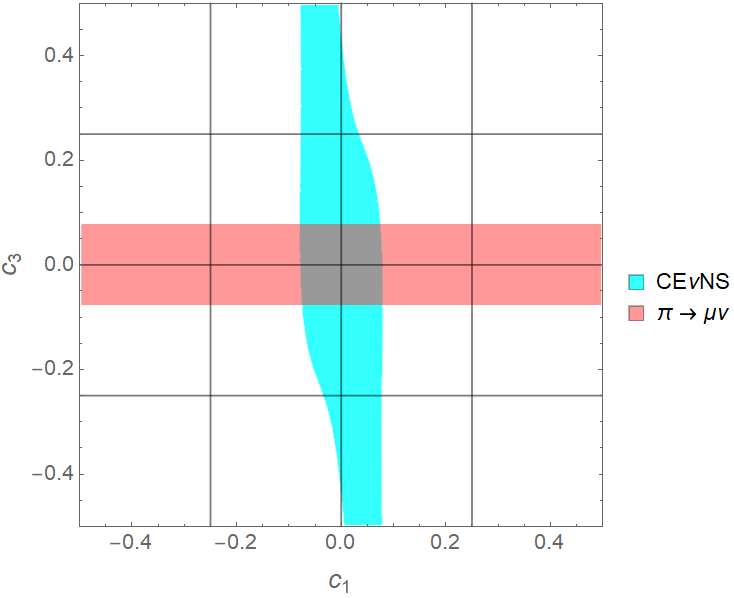}
\caption{90\% CL allowed regions for NP setups with flavor-violating Wilson coefficients in production and detection from the WEFT (left panel) and from the SMEFT (right panel) as probed by COHERENT data. In these fits only two couplings are allowed to be present at a time. Finally, we also include the constraint from the ratio $\Gamma(\pi\to e\nu)/\Gamma(\pi \to \mu\nu)$~\cite{ParticleDataGroup:2022pth,PiENu:2015seu,Cirigliano:2007ga}.}
\label{fig:2WCfitfv}
\end{figure}

For instance, we can study a setup where the NC coefficient $\epsilon^{dd}_{\mu\tau}$ is accompanied by the CC semileptonic Wilson coefficient $\epsilon^{ud}_{\mu\tau} \equiv [\epsilon^{ud}_L]_{\mu\tau}$. The latter affects neutrino production (it generates $\pi\to\mu\,\nu_\tau$), whereas the former affects the detection of muon and tau (anti)neutrinos (it generates $\nu_\mu\,{\cal N}\to\nu_\tau\,{\cal N}, ~\nu_\tau\,{\cal N}\to\nu_\mu\,{\cal N}$ and likewise for antineutrinos). 
Fig.~\ref{fig:2WCfitfv} (left panel) shows the allowed regions when both parameters are present at the same time. Let us stress once again that in our formalism these bounds are obtained without introducing a $\nu_\tau$ flux. As explained in~\cref{sec:ratesubsection}, one would obtain different (and thus incorrect) results if one calculates the event rate in a flux-times-cross-section factorized form. Namely, one would lose all sensitivity to the $\epsilon^{ud}_{\mu\tau}$ parameter. 

The study of simultaneous NP effects in neutrino production and detection is particularly relevant in setups with explicit electroweak symmetry, since neutrinos and charged leptons form $SU(2)_L$ gauge doublets. As a result, non-standard contributions to $\nu_\alpha{\cal N}\to\nu_\beta{\cal N}$ come in general with non-standard effects in leptonic pion decay, $\pi\to\nu_\alpha\ell_\beta$. Let us consider for instance the SMEFT operators $[{\cal O}_{lq}^{(3)}]_{\mu\tau 11} \equiv 
(\bar l_L^2 \gamma_\mu \sigma^k l_L^3)(\bar q_L^1 \gamma^\mu \sigma^k q_L^1)$ and  $[{\cal O}_{lq}^{(1)}]_{\mu\tau 11}\equiv 
 (\bar l_L^2 \gamma_\mu l_L^3)(\bar q_L^1 \gamma^\mu  q_L^1)$ (along with their conjugates so that the Lagrangian is Hermitian), and let us abbreviate their associated Wilson coefficients as $c_3 \equiv 
v^2 [C_{lq}^{(3)}]_{\mu\tau 11}$ and $c_1 \equiv 
v^2 [C_{lq}^{(1)}]_{\mu\tau 11}$.
At tree level they generate the following WEFT coefficients relevant for COHERENT: 
\bea
[\epsilon^{ud}_L]_{\mu\tau} = c_3~, 
\qquad
\epsilon^{uu}_{\mu\tau} = c_1 - c_3~,
\qquad
\epsilon^{dd}_{\mu\tau} = c_1 + c_3~.
\eea
We show in Fig.~\ref{fig:2WCfitfv} (right panel) the bounds that we obtained on the coefficients $c_1$ and $c_3$ using COHERENT data. 

We can also constrain charged-current NSIs using the measurements of leptonic pion decay widths. To make things simpler, we work with the ratio $R_\pi \equiv \Gamma(\pi\to e\nu)/\Gamma(\pi \to \mu\nu)$ that, in the specific cases discussed above, is modified as 
$R_\pi = R_\pi^{\rm{SM}} / (1+ (\epsilon^{ud}_{\mu\tau})^2)$. 
Fig.~\ref{fig:2WCfitfv} shows the interplay between this constraint and the one obtained from COHERENT data.

\section{Comparison and combination with other precision observables} \vspace{0.4cm}
\label{sec:EWPO}

In this section we discuss the place of the COHERENT experiment in the larger landscape of electroweak precision observables. 
To this end we will employ the SMEFT framework~\cite{Buchmuller:1985jz, Grzadkowski:2010es}, which will allow us to combine information from COHERENT and other experiments below the Z-pole,  with that obtained by the high-energy colliders at or above the $Z$-pole.\footnote{See Refs.~\cite{Terol-Calvo:2019vck,Skiba:2020msb,Crivellin:2021bkd} for previous SMEFT analyses that included some COHERENT observables.} 
It will also allow us to combine the information from NC and CC processes, which are related by the $SU(2)_L$ gauge symmetry. 
We consider operators up to dimension six, using the standard SMEFT power counting where the corresponding Wilson coefficients are $\cO(\Lambda^{-2})$ in the new physics scale $\Lambda$. 
Consequently, we expand observables to order $1/\Lambda^2$, ignoring higher order corrections.
This implies that we only need the linearized version of the COHERENT results obtained in the WEFT formalism, cf. \eref{epsilonbounds}. 

The COHERENT experiment probes contact 4-fermion interactions between the left-handed lepton doublets $l_L$ and quark doublets $q_L$, and singlets $u_R$, $d_R$. The relevant dimension-6 operators are~\cite{Grzadkowski:2010es}   
\begin{align}
\label{eq:SMEFT_4fermion}
{\cal L}_{\rm SMEFT} \supset & 
C_{lq}^{(1)} (\bar l_L \gamma_\mu l_L)(\bar q_L \gamma^\mu q_L)
+ C_{lq}^{(3)} (\bar l_L \gamma_\mu \sigma^k l_L)(\bar q_L \gamma^\mu \sigma^k q_L)
\nnl + &  C_{lu} (\bar l_L \gamma_\mu l_L)(\bar u_R \gamma^\mu u_R)
+ C_{ld} (\bar l_L \gamma_\mu l_L)(\bar d_R\gamma^\mu d_R). 
\end{align} 
Here, the capital letter Wilson coefficients are dimensionful, $[C_X] = {\rm mass}^{-2}$. 
For the numerical analysis it is more convenient to work with dimensionless objects $c_X = v^2 C_X$, where $v \simeq 246.22$~GeV.  
The SM fields are 3-component vectors in the generation space, however the flavor index is suppressed here to reduce clutter.  
In this section we assume that the Wilson coefficients are flavor universal, more precisely, that they respect the $U(3)^5$ flavor symmetry acting on the three generations of  $q_L, u_R, d_R, l_L, e_R$. 
This is by far the most studied SMEFT setup, especially in the context of global fits, and often described simply as the EWPO fit.
As we will show below, even in this restricted framework, COHERENT has a significant impact on the global fit. 
Later in \aref{SMEFT} we will relax this assumption, in which case the impact of COHERENT will be even more spectacular thanks to lifting degeneracies in the multi-dimensional parameter space of Wilson coefficients. 

From the SMEFT point of view, the COHERENT experiment also probes the coupling strength of the $Z$ boson to quarks and neutrinos. 
For these interactions we will use the Higgs basis parametrization (see \cite{Azatov:2022kbs} for a recent summary):  
\begin{align}
\label{eq:SMEFT_vertex}
{\cal L}_{\rm SMEFT} \supset & 
-  \bigg \{ 
\bigg [ {1 \over 2}  -   {2 \over 3}
\sin^2\theta_W  + \delta g_L^{Zu} \bigg ]  (\bar u_L \gamma^\mu u_L)
+  \bigg [  - {2 \over 3} \sin^2\theta_W  + \delta g_R^{Zu} \bigg ]  (\bar u_R \gamma^\mu d_R)
\nnl & 
+ \bigg [ -{1 \over 2}   + {1 \over 3} \sin^2\theta_W   + \delta g_L^{Zd} \bigg ]  (\bar d_L \gamma^\mu d_L)
+ \bigg [ {1 \over 3} \sin^2\theta_W    + \delta g_R^{Zd} \bigg ]  (\bar d_R \gamma^\mu d_R)
\nnl &  
+ \bigg [ {1 \over 2} + \delta g_L^{W\ell}  +  \delta g_L^{Z e} \bigg ]  (\bar \nu_L \gamma^\mu \nu_L)
\bigg \}  \sqrt{g_L^2 + g_Y^2} Z_\mu ~,
\end{align}
where $g_L$ and $g_Y$ are the gauge couplings of the $SU(2)_L \times U(1)_Y$ local symmetry, $\sin^2\theta_W\equiv g_Y^2/(g_L^2+g_Y^2)$. 
Above, the effects of the dimension-6 SMEFT operators on the $Z$ couplings to quarks are parametrized by the four dimensionless vertex corrections 
$\delta g_{L/R}^{Zu/d}$. 
There is one more parameter $\delta g_L^{Z \nu}$ describing the dimension-6 effects on the $Z$ coupling to neutrinos.  In the Higgs basis it is expressed by other leptonic vertex corrections: $\delta g_L^{Z \nu} =  \delta g_L^{W\ell}  +  \delta g_L^{Z e}$. 

The COHERENT results analyzed in this paper constrain the 4-fermion Wilson coefficients in \eref{SMEFT_4fermion} and the vertex corrections in \eref{SMEFT_vertex}.  
These Wilson coefficients are related to the NC WEFT Wilson coefficients in \cref{eq:LagrangianNC} by~\cite{Falkowski:2017pss}
\begin{align} 
\label{eq:SMEFT_epsilonMap} 
\epsilon_{\alpha\alpha}^{uu}=&  
\delta g_L^{Zu} +\delta g_R^{Zu} +
\left ( 1 - {8 s_\theta^2 \over 3 } \right ) \delta g_L^{Z\nu} 
- \frac{1}{2} \bigg (c_{lq}^{(1)} + c^{(3)}_{lq} + c_{lu} \bigg )
\equiv \epsilon_u ~,
\nnl 
\epsilon_{\alpha\alpha}^{dd}  =& 
\delta g_L^{Zd} +  \delta g_R^{Zd}  
- \left (  1 -   {4 s_\theta^2  \over 3 }  \right ) \delta g_L^{Z\nu}   
-\frac{1}{2} \bigg(  c_{lq}^{(1)} -  c^{(3)}_{lq} + c_{ld} \bigg ) 
\equiv \epsilon_d ~,
\end{align}
where there is no implicit sum over the repeated index $\alpha$. 
Because of our assumption of $U(3)^5$ symmetry, 
the expression is the same for any value of the index $\alpha$, 
that is to say, the quarks interact with the same strength with all flavors of the neutrino. Therefore, it is appropriate to use the results of the constrained WEFT fit in \cref{eq:NUM_epsilonboundsLFU}. 
Translated to the SMEFT Wilson coefficients, the strong constraint in \cref{eq:NUM_epsilonboundsLFUstrong} becomes
\begin{equation}
 0.71 c_{lq}^{(1)} - 0.04 c^{(3)}_{lq} + 0.34 c_{lu} +0.37 c_{ld} + [\delta g]_{\rm{piece}}
= - 0.003 \pm 0.010~,
\end{equation}
where $[\delta g]_{\rm{piece}} \equiv -0.67 (\delta g_L^{Zu} + \delta g_R^{Zu})  - 0.74 (\delta g_L^{Zd} + \delta g_R^{Zd} )   + 0.26 \delta g_L^{Z\nu}$ collects the contributions from vertex corrections.

In the reminder of this section we will compare the strength of the  COHERENT constraints on SMEFT coefficients to that of the other electroweak precision measurements. 
Ref.~\cite{Falkowski:2017pss} compiled the input from experiments sensitive, much as COHERENT, to flavor-conserving $Zff$ vertex corrections and 4-fermion operators with two leptons and two quarks. That analysis included $Z$ and $W$ pole measurements, $e^+e^-\to f\bar{f}$ data, (non-coherent) neutrino scattering on nucleon targets, atomic parity violation (APV), parity-violating electron scattering, and the decay of pions, neutrons, nuclei and tau leptons.  Ref.~\cite{Falkowski:2017pss} also included purely leptonic observables since, in addition to their dependence on SMEFT 4-lepton operators, they are sensitive to some of the vertex corrections  in \cref{eq:SMEFT_vertex}.  
 We use the likelihood for the SMEFT Wilson coefficients constructed in Ref.~\cite{Falkowski:2017pss}, updated to include new theoretical and experimental developments (the full list of observables used and the updates can be found in \cref{app:SMEFT}).  
In the following we compare and combine these constraints with the ones obtained in this paper using the COHERENT data.

\begin{table}[]
    \centering
    \begin{tabular}{|l|c|c|c|c|c|c|c|c|c|c|}
 \hline
 Coefficient & 
$\delta g_L^{W\ell}$ & $\delta g_L^{Z e}$ &
$\delta g_L^{Zu}$ & $\delta g_R^{Zu}$ &  $\delta g_L^{Zd}$ & $\delta g_R^{Zd}$ &   
$c_{lq}^{(1)}$ & $c^{(3)}_{lq}$ & $c_{lu}$ & $c_{ld}$   
\\ \hline
 w/o COHERENT& 
 0.14 & 0.11 & 0.23 & 0.63 & 0.19 & 0.78 & 
 0.81 & 0.26 & 1.5 & 1.8  
 \\ \hline
COHERENT alone  & 
 71 & 71 & 15 & 15 & 14 & 14 & 
 14 & 260 & 30 & 27
  \\ \hline
    \end{tabular}
    \caption{
Uncertainty (in units of $10^{-3}$) on the Wilson coefficients probed by the COHERENT experiment (central values are not given). 
The COHERENT row shows the constraints based on the results obtained in this paper. 
W/o COHERENT refers to a host of electroweak precision observables analyzed in Ref.~\cite{Falkowski:2017pss}, with the updates described in \cref{app:SMEFT}. }
    \label{tab:SMEFT_onebyone}
\end{table}

The first comment is that the impact of COHERENT is negligible if only a single Wilson coefficient appearing in \cref{eq:SMEFT_epsilonMap} is present at a time. 
In \cref{tab:SMEFT_onebyone} we show the uncertainty obtained in such a one-at-a-time fit. 
We can see that the sensitivity of COHERENT is inferior by 1-2 orders of magnitude compared to that achieved by a combination of other electroweak precision measurements. 
This is not surprising, given that the latter contain a number of observables that have been measured with a (sub)permille precision (namely LEP1, APV, or baryon decays), while COHERENT currently offers a percent level precision.  

\begin{figure}
    \centering
    \includegraphics[width=0.47\textwidth]{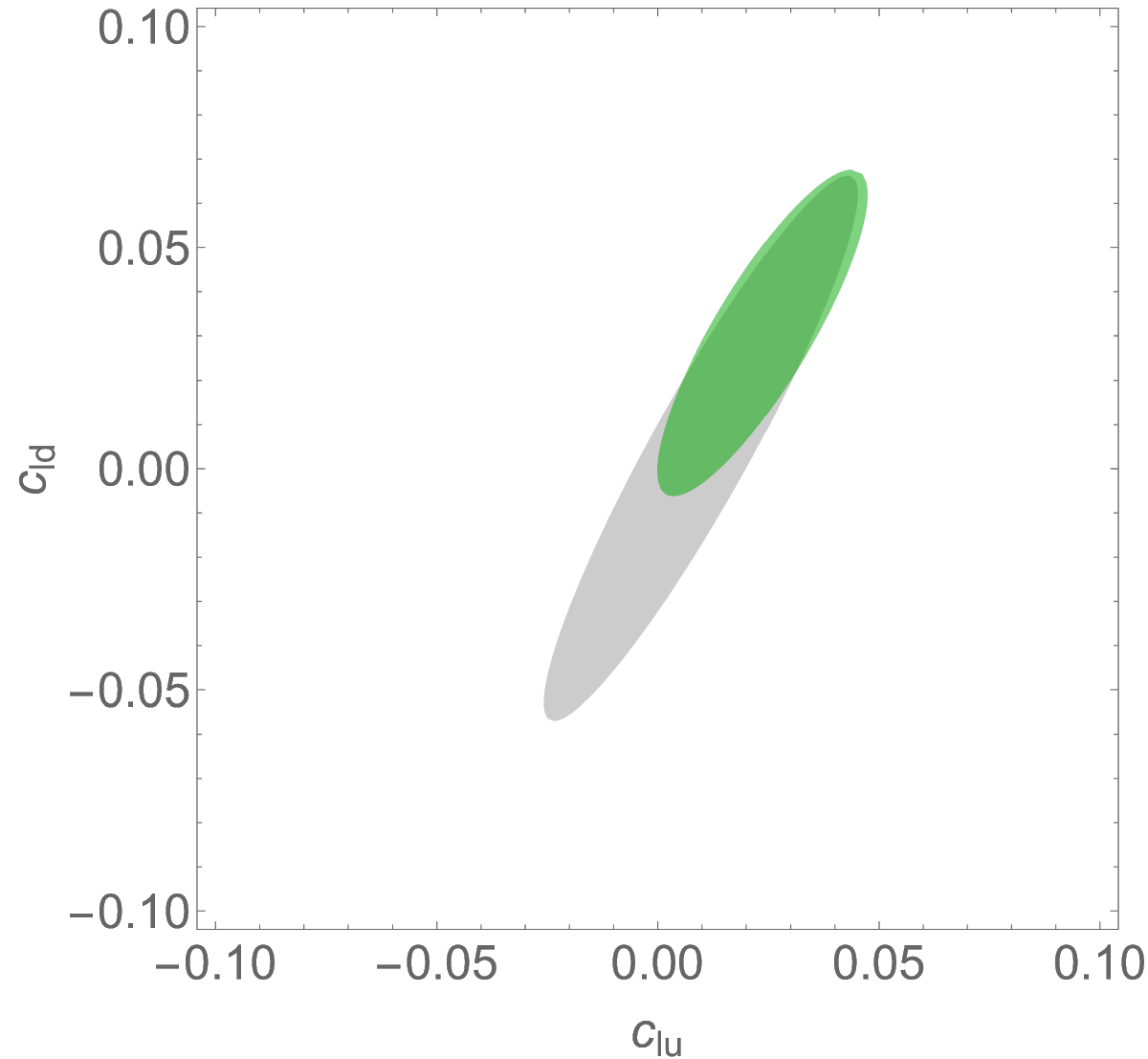}
    \quad 
  \includegraphics[width=0.47\textwidth]{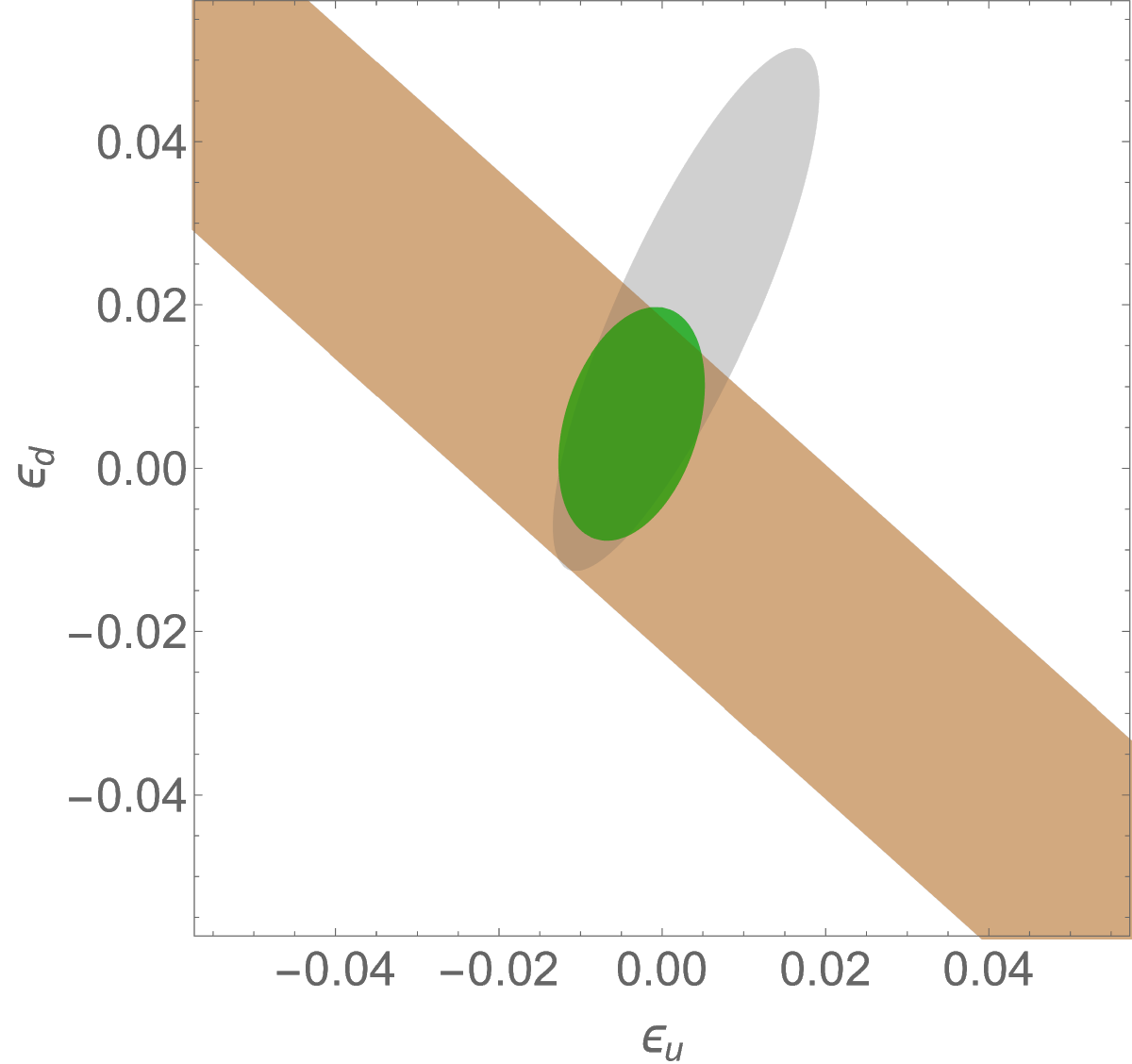}
    \caption{     
    \label{fig:SMEFT_clucld}
     Left panel: Marginalized 1-sigma bounds $(\Delta \chi^2 \simeq 2.3)$ on the 4-fermion SMEFT coefficients $c_{lu}$ and $c_{ud}$ from a global fit to EWPO in the flavor-universal SMEFT without (gray) and with (green) COHERENT data. We stress that the other 16 Wilson coefficients are not set to zero, but marginalized over. 
    Right panel: 1-sigma bounds $(\Delta \chi^2 \simeq 2.3)$ on the 4-fermion lepton-universal WEFT coefficients $\epsilon_u$ and $\epsilon_d$, obtained from COHERENT data (beige), other EWPO (gray) and the combination of both (green). The latter two are obtained in the flavor-universal SMEFT.}
\end{figure}

However, most new physics models generate several operators simultaneously, and thus a global analysis is required to assess the importance of COHERENT data. 
With this in mind, we turn to analyzing the situation when all flavor-universal dimension-6 SMEFT Wilson coefficients are allowed to be present with arbitrary magnitudes within the regime of validity. 
Now we are dealing with a multi-dimensional parameter space, where certain directions may not be constrained by the most precise observables, and where the input from COHERENT may be valuable. 
More precisely, in the flavor universal case  the observables taken into account in our analysis probe 18 independent Wilson coefficients. 
In addition to the six defined in \cref{eq:SMEFT_vertex,eq:SMEFT_vertex}, our analysis probes 11 more four-fermion operators as well as the vertex correction to the Z boson coupling to right-handed leptons. For their definition see \cref{app:SMEFT}, in particular \cref{eq:SMEFT2_2l2q,eq:SMEFT2_4l,eq:SMEFT2_u5params}.  
We find the fully marginalized constraints
\begin{align}
\begin{pmatrix}
\\ \delta g_L^{W\ell} \\ \delta g_L^{Z e}  \\ \delta g_R^{Z e}  \\ \delta g_L^{Zu} \\ \delta g_R^{Zu} \\  \delta g_L^{Zd} \\  \delta g_R^{Zd} \\
c_{lq}^{(1)} \\ c^{(3)}_{lq} \\ c_{lu} \\ c_{ld} \\ 
c_{eq} \\ c_{eu} \\ c_{ed} \\ 
c_{ll}^{(1)} \\ c^{(3)}_{ll} \\ c_{le} \\ c_{ee} \\ 
\end{pmatrix}
= 
\begin{pmatrix}
{\rm w/o \ COHERENT}\\
 -0.27(79) \\
 -0.10(0.21) \\ 
 -0.20(22) \\ 
-1.0(1.6) \\ 
-0.5(3.2) \\
1.5(1.3) \\
12.8(6.7) \\
 -16.6(9.0) \\
-2.4(1.9) \\
{\color{red} 10(23)} \\
{\color{red} 5(41) }\\
{\color{red} -13(22)}\\
7(10)\\
25(18)\\
5.4(3.2)\\
 -0.9(1.6)\\
0.2(1.3)\\
-2.7(3.0)
\end{pmatrix} \times 10^{-3} 
\to
\begin{pmatrix}
{\rm w/ \ COHERENT} \\ 
 -0.26(78) \\
 -0.09(21) \\ 
 -0.17(22) \\ 
-1.3(1.6) \\ 
-1.1(3.1) \\
1.1(1.2) \\
10.4(5.8) \\
 -18.3(8.7) \\
-2.2(1.8) \\
{\color{red} 23(16)}\\
{\color{red} 29(24)}\\
{\color{red} -1(15)}  \\
3.5(9.4)\\
29(17)\\
5.3(3.2)\\
 -0.9(1.6)\\
0.2(1.3)\\
-2.7(3.0)
\end{pmatrix} \times 10^{-3}. 
\label{eq:globalfitresults}
\end{align}
We highlighted the constraints on 
$c_{lu}$, $c_{ld}$, and $c_{eq}$, which improve significantly, by about 30-40\%, after including the COHERENT input. 
The improvement is visualized in the left panel of \cref{fig:SMEFT_clucld}. 
While neutrino scattering experiments have long played an important role in  SMEFT fits of electroweak precision observables, this is the first time {\em coherent} neutrino scattering is included in such a fit. 
In fact, of all neutrino experiments, COHERENT currently makes the largest impact on the flavor-blind SMEFT fit. The correlation matrix for the fit including COHERENT is
\begin{equation}
\tiny
 \hspace{-1cm}\rho = 
\left(
\begin{array}{cccccccccccccccccc}
 1 & -0.581 & 0.211 & 0.077 & 0.103 & 0.019 & -0.011 & -0.056 & 0.478 & 0.122 & -0.003 & 0.034 & 0.122 & -0.037 & -0.46 & 0.978 & -0.008 & -0.052 \\
 \text{} & 1 & 0.294 & -0.098 & -0.074 & -0.118 & -0.186 & 0.11 & -0.252 & -0.168 & -0.059 & -0.092 & -0.102 & -0.095 & 0.203 & -0.565 & 0.002 & 0.113 \\
 \text{} & \text{} & 1 & 0. & -0.032 & -0.275 & -0.251 & 0.1 & 0.267 & -0.084 & -0.101 & -0.08 & 0.002 & -0.211 & -0.19 & 0.208 & -0.013 & 0.079 \\
 \text{} & \text{} & \text{} & 1 & 0.686 & 0.39 & 0.145 & -0.299 & 0.638 & 0.291 & 0.32 & 0.304 & 0.391 & 0.268 & -0.031 & 0.074 & -0.001 & -0.01 \\
 \text{} & \text{} & \text{} & \text{} & 1 & 0.152 & 0.311 & -0.4 & 0.533 & 0.35 & 0.423 & 0.366 & 0.415 & 0.386 & -0.043 & 0.101 & 0. & -0.01 \\
 \text{} & \text{} & \text{} & \text{} & \text{} & 1 & 0.816 & -0.384 & -0.304 & 0.415 & 0.36 & 0.35 & 0.135 & 0.654 & 0.02 & 0.012 & 0.003 & -0.026 \\
 \text{} & \text{} & \text{} & \text{} & \text{} & \text{} & 1 & -0.462 & -0.41 & 0.478 & 0.43 & 0.401 & 0.137 & 0.755 & 0.036 & -0.018 & 0.003 & -0.03 \\
 \text{} & \text{} & \text{} & \text{} & \text{} & \text{} & \text{} & 1 & -0.03 & -0.83 & -0.718 & -0.049 & -0.726 & -0.854 & 0.012 & -0.051 & -0.001 & 0.016 \\
 \text{} & \text{} & \text{} & \text{} & \text{} & \text{} & \text{} & \text{} & 1 & 0.042 & 0.032 & 0.052 & 0.304 & -0.218 & -0.253 & 0.508 & -0.006 & -0.014 \\
 \text{} & \text{} & \text{} & \text{} & \text{} & \text{} & \text{} & \text{} & \text{} & 1 & 0.897 & 0.48 & 0.405 & 0.872 & -0.045 & 0.123 & 0. & -0.021 \\
 \text{} & \text{} & \text{} & \text{} & \text{} & \text{} & \text{} & \text{} & \text{} & \text{} & 1 & 0.656 & 0.262 & 0.796 & 0.016 & -0.011 & 0.001 & -0.011 \\
 \text{} & \text{} & \text{} & \text{} & \text{} & \text{} & \text{} & \text{} & \text{} & \text{} & \text{} & 1 & -0.411 & 0.347 & -0.006 & 0.033 & 0.001 & -0.013 \\
 \text{} & \text{} & \text{} & \text{} & \text{} & \text{} & \text{} & \text{} & \text{} & \text{} & \text{} & \text{} & 1 & 0.495 & -0.054 & 0.122 & -0.001 & -0.01 \\
 \text{} & \text{} & \text{} & \text{} & \text{} & \text{} & \text{} & \text{} & \text{} & \text{} & \text{} & \text{} & \text{} & 1 & 0.043 & -0.047 & 0.003 & -0.02 \\
 \text{} & \text{} & \text{} & \text{} & \text{} & \text{} & \text{} & \text{} & \text{} & \text{} & \text{} & \text{} & \text{} & \text{} & 1 & -0.471 & -0.176 & -0.578 \\
 \text{} & \text{} & \text{} & \text{} & \text{} & \text{} & \text{} & \text{} & \text{} & \text{} & \text{} & \text{} & \text{} & \text{} & \text{} & 1 & -0.008 & -0.051 \\
 \text{} & \text{} & \text{} & \text{} & \text{} & \text{} & \text{} & \text{} & \text{} & \text{} & \text{} & \text{} & \text{} & \text{} & \text{} & \text{} & 1 & -0.234 \\
 \text{} & \text{} & \text{} & \text{} & \text{} & \text{} & \text{} & \text{} & \text{} & \text{} & \text{} & \text{} & \text{} & \text{} & \text{} & \text{} & \text{} & 1 \\
\end{array}
\right). 
\end{equation}
In the $U(3)^5$-invariant SMEFT, one can obtain bounds on 10 new (combinations of) Wilson coefficients from diboson production at LEP2 and Higgs measurements~\cite{Falkowski:2015jaa}. The rest of $U(3)^5$-invariant SMEFT coefficients, which are not probed by these 2 global fits, are made up of only Higgs doublets, only quarks, or only gluons, or violate CP.

Another way to illustrate the impact of COHERENT is to consider global constraints on the combinations $\epsilon_u$ and $\epsilon_d$ of the SMEFT Wilson coefficients, defined in \cref{eq:SMEFT_epsilonMap}.
Let us recall that COHERENT alone constrains one linear combination at a percent level: $0.67 \epsilon_u + 0.74 \epsilon_d = 0.003(10)$, while the orthogonal combination is poorly constrained. 
This is shown in the right panel of \cref{fig:SMEFT_clucld} as the diagonal beige band (this is simply a zoomed in version of~\cref{fig:2WCfitfv}). 
We can use the results of our flavor-diagonal global SMEFT fit of electroweak precision observables in~\cref{eq:globalfitresults} (without COHERENT) to constrain the $\epsilon_{u,d}$ coefficients, finding 
$\epsilon_u = 0.003(10)$ and $\epsilon_d =0.019(21) $. 
This constraint is also percent level, indicating that COHERENT has an important impact on the global fit. 
Indeed, the combination of the COHERENT results with other precision observables leads to 
$\epsilon_u = -0.0037(54)$, $\epsilon_d = 0.0054(93)$, which represents a factor of two improvement. 
These results are represented in the right panel of~\cref{fig:SMEFT_clucld}.

All in all, the results of this section demonstrate  that COHERENT has become an indispensable ingredient in the family of electroweak precision observables constraining the SMEFT Wilson coefficients.

\section{Conclusions and Discussion} \vspace{0.4cm}
\label{sec:conclusions} 

In this paper, we have laid down a new theoretical framework based on effective field theories to describe non-standard effects in coherent neutrino scattering on nuclei. The framework is very versatile, allowing us to handle simultaneous new physics contributions to neutrino production and detection, non-linear effects of non-standard Wilson coefficients, different input schemes for the SM parameters, as well as an arbitrary flavor structure of neutrino-matter CC and NC interactions. 
It can also be readily applied to EFTs at different energy scales (nucleon- or quark-level, below or above the electroweak scale...).
This generalizes the NSI language that is followed by the greater part of the literature, and facilitates connection to specific BSM models with new particles heavier than the characteristic experimental scale.   

An important element of our analysis is that we have included new physics effects coming from both the neutrino production and detection processes.
There exists a nontrivial interplay between these two pieces, which cannot be reduced to a simple factorization of the neutrino flux and the detection cross section. 
We remark that correlated effects in production and detection are rather generic in new physics models. In particular, in the SMEFT framework several dimension-6 operators contribute to both,  due to  the $SU(2)_L$ gauge symmetry relating the CC and NC interactions. 
The consideration of these correlated effects opens the door to using the measurements at COHERENT to probe new physics models that relate CC and NC interactions.

We have introduced {\em three} generalized weak charges $\tilde Q_f$, which, in a certain sense, can be associated to the production and  scattering of $\nu_e$, $\nu_\mu$ and $\bar{\nu}_\mu$ on the nuclear target, cf.~\eref{Qdefs}.\footnote{This association should be interpreted with care, since the generalized charges also include the contribution from, {\it e.g.} tau neutrinos. The association is strictly correct for flavor-diagonal interactions. It is also correct in general in the practical sense of~\eref{THprediction2}. See~\cref{sec:ratesubsection} for more details.} 
These can be extracted from the COHERENT data, given the recoil energy and time distribution of the nuclear recoil events.  
They contain full information about the contributions of new physics affecting the effective contact interactions between neutrinos and matter. 
The framework can be simplified by folding in further assumptions. 
If new physics affects only detection, or if one restricts to a linear order in the EFT Wilson coefficients, there remain two independent generalized weak charges (electronic and muonic), similarly as in the prior analyses within the NSI framework. 
In the SM limit there is a single nuclear weak charge that needs to be extracted from experiment.

We have applied this framework to obtain novel constraints on new physics based on the results of the COHERENT experiment. We have examined the full dataset available at the moment, which includes distributions of energy and time of recoils measured in cesium iodine and argon nuclear targets. 
The central result of our numerical analysis is given in \cref{eq:3qAr,eq:3qCsI}, where we give the confidence intervals for the three generalized weak charges, including the correlations. 
These encode the state-of-the-art description of the COHERENT constraints on EFTs. 
Let us stress the approximate Gaussianity of the results when the squared charges are used. This means that our entire analysis of LAr and CsI data, which contains 664 bins, with thousands of associated backgrounds, and 13 nuisance parameters, can be expressed in terms of three central values, three diagonal errors and one correlation matrix (per target). Any BSM practitioner can then trivially apply these results to their particular NSI setup, EFT approach, or New Physics model. We encourage the community to provide the result of their analyses of COHERENT data in this convenient form.

We have recast these limits into bounds on EFT Wilson coefficients, both in the WEFT and in the SMEFT. 
We find that the COHERENT data so far provide percent level constraints on two particular combinations of Wilson coefficients in the EFT parameter space.
We demonstrate that these constraints are a valuable ingredient in the grander scheme of electroweak precision observables. The impact of the coherent neutrino scattering information is most relevant when performing a global fit, both in the constrained flavor-blind ($U(3)^5$-symmetric) setup, and in the completely generic scenario. 
From the point of view of the SMEFT fits, COHERENT is the most relevant neutrino-detection experiment, clearly superior in comparison to previous neutrino scattering experiments at higher energies.

All in all, in this work we have carried out a complete study of New Physics effects at the COHERENT experiment within an EFT approach. This includes for the first time simultaneous effects in neutrino production and detection, and its addition to the global SMEFT fit of electroweak precision data.  
Our work enables the study of the impact of COHERENT to new BSM setups and opens exciting possibilities for future developments. For instance, our work can be extended to other scenarios with light exotic particles, or to the study of additional experiments in the broad and flourishing \cevns landscape.

\begin{acknowledgments}
We thank Luis Álvarez-Ruso, Sergio de la Cruz, Valentina de Romeri and Santiago Villodre for enlightening discussions and comments, and IJCLab for hospitality. 
VB is supported by Ministerio de Ciencia, Innovación y Universidades, Spain [grant FPU18/01340]. AF has received funding from the Agence Nationale de la Recherche (ANR) under grant ANR-19-CE31-0012 (project MORA) and from the European Union’s Horizon 2020 research and innovation programme under the Marie Skłodowska-Curie grant agreement No 860881-HIDDeN. 
MGA is supported by the {\it Generalitat Valenciana} (Spain) through the {\it plan GenT} program (CIDEGENT/2018/014).
KMP is supported by PROMETEO/2017/053 and PROMETEO/2021/071 (GV). This work was supported by the MCIN with funding from the EU {\it NextGenerationEU} and {\it Generalitat Valenciana} in the PRTR 2022 call (Project QPheno, ASFAE/2022/009), and by
MCIN/AEI/10.13039/501100011033 Grant No. PID2020-114473GB-I00.
\end{acknowledgments}

\appendix

\section{Details of the calculation of the event rate} \vspace{0.4cm}
\label{app:fluxesAndXsections}

In this section we elaborate on some details of our calculation of the CE$\nu$NS event rate. First, we show the expressions for the $f_f(T)$ functions, which contain part of the kinematic dependence of the predicted prompt and delayed number of events in Eq.~\eqref{eq:THprediction}:
\label{app:ratedetails}
\begin{eqnarray}
f_{\mu}(T) &=&
    1-\frac{\left(m_{\cal N}+2E_{\nu,\pi}\right)T}{2E_{\nu,\pi}^2}~,
    \nn
    f_{\bar{\mu}}(T) &=& 16\left( \frac{E_{\nu}^3}{m_{\mu}^3}-\frac{E_{\nu}^4}{m^4_{\mu}}+\frac{4E_{\nu}^3 T}{3m_{\mu}^4}-\frac{3 E_{\nu}^2 T}{2m_{\mu}^3}-\frac{3E_{\nu}m_{\cal N} T}{2m_{\mu}^3}+\frac{E_{\nu}^2 m_{\cal N} T}{m_{\mu}^4}\right)\bigg|^{m_{\mu}/2}_{E_\nu^{min}(T)}
    ~,
    \nn
    f_e(T) &=& 16\left( \frac{2E_{\nu}^3}{m_{\mu}^3} -\frac{3\,E_{\nu}^4}{m_{\mu}^4} +\frac{4E_{\nu}^3 T}{m_{\mu}^4} -\frac{3\,E_{\nu}^2 T}{m_{\mu}^3} -\frac{3\,E_{\nu}m_{\cal N} T}{m_{\mu}^3} +\frac{3\,E_{\nu}^2 m_{\cal N} T}{m_{\mu}^4}\right)\bigg|^{m_{\mu}/2}_{E_\nu^{min}(T)}~, \quad 
\label{eq:ffunctions}
\end{eqnarray}
where, as was mentioned in the main text, 
$E_\nu^{min}(T)$ denotes the minimum energy of the (anti)neutrino required to produce \cevns with a recoil energy $T$, which is given by
\begin{eqnarray}
\label{eq:Enumumin}
    E_\nu^{min}(T) &=& \frac{T}{2}\left(1+\sqrt{1+2\frac{m_{\cal N}}{T}}\right) ~.
\end{eqnarray}

Next, we expand the information presented in Eq.~\eqref{eq:THprediction2}, which we reinstate here: 
\begin{eqnarray}
      \frac{dN^{\rm prompt}}{dT} 
      &=& 
      N_T\, \int dE_\nu\,\frac{d\Phi_{\nu_\mu}}{dE_\nu}
      \,\frac{d\tilde{\sigma}_{\nu_\mu}}{dT}~, 
\nn
    \frac{dN^{\rm delayed}}{dT}
     &=& 
     N_T\, \int dE_\nu\,\Bigg( 
     \frac{d\Phi_{\nu_e}}{dE_\nu} \frac{d\tilde{\sigma}_{\nu_e}}{dT}
     + \frac{d\Phi_{\bar{\nu}_\mu}}{dE_\nu} \frac{d\tilde{\sigma}_{\bar{\nu}_\mu}}{dT} 
     \Bigg)~.
\end{eqnarray}
Here, the fluxes are defined in the usual form:
\begin{align}
\frac{d\phi_{\nu_\mu}}{dE_\nu} &= \frac{N_{\nu_\mu}}{4 \pi L^{2}} \delta (E_\nu - E_{\nu,\pi})~,\nn   
\frac{d\phi_{\nu_e}}{dE_\nu} &= \frac{N_{\nu_e}}{4 \pi L^{2}} \frac{192 E_{\nu}^2}{m_{\mu}^3}\left(\frac{1}{2}-\frac{E_{\nu}}{m_{\mu}} \right)~,\nn    
\frac{d\phi_{\bar{\nu}_\mu}}{dE_\nu} &= \frac{N_{\bar{\nu}_\mu}}{4 \pi L^{2}} \frac{64 E_{\nu}^2}{m_{\mu}^3}\left(\frac{3}{4}-\frac{E_{\nu}}{m_{\mu}} \right)~,
\end{align}
where
\begin{align}
N_{\nu_\mu}         &= n_{\rm POT}\, r_{\pi/p}\, BR(\pi\to \mu\nu)_{\rm exp}~,\nn
N_{\bar{\nu}_\mu}   &= N_{\nu_e} = n_{\rm POT}\, r_{\mu/p}\, BR(\mu\to e\bar{\nu}\nu)_{\rm exp}~.
\end{align}
As for the cross sections, they are also defined in the usual form but using the generalized charges $\tilde{Q}_f$, that is
\begin{eqnarray}
\frac{d\tilde{\sigma}_f}{dT} = (m_{\cal N} + T) \frac{({\cal F}(T))^2}{8 v^4\,\pi}    \bigg ( 1   -  { (m_{\cal N}   + 2 E_{\nu}) \,T \over 2  E_\nu^2} \bigg )  \tilde{Q}_f^2~.
\end{eqnarray}

\section{Numerical analysis: further details} \vspace{0.4cm}
\label{app:simdetails}
\vspace{0.2cm}
In this section we describe in detail the input used in our numerical analysis, which is chosen in every case following closely the corresponding COHERENT prescription.

Before discussing the details that are specific to each measurement, let us show the expression that we use for the form factor, ${\cal F}(q^2)$, since that is a common input to all cases. 
We use the Helm parametrization~\cite{Helm:1956zz}, which gives the following expression for the neutron and proton form factors:
\begin{eqnarray}
{\cal F}_{p/n}(q^2)=3\frac{j_1(q\,R_{0,p/n})}{q\,R_{0,p/n}} e^{-q^2 s^2/2}~.
\end{eqnarray}
Here $j_1(x)$ is the order-1 spherical Bessel function of the first kind, $s=0.9$ fm is the nuclear skin thickness~\cite{Lewin:1995rx} and $R_{0,p/n}$ is a function of $s$ and the proton/neutron root-mean-square (rms) radius $R_{p/n}$ given by
\begin{eqnarray}
R_{0,p/n}^2=\frac{5}{3} R_{p/n}^2 - 5s^{2}~.
\end{eqnarray}
The proton and neutron rms radii for the studied nuclei are taken from Refs.~\cite{ANGELI201369, FRICKE1995177, Hoferichter:2020osn}
\begin{equation}
\begin{split}
R_p(\text{Cs}) &= 4.821(5) \text{ fm}, \qquad R_p(\text{I}) = 4.766(8) \text{ fm},\qquad R_p(\rm Ar) = 3.448(2) \text{ fm}, \\
R_n(\text{Cs}) &= 5.09 \text{ fm}, \qquad \quad \, \, \, \,  R_n(\text{I}) = 5.03 \text{ fm}, \qquad \quad \,  \, \, \, R_n(\rm Ar) = 3.55 \text{ fm}.
\end{split}
\end{equation}
Following the COHERENT prescription, the uncertainty associated to this description of the form factor is included in our analysis through a nuisance parameter, as described below.

For the CsI analysis, we will take the average of $R_{p/n}(\text{Cs})$ and $R_{p/n}(\text{I})$ and of the nuclei masses. Moreover, as discussed in the main text, in both the LAr and the CsI cases we approximate neutron and proton form factors to be equal by taking the average of $R_p$ and $R_n$. This simplifies significantly the presentation of intermediate results and it is not expected to have any impact in the final results for the WEFT Wilson coefficients, taking into account current COHERENT uncertainties.

Finally, we do not consider the contribution from neutrino electron scattering, which acts as an additional background source. This contribution was separated from the signal in the LAr analysis, but not in the CsI analysis. In both cases, its effect on the SM event rate is negligible, while for heavy NP scenarios its impact is also irrelevant~\cite{Coloma:2022avw, DeRomeri:2022twg}.

\subsection{LAr measurement} \vspace{0.4cm}
The LAr dataset consists of a 3D distribution in recoil energy, time and the fraction of integrated amplitude within the first 90 ns after trigger $\left(\text{F}_{90}\right)$~{\cite{COHERENT:2020iec}. The latter plays no direct role in our analysis, so we integrate over F$_{90}$ and work with the resulting 2D recoil and time distribution.
Our analysis covers  the range $0 < T^{rec}_{ee} (\text{keV}) < 40$ and $0<t(\mu s)<5$, using 4x10 bins of equal width. These are the bins with a significant amount of CE$\nu$NS events.

The expected number of events per bin is given by
\begin{eqnarray}
N_{ij}^{\rm th} \left( \vec{Q}_{\rm Ar}^2;\vec{x}\right) &=& N_{ij}^{\rm signal} \left( \vec{Q}_{\rm Ar}^2\right) \left(1+\alpha+ \sum_{z=1,2} \alpha_{z,ij}(\epsilon_z)\right) 
\\
&&+\,N_{ij}^{\rm bkg,pBRN} \left(1+\beta_{\rm pBRN}+ \sum_{z=3,4,5} \alpha_{z,ij} (\epsilon_z)\right)
+ \sum_{a=\rm dBRN,SS} N_{ij}^{\text{bkg},a} \left(1+\beta_a\right)~,
\nonumber
\end{eqnarray}
where $\rm pBRN,dBRN,SS$ correspond to prompt BRN, delayed BRN and SS backgrounds (NIN contribution is neglected in this analysis). Their predicted values, $N_{ij}^{\text{bkg},a}$, are readily provided in the LAr measurement data release~\cite{COHERENT:2020ybo}, with efficiencies already applied to them.

Thus, the nuisance parameters in this analysis are $\vec{x}=\{ \alpha, \beta_{\rm pBRN}, \beta_{\rm dBRN}, \beta_{\rm SS}, \epsilon_z\}$, with $1\le z\le 5$, with uncertainties equal to $\{ 13\%, 32\%, 100\%, 0.8\%\}$ and $100\%$ for all five $\epsilon_z$ parameters. We see that the generic $h_{ij}$ functions introduced in~\eref{NiTH} are in this case $h_{ij}^{signal\,/\,bkg,a}(\vec{x}) = \{ \alpha, \beta_{\rm pBRN}, \beta_{\rm dBRN}, \beta_{\rm SS}, \alpha_{z,ij} (\epsilon_z)\}$. 
We have nuisance parameters associated to the systematic uncertainties of the overall normalization of backgrounds ($\beta_a$) and signal ($\alpha$). The latter includes errors associated to detector efficiency, energy calibration, $\text{F}_{90}$ calibration, quenching factor, nuclear form factor and neutrino flux. In addition, this analysis includes systematic uncertainties affecting the shape of the distributions ($\epsilon_z$). In particular, we have two systematic errors affecting the signal distribution, coming from the energy dependence of the $\text{F}_{90}$ distribution and the trigger time mean, and three systematic errors affecting the prompt BRN distribution, with the energy distribution, the trigger time mean and the trigger time width as their sources. These bin-dependent systematics are included through the following functions:
\begin{eqnarray}
\alpha_{z,ij}(\epsilon_z) =  \frac{N^{1\sigma}_{z,ij}-N^{CV}_{ij}}{N^{CV}_{ij}}\, \epsilon_z~,
\end{eqnarray}
where $N^{1\sigma}_{z,ij}$ is the predicted number of events with a 1-$\sigma$ shift due exclusively to the $z$-th systematic error and $N^{CV}_{ij}$ is the predicted central value. Note that these quantities include the total number of events and not only the (signal/pBRN) events affected by the $z$-th systematic error. We take the five 1-$\sigma$ distributions ($1\le z\le 5$) and the three background CV's (pBRN, dBRN,SS) from the COHERENT data release~\cite{COHERENT:2020ybo}. 

The number of expected \cevns events is obtained using~\eref{NijTH}, which we repeat here 
\begin{eqnarray}
N_{ij}^{\rm signal} &=& g_j^{\rm prompt} N_i^{\rm prompt} + g_j^{\rm delayed} N_i^{\rm delayed}~.\nonumber
\label{eq:}
\end{eqnarray}
The timing information (i.e., the $g_j$ factors) is extracted from the neutrino flux characterization presented in the COHERENT data release~\cite{COHERENT:2020ybo,Picciau:2022xzi}. The energy distributions, $N_i$, are calculated using~\eref{NumberofEventsWithExperimentalEffects}, which involves an efficiency function, energy resolution and quenching factor that we describe below. 

The quenching factor is parametrized through a polynomial expression, given by 
\begin{eqnarray}
T\times\,\text{QF}(T)=a_{\text{QF}}T+b_{\text{QF}}T^2,
\end{eqnarray}
where $a_{\text{QF}}$ = 0.246 and $b_{\text{QF}}$ = 0.00078 keV$^{-1}$. 

The detector resolution function is 
\begin{eqnarray}
    \mathcal{R}\left(T_{ee}^{\text{rec}},T_{ee} \right) =
    \frac{1}{\sqrt{2 \pi} \sigma_{ee}} e^{-\frac{-\left(T_{ee}^{\text{rec}} - T_{ee}\right)^2}{2\sigma^2_{ee}}},
\end{eqnarray}
where the $T_{ee}$-dependent width is given by $\sigma_{ee} = 0.58\, \text{keV} \sqrt{T_{ee}/\text{keV}}$. For the lower limit of the $T$ integration in~\eref{NumberofEventsWithExperimentalEffects} we do not use zero but $T_{min} = 79$ eV (average energy to produce a scintillation photon in Ar~\cite{creus2013light}), but this has a negligible impact in our results.

The efficiency function, $\epsilon(T_{ee}^{\text{rec}})$, used in the calculation of the \cevns events is given by COHERENT as a $T_{ee}^{\text{rec}}$ bin dependent quantity~\cite{COHERENT:2020ybo}.

\subsection{CsI measurement} \vspace{0.4cm}

Our CsI analysis uses the 2D distribution in recoil energy and time covering the ranges $8< \text{PE}< 60$ and $0<t(\mu s)<6$ with 1 PE and 0.5 $\mu s$ of width respectively,  which yields a total of 52x12 bins. This is the same bin set as the one used in the original COHERENT analysis~\cite{COHERENT:2021xmm}. 

The formula for the expected number of events per bin is given by
\begin{eqnarray}
N_{ij}^{\rm th} \left( \vec{Q}_{\rm CsI}^2;\vec{x}\right) =
N_{ij}^{\rm signal}\left( \vec{Q}_{\rm CsI}^2\right) \left(1+\alpha\right) 
+ \sum_a N_{ij}^{\text{bkg},a} \left(1+\beta_a\right)~,
\end{eqnarray}
where $a=$ BRN, NIN, SS are the three background sources considered in this analysis. Thus, the nuisance parameters are $\vec{x}=\{ \alpha, \beta_{\rm BRN}, \beta_{\rm NIN}, \beta_{\rm SS}\}$, with uncertainties equal to $\vec{\sigma}=\{ 12\%, 25\%, 35 \%, 2.1\%\}$ and the $h_{ij}$ nuisance functions are simply given by 
$h_{ij}^{signal\,/\,bkg,a}(\vec{x}) = \vec{x}$. 
The $\alpha$ parameter encodes the systematic uncertainties associated to the signal (due to the QF, neutrino flux and form factor), whereas the $\beta$ ones encode the uncertainty associated 
to the normalization of the backgrounds. 

As in the LAr case, the expected number of \cevns events, $N_{ij}^{\rm signal}$, is calculated using~\eref{NijTH} (repeated above), and the prompt and delayed energy distributions, $N^a_i$, are obtained using~\eref{NumberofEventsWithExperimentalEffects}. The latter involves an efficiency function, energy resolution and QF that we describe below. Finally, the time information, i.e., the $g_j$ factors in~\eref{NijTH}, is described afterwards.

The COHERENT prescription for the QF in this analysis is the following~\cite{COHERENT:2021xmm}
\begin{eqnarray}
T\times\,\text{QF}(T) = a_{\text{QF}}T + b_{\text{QF}}T^2 + c_{\text{QF}}T^3 + d_{\text{QF}}T^4,
\end{eqnarray}
where $a_{\text{QF}}=0.0554628$, $b_{\text{QF}}=4.30681\,\text{MeV}^{-1}$, $c_{\text{QF}}=-111.707\,\text{MeV}^{-2}$ and $d_{\text{QF}}=840.384\,\text{MeV}^{-3}$. The measured events are organized in PE bins, so we apply the light yield, given in this case by LY=13.35 PE/$\text{keV}$.

The energy resolution function is given by
\begin{equation}
    \mathcal{R}(\text{PE},T_{ee})=\frac{\left(a\left(1+b\right)\right)^{1+b}}{\Gamma(1+b)}(\text{PE})^b e^{-a(1+b)\text{PE}},
\end{equation}
where $a$ and $b$ encode the $T_{ee}$ dependence:  $a=0.0749\,\text{keV}/T_{ee}$ and $b=9.56\,\text{keV}^{-1} \times T_{ee}$.

The energy-dependent efficiency applied in this measurement is
\begin{eqnarray}
\epsilon(\text{PE})&=&\frac{a}{1+e^{-b((\text{PE}/\text{LY})-c)}}+d~,
\label{eq:efficiencyCsI2}
\end{eqnarray}
where $a=1.320 \pm 0.023$, $b=(0.28598\pm 0.00061)\,\text{PE}^{-1}$, $c=(10.9\pm1.0)\,\text{PE}$, $d=-0.333\pm0.023$. We have checked that these uncertainties have a negligible effect in our fit and thus we have neglected them.

As for the timing  information ($g_j$ factors),  we extract them from the information about the flux for every neutrino flavor, which is binned in time and recoil energy and provided in the data release~\cite{COHERENT:2020ybo}. Integrating over the recoil energy we obtain the prompt and delayed distributions. Then we take into account the timing efficiency of the detector, which is available for this measurement. Namely
\begin{eqnarray}
\epsilon_t(t) = \left\{ \begin{matrix} 1 & \,\; t < a \\  e^{-b(t-a)} & \, \, \, t \geq a \end{matrix} \right. ,
\label{eq:timeefficiency}
\end{eqnarray}
where $a=0.52\,\mu s$ and $b=0.0494\, \mu s^{-1}$. Once again, the impact of the uncertainties of these parameters in our fits is negligible. 
We apply this timing efficiency to the projected time distributions, and then we normalize each of them to recast them as probability distribution functions. The $\nu_\mu$ flux gives the prompt distribution, and the $\nu_e$ (or $\bar{\nu}_\mu$) flux can be used to obtain the delayed distribution. 

Concerning the backgrounds, the BRN and NIN distributions are provided in the COHERENT data release~\cite{COHERENT:2020ybo}. They can be normalized to produce 1D distributions in the recoil energy and timing directions, denoted by $g_a(PE)$ and $f_a(t)$ respectively ($a=$BRN, NIN). The full 2D distributions are obtained just by taking $N_{tot}^a\,g_a(\text{PE})\,f_a(t)\,\epsilon_t(t)$, where $N_{tot}^{\text{BRN}}=18.4$ and $N_{tot}^{\text{NIN}}=5.6$ are the total number of predicted BRN and NIN events.

The SS background can be estimated from the anti-coincidence data (AC), which is also given in a 2D distribution in recoil energy and time.  The projection onto the PE axis provides directly the recoil-energy distribution, while for the description of the time evolution of this background, the collaboration advises the use of an exponential model. 
That exponential is fitted to the projection of the AC data on the time axis and then normalized, yielding an expression $f_{\text{SS}}(t) \propto e^{-a_{\text{SS}} t}$, with $a_{\text{SS}} = -0.0494\, \mu s^{-1}$. This procedure (instead of working directly with the 2D AC distribution) avoids possible biases in the fit due to limited statistics in the sample~\cite{COHERENT:2017ipa}. Since this distribution is inferred directly from the data, it is not necessary to apply efficiencies here.

\section{Details of the SMEFT analysis} \vspace{0.4cm}
\label{app:SMEFT}

In this section we provide additional details regarding our SMEFT fit combining the COHERENT results with other low- and high-energy precision measurements. 
We list the observables used in the fit, and discuss the updates with respect to the analogous fit in Ref.~\cite{Falkowski:2017pss}. 
We also present the global fit to the Wilson coefficients without assuming the $U(3)^5$ flavor symmetry.  

\subsection{Operators} \vspace{0.4cm}

If the BSM particles are not only heavier than the characteristic scale of COHERENT but also heavier than the electroweak scale, then above that scale their dynamics can be described by the effective theory called SMEFT~\cite{Buchmuller:1985jz, Grzadkowski:2010es}. The latter contains all the SM fields (including the heaviest ones, such as the Higgs and the top quark) and the $SU(3)\times SU(2) \times U(1)$ gauge symmetry is (linearly) realized. 
The leading effects of lepton-number-conserving BSM particles are encoded in the dimension-6 operators in the SMEFT Lagrangian.
We parametrize the dimension-6 operators using the so-called Higgs basis~\cite{LHCHiggsCrossSectionWorkingGroup:2016ypw} (see \cite{Azatov:2022kbs} for a recent review) as it is very convenient for the sake of electroweak precision fits.  
The operators involved in our fit are electroweak gauge boson masses and a subset of 4-lepton and 2-lepton-2-quark operators, as well as vertex corrections to electroweak gauge bosons interactions with fermions: 
\begin{align}
\label{eq:SMEFT2_lagrangian}
{\cal L}_{\rm SMEFT} \supset 
 {\cal L}_{2l2q} + {\cal L}_{4l} + {\cal L}_{\rm vertex} 
 + {\cal L}_{\rm VV} .   
\end{align}
The operators in the three categories entering our fit are defined as 
\begin{align}
\label{eq:SMEFT2_2l2q}
{\cal L}_{2l2q} = & 
[C_{lq}^{(1)}]_{JKMN} (\bar l_L^J \gamma_\mu l_L^K)(\bar q_L^M \gamma^\mu q_L^N)
+ [C_{lq}^{(3)}]_{JKMN} (\bar l_L^J \gamma_\mu \sigma^k l_L^K)(\bar q_L^M \gamma^\mu \sigma^k q_L^N)
\nnl + & 
 [C_{lu}]_{JKMN} (\bar l_L^J \gamma_\mu l_L^K)(\bar u_R^M \gamma^\mu u_R^N)
+ [C_{ld}]_{JKMN} (\bar l_L^J \gamma_\mu l_L^K)(\bar d_R^M \gamma^\mu d_R^N)
\nnl + & 
 [C_{eu}]_{JKMN} (\bar e_R^J \gamma_\mu e_R^K)(\bar u_R^M \gamma^\mu u_R^N)
+ [C_{ed}]_{JKMN} (\bar e_R^J \gamma_\mu e_R^K)(\bar d_R^M \gamma^\mu d_R^N) 
\nnl + & 
[C_{eq}]_{JKMN} (\bar e_R^J \gamma_\mu e_R^K)(\bar q_L^M \gamma^\mu q_L^N) 
+ \left\{ \right.[C_{lequ}^{(1)}]_{JKMN}(\bar l_{L,j}^J e_R^K) \epsilon^{jk} (\bar q_{L,k}^M  u_R^N) 
\nnl + & 
 [C_{lequ}^{(3)}]_{JKMN}(\bar l_{L,j}^J \sigma^{\mu\nu} e_R^K) \epsilon^{jk} (\bar q_{L,k}^M \sigma^{\mu\nu}  u_R^N) 
+ [C_{ledq}]_{JKMN}(\bar l_L^J e_R^K) (\bar d_R^M q_L^N) \left. + \rm{h.c.} \right\}, \\[10pt]
\label{eq:SMEFT2_4l}
{\cal L}_{4l} = & 
{1 \over 2} [C_{ll}]_{IJKL} (\bar l_L^I \gamma_\mu l_L^J)(\bar l_L^K \gamma^\mu l_L^L)
+ {1 \over 2} [C_{ee}]_{IJKL} (\bar e^I_R \gamma_\mu e^J_R)(\bar e^K_R \gamma^\mu e^L_R)
\nnl + & 
 [C_{le}]_{IJKL} (\bar l_L^I \gamma_\mu l_L^J)(\bar e^K_R \gamma^\mu e^L_R), 
\end{align}
\begin{align}
\label{eq:SMEFT2_vertex}
{\cal L}_{\rm vertex} = & 
- \sqrt{g_L^2 + g_Y^2} Z_\mu \bigg \{ 
\bigg [ {1 \over 2} \mathbb{1}    + \delta g_L^{W\ell}  +  \delta g_L^{Z e} \bigg ]_{JK}  (\bar \nu_L^J \gamma^\mu \nu_L^K)
\nnl + & 
\bigg [ \bigg (- {1 \over 2}  + \sin^2 \theta_W \bigg )\mathbb{1} +  \delta g_L^{Z e} \bigg ]_{JK}  (\bar e_L^J \gamma^\mu e_L^K)
+\big [ \sin^2 \theta_W  \mathbb{1} +  \delta g_R^{Z e} \big ]_{JK}  (\bar e_R^J \gamma^\mu e_R^K)
\nnl + & 
\bigg [ \bigg ({1 \over 2}  -  {2 \over 3} \sin^2 \theta_W \bigg )\mathbb{1}  + \delta g_L^{Zu}  \bigg ]_{MN}  (\bar u_L^M \gamma^\mu  u^N_L)
+  \bigg [  - {2 \over 3} \sin^2 \theta_W  \mathbb{1}  
+ \delta g_R^{Zu} \bigg ]_{MN}  (\bar u_R^M \gamma^\mu u_R^N)
\nnl + & 
 \bigg [ \bigg (-{1 \over 2} 
 + \frac{1}{3}\sin^2 \theta_W \bigg )\mathbb{1}  + \delta g_L^{Zd} \bigg ]_{MN}  (\bar d_L^M \gamma^\mu  d_L^N)
+ \bigg [ {1 \over 3} \sin^2 \theta_W  \mathbb{1}  + \delta g_R^{Zd} \bigg ]_{MN}  (\bar d_R^M  \gamma^\mu  d_R^N)
\bigg \} 
\nnl + & 
 {g_L \over \sqrt 2}  \bigg [ 
  W^{\mu+} \big [ \mathbb{1} + \delta g^{Wl}_L \big]_{JK}  (\bar \nu_L^J   \gamma_\mu  e^K_L)
  +W^{\mu+} \big [ V + \delta g^{Z u}_L - \delta g^{Z d}_L \big]_{MN}  (\bar u_L^M   \gamma_\mu   d_L^N) 
  \nnl + & 
  W^{\mu+} \big [\delta g^{W q}_R \big]_{MN}  (\bar u_R^M   \gamma_\mu  d_R^N) +\hc \bigg ]~, \\[10pt]
\label{eq:SMEFT2_vv}
{\cal L}_{\rm VV} ~= & 
~ {(g_L^2 + g_Y^2) v^2 \over 8} Z_\mu Z^\mu +  {g_L^2 v^2 \over 4} \left (1 + 2 \delta m \right) W_\mu^+ W^{-\,\mu} . 
\end{align}
Here, $l_L^J \equiv (\nu_L^J,e_L^J)$ and $q_L^M \equiv (V^\dagger u_L^M,d_L^M)$ are the lepton and  quark  doublets of the $SU(2)_L$ gauge group, 
while $e_R^J$,  $u_R^M$, $d_R^M$ are $SU(2)_L$ singlets.
The generation indices, denoted by  capital Latin letters,  $I,J,K,L=e,\mu,\tau$, 
$M,N=1,2,3$, are implicitly summed over.
All fields except for $q_L^M$ are in the mass eigenstate basis, while for $q_L^M$ we use the down-type basis where $d_L^M$ and $u_L^M$ are mass eigenstates and $V$ is the CKM matrix.
In our power counting the off-diagonal CKM elements multiplying dimension-6 Wilson coefficients are treated as zero (in this approximation, there is actually no practical difference between the down-type and up-type bases).  
In the 4-lepton Lagrangian we identify the Wilson coefficients of identical operators: 
$[C_X]_{IJKL} = [C_X]_{KLIJ}$ for $X=ll,ee$.\footnote{
In our conventions, the normalization of the $[C_{ll}]_{JJJJ}$ and $[C_{ee}]_{JJJJ}$ Wilson coefficients differs by a factor of 1/2 from most of the SMEFT literature. The motivation for our normalization is that it makes it  more straightforward to impose the   $U(3)^5$ symmetry.} 
Moreover, $[C_{ee}]_{IJJI}=0$ (for $J\neq I$) and $[C_X]_{IJKL}  = [C_X]_{JILK}^*$ (for $X=ll,ee,l e$). 
Finally, the $W$ mass correction is connected to the other parameters by 
$\delta m = {[\delta g_L^{Wl}]_{ee}+ [\delta g_L^{Wl}]_{\mu\mu} \over 2} - {v^2[C_{ll}]_{e\mu\mu e} \over 4}$.

The four-fermion pieces, ${\cal L}_{2l2q}$ and ${\cal L}_{4l}$, are manifestly invariant under the SM gauge group, 
and are actually the same as the analogous terms in the Warsaw basis. 
Concerning ${\cal L}_{\rm vertex}$, 
although it is manifestly invariant only under $SU(3)_C \times U(1)_{\rm em}$, it is obtained  from an $SU(3)_C \times SU(2)_L \times U(1)_{Y}$ invariant Lagrangian after electroweak symmetry breaking.  
In this case the imprints of the full SM gauge symmetry are the correlations between the $W$ and $Z$ couplings to fermions.
The map between the vertex corrections and the 
 Wilson coefficients of the manifestly SM gauge-invariant Warsaw basis is 
\begin{align}
\label{eq:dgtowarsaw}
\delta g^{W\ell}_L  = &   c^{(3)}_{H \ell} + f(1/2,0) - f(-1/2,-1), 
\nnl
\delta g^{Ze}_L  = &    - {1 \over 2} c^{(3)}_{H\ell} - {1\over 2} c_{H\ell}    +   f(-1/2, -1), 
\nnl
\delta g^{Ze}_R  = &    - {1\over 2} c_{He}   +  f(0, -1),  
\nnl 
\delta g^{Zu}_L  = &   {1 \over 2}  c^{(3)}_{Hq} - {1\over 2} c_{Hq}^{(1)}   + f(1/2,2/3), 
\nnl
\delta g^{Zd}_L = &    -{1 \over 2}  c^{(3)}_{Hq}    - {1\over 2}  c_{Hq}^{(1)}    + f(-1/2,-1/3),
\nn
\delta g^{Zu}_R  = &    - {1\over 2} c_{Hu}   +  f(0,2/3),  
\nn
\delta g^{Zd}_R  = &    - {1\over 2} c_{Hd}  +  f(0,-1/3), 
\nnl
\delta g^{Wq}_R  = &   - {1 \over 2 } c_{Hud},
\end{align}
where 
\begin{align}
f(T^3,Q) = & 
-     Q   {g_L g_Y \over g_L^2 - g_Y^2} c_{HWB} \mathbb{1}
\\ + & 
\left ( {1 \over 4 }[c_{ll}]_{e\mu\mu e}  - {1 \over 2}  [c^{(3)}_{H l } ]_{ee}  -  {1 \over 2} [c^{(3)}_{H l } ]_{\mu \mu} 
- {1 \over 4} c_{HD}  \right )  \left ( T^3 + Q {g_Y^2 \over g_L^2 - g_Y^2} \right ) \mathbb{1} ,  
\nonumber
\end{align}

In the $U(3)^5$ symmetric limit, assumed in the main body of this paper, our Wilson coefficients reduce to 
\begin{equation}
\label{eq:SMEFT2_u5params}
\begin{pmatrix}
[\delta g^{W\ell}_L]_{JJ} \\ 
 [\delta g^{Ze}_L]_{JJ} \\ 
[\delta g^{Ze}_R]_{JJ}  \\ 
 [\delta g^{Zu}_L]_{NN} \\ 
 [\delta g^{Zu}_R]_{NN}  \\ 
[\delta g^{Zd}_L]_{NN} \\ 
 [\delta g^{Zd}_R]_{NN} 
\end{pmatrix}
= 
\begin{pmatrix}
\delta g^{W\ell}_L \\ 
 \delta g^{Ze}_L \\ 
 \delta g^{Ze}_R  \\ 
 \delta g^{Zu}_L \\ 
\delta g^{Zu}_R  \\ 
\delta g^{Zd}_L \\ 
 \delta g^{Zd}_R 
\end{pmatrix}
, \, 
\begin{pmatrix}
  [c_{l l}]_{JJJJ} \\ 
  [c_{l l}]_{IJJI}, I\neq J \\ 
   [c_{l l}]_{IIJJ},I\neq J \\ 
  [c_{l e}]_{IIJJ} \\     
 [c_{e e}]_{IIJJ}    
\end{pmatrix}
=  
\begin{pmatrix}
 c_{l l}^{(1)}  +  c_{l l}^{(3)}  \\
 2  c_{l l}^{(3)} \\ 
c_{l l}^{(1)}  -c_{l l}^{(3)}   \\
 c_{l e} \\     
c_{e e} 
\end{pmatrix}
, \,
\begin{pmatrix}
 [c_{lq}^{(3)}]_{JJNN} \\  
 [c_{lq}^{(1)}]_{JJNN} \\   
 [c_{e q}]_{JJNN} \\
 [c_{l u}]_{JJNN} \\   
 [c_{l d}]_{JJNN}  \\   
 [c_{eu}]_{JJNN}  \\   
 [c_{ed}]_{JJNN}
\end{pmatrix}
=
\begin{pmatrix}
c_{lq}^{(3)} \\  c_{lq}^{(1)} \\   c_{e q} \\
   c_{l u} \\   c_{l d}  \\   c_{eu}  \\   c_{ed}
\end{pmatrix}, 
\end{equation}
while the remaining Wilson coefficients in \cref{eq:SMEFT2_2l2q,eq:SMEFT2_4l,eq:SMEFT2_vertex} should be set to zero.

\subsection{Experimental input} \vspace{0.4cm}

Here we list the observables that are included in our update of the global SMEFT fit carried out in Ref.~\cite{Falkowski:2017pss}, with special emphasis on the changes with respect to that work:
\begin{itemize}
\item $e^+ e^-$ collisions at energies above~\cite{Electroweak:2003ram,ALEPH:2013dgf}, at~\cite{ALEPH:2005ab,SLD:2000jop,ParticleDataGroup:2022pth}, and below~\cite{VENUS:1993pob,VENUS:1997cjg,TOPAZ:2000evx} the $Z$-pole.  
Concerning the Z-pole results we follow Ref.~\cite{Breso-Pla:2021qoe}, where we took into account recent theoretical calculations~\cite{Janot:2019oyi,dEnterria:2020cgt} that lead to minor modifications of the $Z$ width, the hadronic cross section,  and the forward-backward asymmetry of b quarks.
\item For W boson data we follow Ref.~\cite{Breso-Pla:2021qoe} as well: we include the mass and total width~\cite{ParticleDataGroup:2022pth}, leptonic branching ratios from LEP~\cite{ALEPH:2013dgf}, Tevatron~\cite{D0:1999bqi} and LHC~\cite{LHCb:2016zpq,ATLAS:2016nqi,ATLAS:2020xea}, and the ratio 
$R_{Wc}\equiv \Gamma(W \!\to\! c s) / \Gamma(W \!\to\! u d, c s)$~\cite{ParticleDataGroup:2022pth}, which has a limited precision, but helps to break a flat direction. 
These data include three significant updates with respect to the fit in Ref.~\cite{Falkowski:2017pss}: (i) the recent LHC results concerning the leptonic branching ratios~\cite{LHCb:2016zpq,ATLAS:2016nqi,ATLAS:2020xea}, which play an important role in removing certain LEP-2 tensions and improving the constraints on the $W\ell\nu$ vertices; (ii) the $W$ boson mass, which is updated using the current PDG combination $m_W = 80.377(12)$~GeV~\cite{LHCb:2021bjt,ATLAS:2017rzl,CDF:2012gpf,D0:2012kms,ALEPH:2013dgf,ParticleDataGroup:2022pth}\footnote{We note this average does {\em not} include the recent CDF result~\cite{CDF:2022hxs}, which is in tension with the other most precise measurements.}; (iii) we no longer use Ref.~\cite{CMS:2014mgj} to extract the $Wt_Lb_L$ vertex (and hence $[\delta g^{Zu}_L]_{33}$), as this measurement is also sensitive to other operators in the top sector.
\item 
Forward-backward asymmetries in $\ell^+\ell^-$ production at the LHC~\cite{ATLAS:2018gqq} and D0~\cite{D0:2011baz}. The LHC bounds, which were obtained in Ref.~\cite{Breso-Pla:2021qoe}, represent a novel addition with respect to the fit in Ref.~\cite{Falkowski:2017pss} and they tighten the LEP constraints on the Z boson couplings to first generation quarks.
\item Electron-neutrino scattering on nuclei by the CHARM experiment~\cite{CHARM:1986vuz}. 
\item  Muon-neutrino scattering on nuclei in  the CHARM~\cite{CHARM:1987pwr}, CDHS~\cite{Blondel:1989ev}, and CCFR~\cite{CCFR:1997zzq} experiments. 
As in Ref.~\cite{Falkowski:2017pss}, we use the PDG combination of these data, which includes as well additional experimental input (with very limited precision) from elastic neutrino-proton scattering and neutrino-induced coherent neutral-pion production from nuclei.
\item Parity violation in atoms and in scattering: 
(i) measurements of atomic parity violation in cesium~\cite{Wood:1997zq} and thallium~\cite{Edwards:1995zz,Vetter:1995vf} atoms; 
(ii) the weak charge of the proton measured in scattering of low-energy polarized electrons by QWEAK~\cite{Qweak:2018tjf};
and (iii) deep-inelastic scattering of polarized electrons on deuterium by the PVDIS experiment~\cite{PVDIS:2014cmd}. 
As in Ref.~\cite{Falkowski:2017pss}, we use the PDG combination of these data, supplemented by the SAMPLE measurement of the scattering of polarized electrons on deuterons in the quasi-elastic kinematic regime~\cite{Beise:2004py}. The only difference with Ref.~\cite{Falkowski:2017pss} is that we use the 2022 PDG combination (Table~10.9 of Ref.~\cite{ParticleDataGroup:2022pth}), which includes the updated measurement of the proton weak charge by the QWEAK experiment~\cite{Qweak:2018tjf}.
\item  Deep-inelastic scattering of polarized  muons on carbon at the CERN SPS~\cite{Argento:1982tq}.
\item Various (semi-)leptonic hadron decays (nuclear beta, pion, kaon) mediated by the quark-level process $d(s)\to u \ell \bar{\nu}_\ell$. The global fit in Ref.~\cite{Falkowski:2017pss} used the bounds obtained in Ref.~\cite{Gonzalez-Alonso:2016etj}, whereas here we use instead the updated version of Ref.~\cite{Cirigliano:2021yto}. These updates concern new nuclear beta decay input, lattice calculations and a refined analysis of radiative pion decay $\pi^{-}\rightarrow e^{-}\bar{\nu}_{e}\gamma$~\cite{Falkowski:2020pma,Cirigliano:2021yto}. The new measurements of the neutron lifetime~\cite{UCNt:2021pcg} and beta asymmetry~\cite{UCNA:2017obv,Markisch:2018ndu} are particularly important, entailing a significant improvement in the tensor 4-fermion interaction $[c^{(3)}_{lequ}]_{ee11}$.\footnote{There is a nonzero correlation between up-down effective couplings (the focus of this work) and up-strange couplings (which we
marginalized over). This must be taken into account when going to specific scenarios, such as the one-a-time results in~\tref{SMEFT_onebyone} or the flavor-blind fit in~\eref{globalfitresults}. The full likelihood is
available in Ref.~\cite{Cirigliano:2021yto}.}  
\item 
Muon-neutrino scattering on electrons~\cite{CHARM:1988tlj,Ahrens:1990fp,CHARM-II:1994dzw}. 
As in Ref.~\cite{Falkowski:2017pss}, we use the PDG combination for the low-energy $\nu_\mu-e$ couplings~\cite{ParticleDataGroup:2022pth}.
\item Parity-violating scattering of electrons at low energies in the SLAC~E158 experiment~\cite{SLACE158:2005uay}. 
As in Ref.~\cite{Falkowski:2017pss}, we use the PDG combination for the low-energy parity-violating electron self-coupling~\cite{ParticleDataGroup:2022pth}.
\item Trident production $\nu_\mu \gamma^* \to \nu_\mu \mu^+ \mu^-$ in the CHARM-II~\cite{CHARM-II:1990dvf} and CCFR~\cite{CCFR:1991lpl} experiments. 
\item 
Leptonic decays of taus and muons~\cite{ParticleDataGroup:2022pth}.
Unlike \cite{Falkowski:2017pss}, we no longer use the ratio of effective Fermi constants  
$G_{\tau \mu}/G_F$ from the {\it $\tau$-Lepton Decay Parameters} review in Ref.~\cite{ParticleDataGroup:2016lqr}, since its large correlation with the poorly known $[c_{le}]_{\mu\tau\tau\mu}$ coefficient was not provided.  
Instead, we directly use the measured $\tau\to e\nu\bar{\nu}$ branching fraction and the associated Michel parameter~\cite{ParticleDataGroup:2022pth}. 
For completeness we also include the measurement of the Michel parameter in muon decay~\cite{ParticleDataGroup:2022pth}. These modifications allow us to target additional 4-fermion semileptonic operators, not constrained by the fit in Ref.~\cite{Falkowski:2017pss}, namely the $[c_{le}]_{\mu\tau\tau\mu}$ and $[c_{le}]_{e\mu\mu e}$ coefficients.
\item 
We include a set of hadronic tau decay observables described in Ref.~\cite{Cirigliano:2021yto}. These constraints, which were obtained subsequently to the global fit of Ref.~\cite{Falkowski:2017pss}, give access to contact interactions between first generation quarks and third generation leptons. They also improve our knowledge of the vertex corrections.
\end{itemize}
We refer to Refs.~\cite{Falkowski:2017pss,Cirigliano:2021yto,Breso-Pla:2021qoe} and to the original experimental papers for the central values and errors used in our fit. 
We calculate these observables at tree level in the SMEFT  in terms of the Wilson coefficients in \cref{eq:SMEFT2_2l2q,eq:SMEFT2_4l,eq:SMEFT2_vertex}, and expand them to linear order. In other words, we keep track of the $\cO(1/\Lambda^2)$ effects in the SMEFT power counting, while ignoring $\cO(1/\Lambda^4)$ effects due to dimension-8 operators and squares of dimension-6 operators. 
We also ignore all loop effects (except for QCD running of certain WEFT operators below the electroweak scale).   

In addition to the inputs above, we consider the COHERENT constraints, which represent the main topic of this paper. Below we incorporate them for first time in this kind of global SMEFT fit. For that, the constraints on the WEFT Wilson coefficients listed in \cref{eq:epsilonbounds} are translated into constraints on the SMEFT Wilson coefficients using the map
\begin{align}
\label{eq:SMEFT2_epsilonToHiggsBasis} 
\epsilon_{\alpha\alpha}^{uu} ~=~& 
\delta g_L^{Zu} + \delta g_R^{Zu} 
+\left ( 1 - {8 s_\theta^2 \over 3 }  \right ) \delta g_L^{Z\nu_\alpha} 
- \frac{1}{2} [ c^{(1)}_{lq} + c^{(3)}_{lq} + c_{lu} ]_{\alpha\alpha11} ~,
\nnl 
\epsilon_{\alpha\alpha}^{dd} 
~=~& 
 \delta g_L^{Zd} +  \delta g_R^{Zd}  
- \left (  1 -   {4 s_\theta^2  \over 3 }  \right ) \delta g_L^{Z\nu_\alpha} 
-\frac{1}{2} [ c^{(1)}_{lq} -  c^{(3)}_{lq} + c_{ld}]_{\alpha\alpha11} ~.
\end{align}

\subsection{Fit results} \vspace{0.4cm}

As discussed at length in Ref.~\cite{Falkowski:2017pss}, not all linear combinations of the Wilson coefficients in \cref{eq:SMEFT2_2l2q,eq:SMEFT2_4l,eq:SMEFT2_vertex} can be constrained by the observables we consider. 
Some of the Wilson coefficients are not constrained at all by these observables (at least in the tree-level approximation we use), while others display flat directions (only certain linear combinations of Wilson coefficients are constrained, but not all of them independently). 
One can show that adding the COHERENT results, while improving some constraints, does not lift any flat directions in the fit of Ref.~\cite{Falkowski:2017pss}. 
On the other hand, the new observables we add (hadronic tau decays, Michel parameters) lead to new Wilson coefficients being constrained, but again do not lift any existing flat directions.
In order to isolate the flat directions we define the hatted variables 
\begin{align}
\label{eq:SMEFT2_chats}
  \, [\hat c_{e q}]_{ee11}   = &   [c_{e q}]_{ee11} +[c_{lq}^{(1)}]_{ee11}, 
     \nnl 
  \, [\hat c_{l u}]_{ee11}  = & [c_{l u}]_{ee11} + [c_{lq}^{(1)}]_{ee11}  -   [\hat c_{e q}]_{ee11}, 
     \nnl 
   \,   [\hat c_{l d}]_{ee11}    = &   [c_{l d}]_{ee11}+[c_{lq}^{(1)}]_{ee11}   -   [\hat c_{e q}]_{ee11} , 
   \nnl
     \,   [\hat c_{eu}]_{ee11}   = &     [c_{eu}]_{ee11} -   [c_{lq}^{(1)}]_{ee11} ,
       \nnl 
    \,  [\hat c_{ed}]_{ee11}    = &  [c_{ed}]_{ee11} -   [c_{lq}^{(1)}]_{ee11} ,
     \nnl 
  \,  [\hat c_{l q}^{(3)}]_{ee22}   = &      [c_{l q}^{(3)}]_{ee22} -  [c_{lq}^{(1)}]_{ee22},
     \nnl 
   \, [\hat c_{l d}]_{ee22}   = &    [c_{l d}]_{ee22}  
   +   \left (5 -  {3 g_L^2 \over g_Y^2} \right ) [c_{l q}^{(1)}]_{ee22}  - [\hat c_{e q}]_{ee11}  ,
     \nnl 
   \,  [\hat c_{ed}]_{ee22}    = &      [c_{ed}]_{ee22}  
     -  \left (3 -  {3 g_L^2 \over g_Y^2} \right )[c_{lq}^{(1)}]_{ee22}  - [\hat c_{eq}]_{ee11}  ,
\nnl 
 \,  [\hat c_{l q}^{(3)}]_{ee33}   = &    [c_{l q}^{(3)}]_{ee33}  +  [c_{l q}^{(1)}]_{ee33}, 
 \nnl
 \, [\hat c_{eq}]_{\mu\mu 11}   =&    [c_{eq}]_{\mu\mu 11} + [c_{ed}]_{\mu\mu 11} - 2 [c_{eu}]_{\mu\mu 11},
 \nnl 
 \epsilon_P^{d \ell} (2~\rm{GeV})  = & 0.86 [c_{ledq}]_{\ell\ell 11} - 0.86 [c_{lequ}^{(1)}]_{\ell\ell 11} + 0.012  [c_{lequ}^{(3)}]_{\ell\ell 11}, 
 \nnl 
\, [\hat c_{l l}]_{\mu\mu\mu\mu} = &  [c_{l l}]_{\mu\mu\mu\mu} 
 + {2 g_Y^2 \over g_L^2 + 3 g_Y^2} [c_{l e}]_{\mu\mu\mu\mu}.  
\end{align}
Using these variables, the global likelihood depends on the Wilson coefficients 
on the right-hand sides of Eqs.~(\ref{eq:SMEFT2_chats}) only via the $\hat c$ and $\eps_P^{d \ell}(2~\rm{GeV})$ combinations.
Let us stress that the Wilson coefficients in the r.h.s. of $\epsilon_P^{d \ell}(2~\rm{GeV})$ in \cref{eq:SMEFT2_chats} are defined at $\mu=m_Z$. 

All in all, we simultaneously fit 65 independent (combinations of) SMEFT Wilson coefficients, including all correlations. 
We find the following $1\sigma$ intervals
\begin{equation}
\nonumber
\begin{pmatrix}
\, [\delta g^{Wl}_L]_{ee} \\ 
\, [\delta g^{Wl}_L]_{\mu\mu} \\ 
\, [\delta g^{Wl}_L]_{\tau\tau} \\ 
\, [\delta g^{Ze}_L]_{ee} \\ 
\, [\delta g^{Ze}_R]_{ee}  \\ 
\, [\delta g^{Ze}_L]_{\mu\mu} \\ 
\, [\delta g^{Ze}_R]_{\mu\mu} \\ 
\, [\delta g^{Ze}_L]_{\tau\tau} \\ 
\, [\delta g^{Ze}_R]_{\tau\tau} \\ 
\, [\delta g^{Wq}_R]_{11}  \\
\, [\delta g^{Zu}_L]_{11} \\ 
\, [\delta g^{Zu}_R]_{11}  \\ 
\, [\delta g^{Zd}_L]_{11} \\ 
\, [\delta g^{Zd}_R]_{11} \\ 
\, [\delta g^{Zu}_L]_{22} \\ 
\, [\delta g^{Zu}_R]_{22} \\ 
\, [\delta g^{Zd}_L]_{22} \\
\, [\delta g^{Zd}_R]_{22} \\ 
\, [\delta g^{Zd}_L]_{33} \\
\, [\delta g^{Zd}_R]_{33} \\
\end{pmatrix}
= 
\begin{pmatrix}
-1.8(2.6) \\
-0.6(2.2) \\
0.2(3.5)  \\
-0.21(28) \\
-0.42(27) \\
 0.2(1.2) \\
 0.0(1.4) \\
-0.09(59) \\
0.61(62) \\
-3.8(8.1) \\ 
-7(22) \\
4(29) \\
-13(35)\\
10(120) \\
-1.5(3.6) \\
-3.3(5.3) \\
14(27)\\
34(46) \\
3.2(1.7) \\
22(8.8) \\
\end{pmatrix}\times 10^{-3} , 
\quad
\begin{pmatrix}
\, [c_{l l}]_{eeee} \\ 
 \, [c_{l e}]_{eeee} \\     
\,  [c_{e e}]_{eeee}   \\    
\, [c_{l l}]_{e\mu\mu e} \\ 
\, [c_{l l}]_{ee\mu\mu} \\ 
 \, [c_{l e}]_{e\mu\mu e} \\ 
 \, [c_{l e}]_{ee\mu\mu} \\   
 \, [c_{l e}]_{\mu\mu ee} \\    
\,  [c_{e e}]_{ee \mu\mu}  \\     
\, [c_{l l}]_{e\tau\tau e} \\ 
\, [c_{l l}]_{ee\tau\tau} \\ 
 \, [c_{l e}]_{ee \tau\tau} \\
 \,  [c_{l e}]_{\tau\tau ee} \\   
\,  [c_{e e}]_{ee \tau\tau}     \\ 
\, [\hat c_{l l}]_{\mu\mu\mu\mu} \\ 
\, [c_{l l}]_{\mu\tau\tau\mu} \\ 
\, [c_{l e}]_{\mu\tau\tau\mu}
\end{pmatrix}
= 
\begin{pmatrix}
1.03(38) \\
-0.22(22)\\
 0.19(38) \\
-0.56(80) \\
 0.1(2.0) \\
 11.4(6.8) \\
0.3(2.2) \\
-0.2(2.1) \\
0.2(2.3) \\
-0.60(68) \\
 2(11) \\
 -2.3(7.2) \\
1.7(7.2) \\
-1(12) \\ 
 2(21) \\
1.5(1.9) \\
19(15)
\end{pmatrix}
\times 10^{-2},
\end{equation}
\begin{equation}
    \label{eq:SMEFT2_alloutDeltaG}
    \vspace{-10pt}
\end{equation}
\begin{equation}
\nonumber
\qquad
\begin{pmatrix}
\, [c_{lq}^{(3)}]_{ee11}  \\  
\, {\color{red} [\hat c_{e q}]_{ee11} } \\
\, [\hat c_{l u}]_{ee11} \\  
\,   [\hat c_{l d}]_{ee11}  \\  
\,   [\hat c_{eu}]_{ee11} \\   
\,  [\hat c_{ed}]_{ee11} \\ 
 \, [c_{l e q u}^{(1)}]_{ee11} \\ 
\, [c_{l e d q}]_{ee11}  \\
\,[c^{(3)}_{l e q u}]_{ee11}  \\ 
\,  [\hat c_{lq}^{(3)}]_{ee22} \\ 
\,  [c_{l u}]_{ee22}  \\  
\,   [\hat c_{l d}]_{ee22} \\  
\,    [c_{eq}]_{ee22}  \\   
\,  [c_{eu}]_{ee22} \\  
\,  [\hat c_{ed}]_{ee22} \\ 
 \,  [\hat c_{lq}^{(3)}]_{ee33}  \\ 
 \,  [c_{l d}]_{ee33} \\  
 \,  [c_{eq}]_{ee33} \\ 
 \,   [c_{ed}]_{ee33}
\end{pmatrix}
= 
\begin{pmatrix}
0.1(2.8)_{\phantom{e}}^{\phantom{(}}  \\  
 {\color{red} -4(30)_{\phantom{e}}} \\
 -2.5(8.7)_{\phantom{e}} \\  
  -2(18)_{\phantom{e}} \\  
  -3.1(9.4)_{\phantom{e}} \\   
-2(17)_{\phantom{e}} \\ 
 -0.017(60)_{\phantom{e}}^{\phantom{(}} \\ 
 -0.018(57)_{\phantom{e}} \\
 0.023(66)_{\phantom{e}}^{\phantom{(}}  \\ 
  -61(32)_{\phantom{e}}^{\phantom{(}} \\ 
  2.4(8.0)_{\phantom{e}}  \\  
  -300(130)_{\phantom{e}} \\  
   -21(28)_{\phantom{e}}  \\   
  -87(46)_{\phantom{e}} \\  
  250(140)_{\phantom{e}} \\ 
  -8.5(8.0)_{\phantom{e}}^{\phantom{(}}  \\ 
  -1(10)_{\phantom{e}} \\  
  -3.1(5.1)_{\phantom{e}} \\ 
  18(20)_{\phantom{e}} 
\end{pmatrix} \times 10^{-2}~,
\quad
\begin{pmatrix}
 \, [c_{lq}^{(3)}]_{\mu\mu11} \\ 
 \, [c_{lq}^{(1)}]_{\mu\mu 11} \\  
 \, {\color{red} [c_{l u}]_{\mu\mu 11} }\\   
 \,  {\color{red} [c_{l d}]_{\mu\mu 11} } \\ 
 \, [\hat c_{eq}]_{\mu\mu 11}  \\
 \, \eps_P^{d \mu}(2~\rm{GeV}) \\
  \,  [c_{l q}^{(3)}]_{\tau\tau 11}  \\ 
  \,[c^{(3)}_{l e q u}]_{\tau\tau 11}  \\ 
  \, \eps_P^{d \tau}(2~\rm{GeV}) \\
\end{pmatrix}
= 
\begin{pmatrix}
 3.0(3.5)_{\phantom{e}}^{\phantom{(}} \\ 
 -0.2(5.8)_{\phantom{e}} \\  
 {\color{red} 2.5(6.5)_{\phantom{e}}^{\phantom{(}}}\\   
{\color{red} 5(24)_{\phantom{e}}^{\phantom{(}} } \\ 
 3(41)_{\phantom{e}}  \\
 -0.080(95)_{\phantom{e}}^{\phantom{(}} \\
 -0.3(2.8)_{\phantom{e}}^{\phantom{(}}  \\ 
 -0.3(1.2)_{\phantom{e}}^{\phantom{(}} \\ 
 0.93(85)_{\phantom{e}}^{\phantom{(}}
\end{pmatrix} \times 10^{-2}. 
\end{equation}
We restrain ourselves from copying here the 
$65\times 65$ correlation matrix. 
This, as well as the full Gaussian likelihood function in the Higgs or Warsaw basis, is available in the numerical form on request. 
We highlighted in red the Wilson coefficients whose bounds have improved significantly thanks to including the COHERENT results. 
The improvement is illustrated in \cref{fig:SMEFT2_ellipses}. 
The most spectacular effect is observed for the combination 
$[\hat c_{e q}]_{ee11} = [c_{e q}]_{ee11} +[c_{lq}^{(1)}]_{ee11}$. 
In the fit of Ref.~\cite{Falkowski:2017pss} it was constrained only by the CHARM measurement of electron neutrino scattering on nuclei~\cite{CHARM:1986vuz}, which is however very imprecise. 
The COHERENT results analyzed in this paper reduce the error bars on $[\hat c_{e q}]_{ee11}$ by a factor of 5.

\begin{figure}
    \centering
    \includegraphics[width=0.47\textwidth]{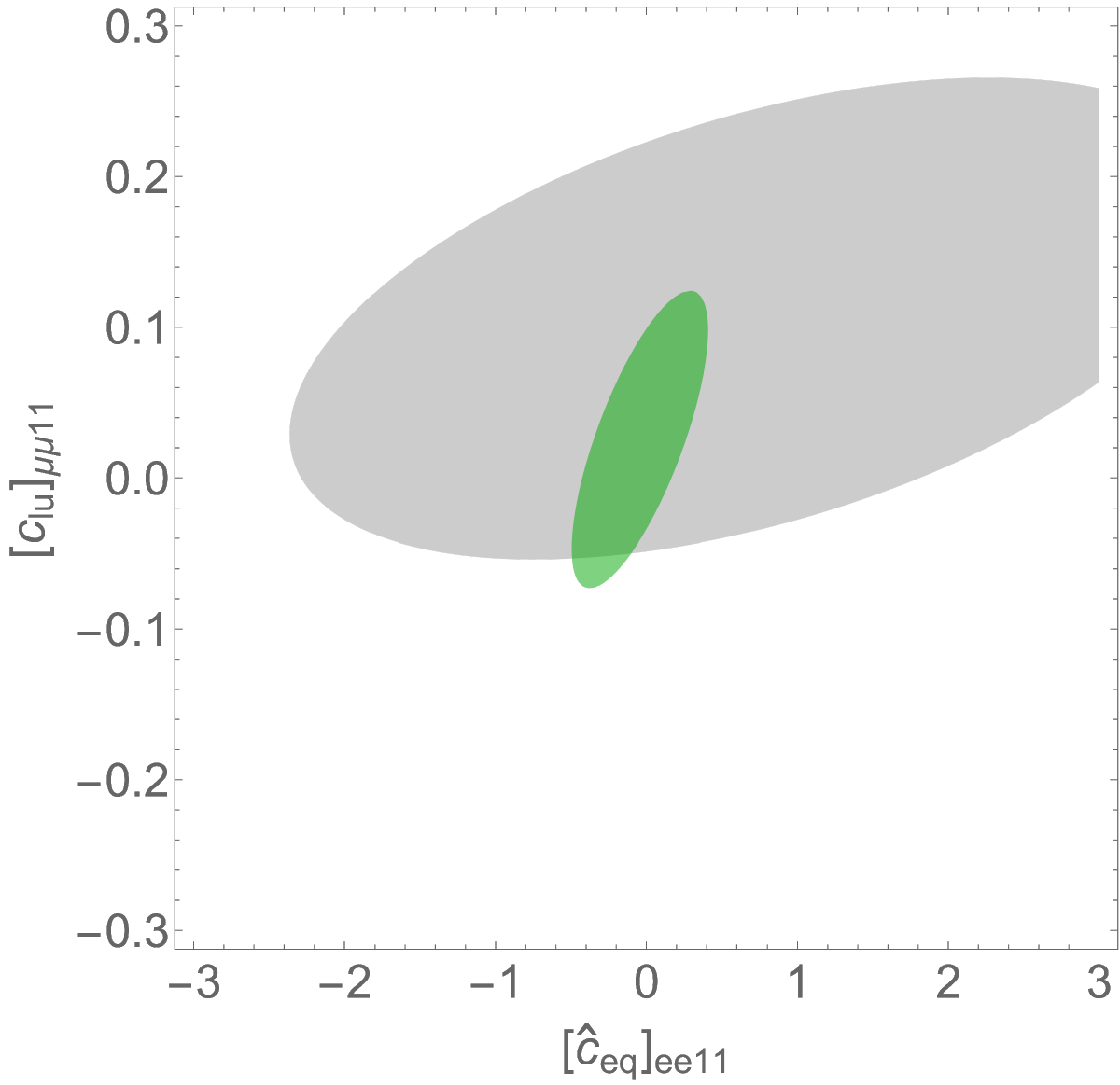}
    \quad 
  \includegraphics[width=0.47\textwidth]{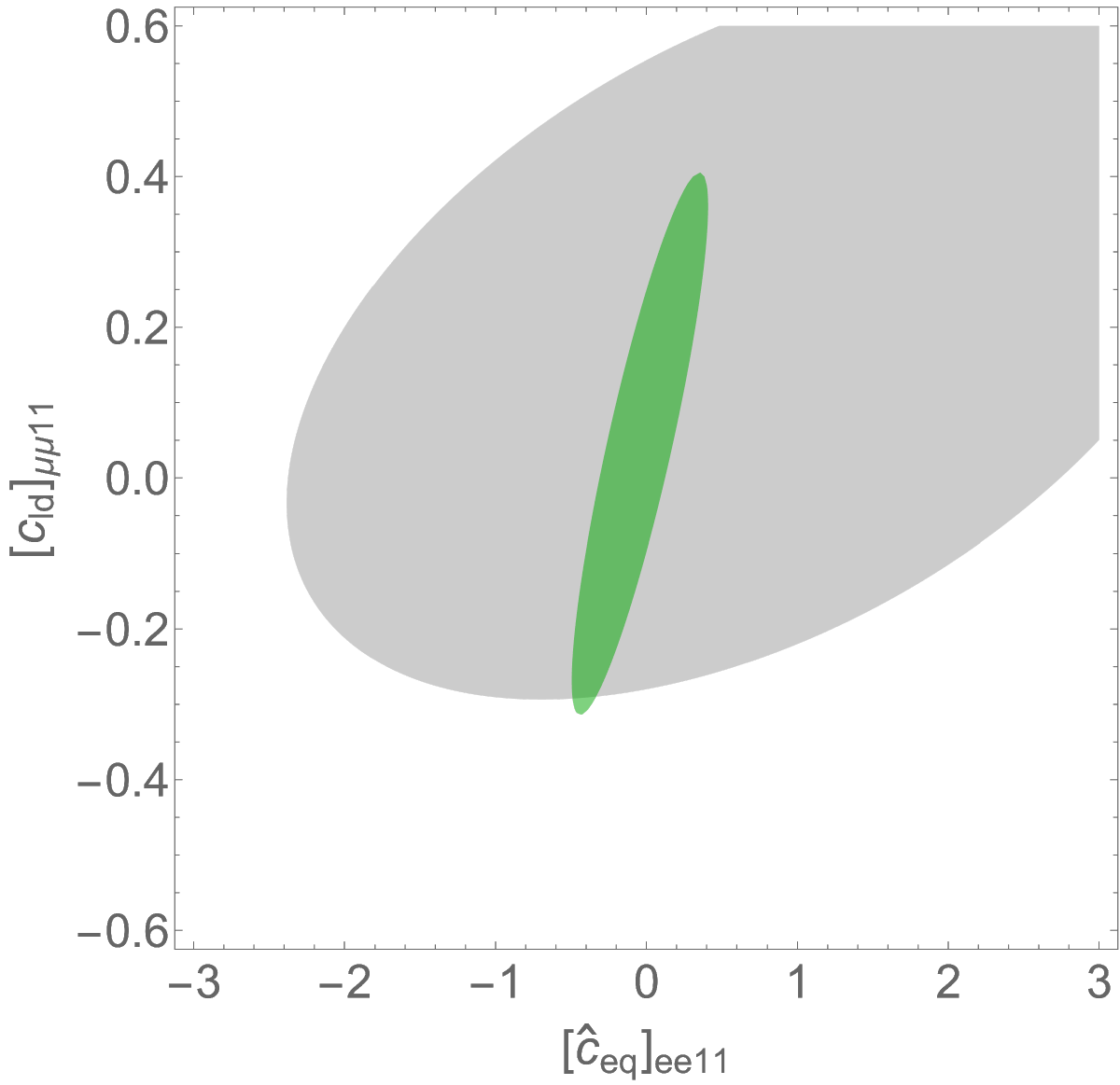}
    \caption{
\label{fig:SMEFT2_ellipses}
Left panel: Marginalized 1-sigma bounds $(\Delta \chi^2 \simeq 2.3)$ on the (combination of) SMEFT Wilson coefficients $[\hat c_{eq}]_{ee11}$ and $[c_{lu}]_{\mu\mu 11}$ from a global fit to EWPO in the flavor-generic SMEFT without (gray) and with (green) COHERENT data.  Right panel: The same for the  $[\hat c_{eq}]_{ee11}$-$[c_{ld}]_{\mu\mu 11}$ pair of Wilson coefficients.     }
\end{figure}

\bibliographystyle{JHEP}
\bibliography{references}

\providecommand{\href}[2]{#2}\begingroup\raggedright\begin{thebibliography}{100}

\bibitem{Freedman:1973yd}
D.~Z. Freedman, {\it {Coherent Neutrino Nucleus Scattering as a Probe of the
  Weak Neutral Current}},  {\em Phys. Rev. D} {\bf 9} (1974) 1389--1392.

\bibitem{Drukier:1984vhf}
A.~Drukier and L.~Stodolsky, {\it {Principles and Applications of a Neutral
  Current Detector for Neutrino Physics and Astronomy}},  {\em Phys. Rev. D}
  {\bf 30} (1984) 2295.

\bibitem{COHERENT:2017ipa}
{\bf COHERENT} Collaboration, D.~Akimov et~al., {\it {Observation of Coherent
  Elastic Neutrino-Nucleus Scattering}},  {\em Science} {\bf 357} (2017),
  no.~6356 1123--1126, [\href{http://arxiv.org/abs/1708.01294}{{\tt
  arXiv:1708.01294}}].

\bibitem{Weinberg:1980wa}
S.~Weinberg, {\it {Effective Gauge Theories}},  {\em Phys. Lett. B} {\bf 91}
  (1980) 51--55.

\bibitem{Weinberg:2021exr}
S.~Weinberg, {\it {On the Development of Effective Field Theory}},  {\em Eur.
  Phys. J. H} {\bf 46} (2021), no.~1 6,
  [\href{http://arxiv.org/abs/2101.04241}{{\tt arXiv:2101.04241}}].

\bibitem{Pich:1998xt}
A.~Pich, {\it {Effective field theory: Course}},  in {\em {Les Houches Summer
  School in Theoretical Physics, Session 68: Probing the Standard Model of
  Particle Interactions}}, pp.~949--1049, 6, 1998.
\newblock \href{http://arxiv.org/abs/hep-ph/9806303}{{\tt hep-ph/9806303}}.

\bibitem{Manohar:2020nzp}
A.~V. Manohar and E.~Nardoni, {\it {Renormalization Group Improvement of the
  Effective Potential: an EFT Approach}},  {\em JHEP} {\bf 04} (2021) 093,
  [\href{http://arxiv.org/abs/2010.15806}{{\tt arXiv:2010.15806}}].

\bibitem{Erler:2013xha}
J.~Erler and S.~Su, {\it {The Weak Neutral Current}},  {\em Prog. Part. Nucl.
  Phys.} {\bf 71} (2013) 119--149, [\href{http://arxiv.org/abs/1303.5522}{{\tt
  arXiv:1303.5522}}].

\bibitem{Scholberg:2005qs}
K.~Scholberg, {\it {Prospects for measuring coherent neutrino-nucleus elastic
  scattering at a stopped-pion neutrino source}},  {\em Phys. Rev. D} {\bf 73}
  (2006) 033005, [\href{http://arxiv.org/abs/hep-ex/0511042}{{\tt
  hep-ex/0511042}}].

\bibitem{Barranco:2005yy}
J.~Barranco, O.~Miranda, and T.~I. Rashba, {\it {Probing new physics with
  coherent neutrino scattering off nuclei}},  {\em JHEP} {\bf 12} (2005) 021,
  [\href{http://arxiv.org/abs/hep-ph/0508299}{{\tt hep-ph/0508299}}].

\bibitem{Jenkins:2017jig}
E.~E. Jenkins, A.~V. Manohar, and P.~Stoffer, {\it {Low-Energy Effective Field
  Theory below the Electroweak Scale: Operators and Matching}},  {\em JHEP}
  {\bf 03} (2018) 016, [\href{http://arxiv.org/abs/1709.04486}{{\tt
  arXiv:1709.04486}}].

\bibitem{Gago:2001xg}
A.~M. Gago, M.~M. Guzzo, H.~Nunokawa, W.~J.~C. Teves, and
  R.~Zukanovich~Funchal, {\it {Probing flavor changing neutrino interactions
  using neutrino beams from a muon storage ring}},  {\em Phys. Rev. D} {\bf 64}
  (2001) 073003, [\href{http://arxiv.org/abs/hep-ph/0105196}{{\tt
  hep-ph/0105196}}].

\bibitem{Campanelli:2002cc}
M.~Campanelli and A.~Romanino, {\it {Effects of new physics in neutrino
  oscillations in matter}},  {\em Phys. Rev. D} {\bf 66} (2002) 113001,
  [\href{http://arxiv.org/abs/hep-ph/0207350}{{\tt hep-ph/0207350}}].

\bibitem{Grossman:1995wx}
Y.~Grossman, {\it {Nonstandard neutrino interactions and neutrino oscillation
  experiments}},  {\em Phys. Lett. B} {\bf 359} (1995) 141--147,
  [\href{http://arxiv.org/abs/hep-ph/9507344}{{\tt hep-ph/9507344}}].

\bibitem{Falkowski:2019kfn}
A.~Falkowski, M.~Gonz\'alez-Alonso, and Z.~Tabrizi, {\it {Consistent QFT
  description of non-standard neutrino interactions}},  {\em JHEP} {\bf 11}
  (2020) 048, [\href{http://arxiv.org/abs/1910.02971}{{\tt arXiv:1910.02971}}].

\bibitem{Buchmuller:1985jz}
W.~Buchmuller and D.~Wyler, {\it {Effective Lagrangian Analysis of New
  Interactions and Flavor Conservation}},  {\em Nucl. Phys. B} {\bf 268} (1986)
  621--653.

\bibitem{Grzadkowski:2010es}
B.~Grzadkowski, M.~Iskrzynski, M.~Misiak, and J.~Rosiek, {\it {Dimension-Six
  Terms in the Standard Model Lagrangian}},  {\em JHEP} {\bf 10} (2010) 085,
  [\href{http://arxiv.org/abs/1008.4884}{{\tt arXiv:1008.4884}}].

\bibitem{Carmona:2021xtq}
A.~Carmona, A.~Lazopoulos, P.~Olgoso, and J.~Santiago, {\it {Matchmakereft:
  automated tree-level and one-loop matching}},  {\em SciPost Phys.} {\bf 12}
  (2022), no.~6 198, [\href{http://arxiv.org/abs/2112.10787}{{\tt
  arXiv:2112.10787}}].

\bibitem{Coloma:2017ncl}
P.~Coloma, M.~C. Gonzalez-Garcia, M.~Maltoni, and T.~Schwetz, {\it {COHERENT
  Enlightenment of the Neutrino Dark Side}},  {\em Phys. Rev. D} {\bf 96}
  (2017), no.~11 115007, [\href{http://arxiv.org/abs/1708.02899}{{\tt
  arXiv:1708.02899}}].

\bibitem{Papoulias:2017qdn}
D.~K. Papoulias and T.~S. Kosmas, {\it {COHERENT constraints to conventional
  and exotic neutrino physics}},  {\em Phys. Rev. D} {\bf 97} (2018), no.~3
  033003, [\href{http://arxiv.org/abs/1711.09773}{{\tt arXiv:1711.09773}}].

\bibitem{Shoemaker:2017lzs}
I.~M. Shoemaker, {\it {COHERENT search strategy for beyond standard model
  neutrino interactions}},  {\em Phys. Rev. D} {\bf 95} (2017), no.~11 115028,
  [\href{http://arxiv.org/abs/1703.05774}{{\tt arXiv:1703.05774}}].

\bibitem{Liao:2017uzy}
J.~Liao and D.~Marfatia, {\it {COHERENT constraints on nonstandard neutrino
  interactions}},  {\em Phys. Lett. B} {\bf 775} (2017) 54--57,
  [\href{http://arxiv.org/abs/1708.04255}{{\tt arXiv:1708.04255}}].

\bibitem{AristizabalSierra:2018eqm}
D.~Aristizabal~Sierra, V.~De~Romeri, and N.~Rojas, {\it {COHERENT analysis of
  neutrino generalized interactions}},  {\em Phys. Rev. D} {\bf 98} (2018)
  075018, [\href{http://arxiv.org/abs/1806.07424}{{\tt arXiv:1806.07424}}].

\bibitem{Denton:2018xmq}
P.~B. Denton, Y.~Farzan, and I.~M. Shoemaker, {\it {Testing large non-standard
  neutrino interactions with arbitrary mediator mass after COHERENT data}},
  {\em JHEP} {\bf 07} (2018) 037, [\href{http://arxiv.org/abs/1804.03660}{{\tt
  arXiv:1804.03660}}].

\bibitem{Esteban:2018ppq}
I.~Esteban, M.~C. Gonzalez-Garcia, M.~Maltoni, I.~Martinez-Soler, and
  J.~Salvado, {\it {Updated constraints on non-standard interactions from
  global analysis of oscillation data}},  {\em JHEP} {\bf 08} (2018) 180,
  [\href{http://arxiv.org/abs/1805.04530}{{\tt arXiv:1805.04530}}]. [Addendum:
  JHEP 12, 152 (2020)].

\bibitem{Khan:2019cvi}
A.~N. Khan and W.~Rodejohann, {\it {New physics from COHERENT data with an
  improved quenching factor}},  {\em Phys. Rev. D} {\bf 100} (2019), no.~11
  113003, [\href{http://arxiv.org/abs/1907.12444}{{\tt arXiv:1907.12444}}].

\bibitem{Giunti:2019xpr}
C.~Giunti, {\it {General COHERENT constraints on neutrino nonstandard
  interactions}},  {\em Phys. Rev. D} {\bf 101} (2020), no.~3 035039,
  [\href{http://arxiv.org/abs/1909.00466}{{\tt arXiv:1909.00466}}].

\bibitem{Arcadi:2019uif}
G.~Arcadi, M.~Lindner, J.~Martins, and F.~S. Queiroz, {\it {New physics probes:
  Atomic parity violation, polarized electron scattering and neutrino-nucleus
  coherent scattering}},  {\em Nucl. Phys. B} {\bf 959} (2020) 115158,
  [\href{http://arxiv.org/abs/1906.04755}{{\tt arXiv:1906.04755}}].

\bibitem{Coloma:2019mbs}
P.~Coloma, I.~Esteban, M.~C. Gonzalez-Garcia, and M.~Maltoni, {\it {Improved
  global fit to Non-Standard neutrino Interactions using COHERENT energy and
  timing data}},  {\em JHEP} {\bf 02} (2020) 023,
  [\href{http://arxiv.org/abs/1911.09109}{{\tt arXiv:1911.09109}}]. [Addendum:
  JHEP 12, 071 (2020)].

\bibitem{Denton:2020hop}
P.~B. Denton and J.~Gehrlein, {\it {A Statistical Analysis of the COHERENT Data
  and Applications to New Physics}},  {\em JHEP} {\bf 04} (2021) 266,
  [\href{http://arxiv.org/abs/2008.06062}{{\tt arXiv:2008.06062}}].

\bibitem{Miranda:2020tif}
O.~G. Miranda, D.~K. Papoulias, G.~Sanchez~Garcia, O.~Sanders, M.~T\'ortola,
  and J.~W.~F. Valle, {\it {Implications of the first detection of coherent
  elastic neutrino-nucleus scattering (CEvNS) with Liquid Argon}},  {\em JHEP}
  {\bf 05} (2020) 130, [\href{http://arxiv.org/abs/2003.12050}{{\tt
  arXiv:2003.12050}}]. [Erratum: JHEP 01, 067 (2021)].

\bibitem{Coloma:2022avw}
P.~Coloma, I.~Esteban, M.~C. Gonzalez-Garcia, L.~Larizgoitia, F.~Monrabal, and
  S.~Palomares-Ruiz, {\it {Bounds on new physics with data of the Dresden-II
  reactor experiment and COHERENT}},  {\em JHEP} {\bf 05} (2022) 037,
  [\href{http://arxiv.org/abs/2202.10829}{{\tt arXiv:2202.10829}}].

\bibitem{AtzoriCorona:2022qrf}
M.~Atzori~Corona, M.~Cadeddu, N.~Cargioli, F.~Dordei, C.~Giunti, Y.~F. Li,
  C.~A. Ternes, and Y.~Y. Zhang, {\it {Impact of the Dresden-II and COHERENT
  neutrino scattering data on neutrino electromagnetic properties and
  electroweak physics}},  {\em JHEP} {\bf 09} (2022) 164,
  [\href{http://arxiv.org/abs/2205.09484}{{\tt arXiv:2205.09484}}].

\bibitem{DeRomeri:2022twg}
V.~De~Romeri, O.~G. Miranda, D.~K. Papoulias, G.~Sanchez~Garcia, M.~T\'ortola,
  and J.~W.~F. Valle, {\it {Physics implications of a combined analysis of
  COHERENT CsI and LAr data}},  \href{http://arxiv.org/abs/2211.11905}{{\tt
  arXiv:2211.11905}}.

\bibitem{COHERENT:2020iec}
{\bf COHERENT} Collaboration, D.~Akimov et~al., {\it {First Measurement of
  Coherent Elastic Neutrino-Nucleus Scattering on Argon}},  {\em Phys. Rev.
  Lett.} {\bf 126} (2021), no.~1 012002,
  [\href{http://arxiv.org/abs/2003.10630}{{\tt arXiv:2003.10630}}].

\bibitem{COHERENT:2021xmm}
{\bf COHERENT} Collaboration, D.~Akimov et~al., {\it {Measurement of the
  Coherent Elastic Neutrino-Nucleus Scattering Cross Section on CsI by
  COHERENT}},  {\em Phys. Rev. Lett.} {\bf 129} (2022), no.~8 081801,
  [\href{http://arxiv.org/abs/2110.07730}{{\tt arXiv:2110.07730}}].

\bibitem{ParticleDataGroup:2022pth}
{\bf Particle Data Group} Collaboration, R.~L. Workman et~al., {\it {Review of
  Particle Physics}},  {\em PTEP} {\bf 2022} (2022) 083C01.

\bibitem{Hoferichter:2020osn}
M.~Hoferichter, J.~Men\'endez, and A.~Schwenk, {\it {Coherent elastic
  neutrino-nucleus scattering: EFT analysis and nuclear responses}},  {\em
  Phys. Rev. D} {\bf 102} (2020), no.~7 074018,
  [\href{http://arxiv.org/abs/2007.08529}{{\tt arXiv:2007.08529}}].

\bibitem{Cirigliano:2009wk}
V.~Cirigliano, J.~Jenkins, and M.~Gonzalez-Alonso, {\it {Semileptonic decays of
  light quarks beyond the Standard Model}},  {\em Nucl. Phys. B} {\bf 830}
  (2010) 95--115, [\href{http://arxiv.org/abs/0908.1754}{{\tt
  arXiv:0908.1754}}].

\bibitem{Falkowski:2017pss}
A.~Falkowski, M.~Gonz\'alez-Alonso, and K.~Mimouni, {\it {Compilation of
  low-energy constraints on 4-fermion operators in the SMEFT}},  {\em JHEP}
  {\bf 08} (2017) 123, [\href{http://arxiv.org/abs/1706.03783}{{\tt
  arXiv:1706.03783}}].

\bibitem{Gonzalez-Alonso:2018omy}
M.~Gonz\'alez-Alonso, O.~Naviliat-Cuncic, and N.~Severijns, {\it {New physics
  searches in nuclear and neutron $\beta$ decay}},  {\em Prog. Part. Nucl.
  Phys.} {\bf 104} (2019) 165--223,
  [\href{http://arxiv.org/abs/1803.08732}{{\tt arXiv:1803.08732}}].

\bibitem{Falkowski:2020pma}
A.~Falkowski, M.~Gonz\'alez-Alonso, and O.~Naviliat-Cuncic, {\it {Comprehensive
  analysis of beta decays within and beyond the Standard Model}},  {\em JHEP}
  {\bf 04} (2021) 126, [\href{http://arxiv.org/abs/2010.13797}{{\tt
  arXiv:2010.13797}}].

\bibitem{Gonzalez-Alonso:2017iyc}
M.~Gonz\'alez-Alonso, J.~Martin~Camalich, and K.~Mimouni, {\it
  {Renormalization-group evolution of new physics contributions to
  (semi)leptonic meson decays}},  {\em Phys. Lett. B} {\bf 772} (2017)
  777--785, [\href{http://arxiv.org/abs/1706.00410}{{\tt arXiv:1706.00410}}].

\bibitem{Tomalak:2020zfh}
O.~Tomalak, P.~Machado, V.~Pandey, and R.~Plestid, {\it {Flavor-dependent
  radiative corrections in coherent elastic neutrino-nucleus scattering}},
  {\em JHEP} {\bf 02} (2021) 097, [\href{http://arxiv.org/abs/2011.05960}{{\tt
  arXiv:2011.05960}}].

\bibitem{Falkowski:2021vdg}
A.~Falkowski, M.~Gonz\'alez-Alonso, A.~Palavri\'c, and
  A.~Rodr\'\i{}guez-S\'anchez, {\it {Constraints on subleading interactions in
  beta decay Lagrangian}},  \href{http://arxiv.org/abs/2112.07688}{{\tt
  arXiv:2112.07688}}.

\bibitem{Giunti:1993se}
C.~Giunti, C.~W. Kim, J.~A. Lee, and U.~W. Lee, {\it {On the treatment of
  neutrino oscillations without resort to weak eigenstates}},  {\em Phys. Rev.
  D} {\bf 48} (1993) 4310--4317,
  [\href{http://arxiv.org/abs/hep-ph/9305276}{{\tt hep-ph/9305276}}].

\bibitem{Helm:1956zz}
R.~H. Helm, {\it {Inelastic and Elastic Scattering of 187-Mev Electrons from
  Selected Even-Even Nuclei}},  {\em Phys. Rev.} {\bf 104} (1956) 1466--1475.

\bibitem{Lindner:2016wff}
M.~Lindner, W.~Rodejohann, and X.-J. Xu, {\it {Coherent Neutrino-Nucleus
  Scattering and new Neutrino Interactions}},  {\em JHEP} {\bf 03} (2017) 097,
  [\href{http://arxiv.org/abs/1612.04150}{{\tt arXiv:1612.04150}}].

\bibitem{Canas:2018rng}
B.~C. Ca\~nas, E.~A. Garc\'es, O.~G. Miranda, and A.~Parada, {\it {Future
  perspectives for a weak mixing angle measurement in coherent elastic neutrino
  nucleus scattering experiments}},  {\em Phys. Lett. B} {\bf 784} (2018)
  159--162, [\href{http://arxiv.org/abs/1806.01310}{{\tt arXiv:1806.01310}}].

\bibitem{Huang:2019ene}
X.-R. Huang and L.-W. Chen, {\it {Neutron Skin in CsI and Low-Energy Effective
  Weak Mixing Angle from COHERENT Data}},  {\em Phys. Rev. D} {\bf 100} (2019),
  no.~7 071301, [\href{http://arxiv.org/abs/1902.07625}{{\tt
  arXiv:1902.07625}}].

\bibitem{Khan:2021wzy}
A.~N. Khan, D.~W. McKay, and W.~Rodejohann, {\it {CP-violating and charged
  current neutrino nonstandard interactions in CE\ensuremath{\nu}NS}},  {\em
  Phys. Rev. D} {\bf 104} (2021), no.~1 015019,
  [\href{http://arxiv.org/abs/2104.00425}{{\tt arXiv:2104.00425}}].

\bibitem{COHERENT:2020ybo}
{\bf COHERENT} Collaboration, D.~Akimov et~al., {\it {COHERENT Collaboration
  data release from the first detection of coherent elastic neutrino-nucleus
  scattering on argon}},  \href{http://arxiv.org/abs/2006.12659}{{\tt
  arXiv:2006.12659}}.

\bibitem{Farzan:2017xzy}
Y.~Farzan and M.~Tortola, {\it {Neutrino oscillations and Non-Standard
  Interactions}},  {\em Front. in Phys.} {\bf 6} (2018) 10,
  [\href{http://arxiv.org/abs/1710.09360}{{\tt arXiv:1710.09360}}].

\bibitem{Escrihuela:2011cf}
F.~J. Escrihuela, M.~Tortola, J.~W.~F. Valle, and O.~G. Miranda, {\it {Global
  constraints on muon-neutrino non-standard interactions}},  {\em Phys. Rev. D}
  {\bf 83} (2011) 093002, [\href{http://arxiv.org/abs/1103.1366}{{\tt
  arXiv:1103.1366}}].

\bibitem{Davidson:2003ha}
S.~Davidson, C.~Pena-Garay, N.~Rius, and A.~Santamaria, {\it {Present and
  future bounds on nonstandard neutrino interactions}},  {\em JHEP} {\bf 03}
  (2003) 011, [\href{http://arxiv.org/abs/hep-ph/0302093}{{\tt
  hep-ph/0302093}}].

\bibitem{Colaresi:2021kus}
J.~Colaresi, J.~I. Collar, T.~W. Hossbach, A.~R.~L. Kavner, C.~M. Lewis, A.~E.
  Robinson, and K.~M. Yocum, {\it {First results from a search for coherent
  elastic neutrino-nucleus scattering at a reactor site}},  {\em Phys. Rev. D}
  {\bf 104} (2021), no.~7 072003, [\href{http://arxiv.org/abs/2108.02880}{{\tt
  arXiv:2108.02880}}].

\bibitem{Salvado:2016uqu}
J.~Salvado, O.~Mena, S.~Palomares-Ruiz, and N.~Rius, {\it {Non-standard
  interactions with high-energy atmospheric neutrinos at IceCube}},  {\em JHEP}
  {\bf 01} (2017) 141, [\href{http://arxiv.org/abs/1609.03450}{{\tt
  arXiv:1609.03450}}].

\bibitem{PiENu:2015seu}
{\bf PiENu} Collaboration, A.~Aguilar-Arevalo et~al., {\it {Improved
  Measurement of the $\pi \to \textrm{e} \nu$ Branching Ratio}},  {\em Phys.
  Rev. Lett.} {\bf 115} (2015), no.~7 071801,
  [\href{http://arxiv.org/abs/1506.05845}{{\tt arXiv:1506.05845}}].

\bibitem{Cirigliano:2007ga}
V.~Cirigliano and I.~Rosell, {\it {pi/K ---\ensuremath{>} e anti-nu(e)
  branching ratios to O(e**2 p**4) in Chiral Perturbation Theory}},  {\em JHEP}
  {\bf 10} (2007) 005, [\href{http://arxiv.org/abs/0707.4464}{{\tt
  arXiv:0707.4464}}].

\bibitem{Terol-Calvo:2019vck}
J.~Terol-Calvo, M.~T\'ortola, and A.~Vicente, {\it {High-energy constraints
  from low-energy neutrino nonstandard interactions}},  {\em Phys. Rev. D} {\bf
  101} (2020), no.~9 095010, [\href{http://arxiv.org/abs/1912.09131}{{\tt
  arXiv:1912.09131}}].

\bibitem{Skiba:2020msb}
W.~Skiba and Q.~Xia, {\it {Electroweak constraints from the COHERENT
  experiment}},  {\em JHEP} {\bf 10} (2022) 102,
  [\href{http://arxiv.org/abs/2007.15688}{{\tt arXiv:2007.15688}}].

\bibitem{Crivellin:2021bkd}
A.~Crivellin, M.~Hoferichter, M.~Kirk, C.~A. Manzari, and L.~Schnell, {\it
  {First-generation new physics in simplified models: from low-energy parity
  violation to the LHC}},  {\em JHEP} {\bf 10} (2021) 221,
  [\href{http://arxiv.org/abs/2107.13569}{{\tt arXiv:2107.13569}}].

\bibitem{Azatov:2022kbs}
A.~Azatov et~al., {\it {Off-shell Higgs Interpretations Task Force: Models and
  Effective Field Theories Subgroup Report}},
  \href{http://arxiv.org/abs/2203.02418}{{\tt arXiv:2203.02418}}.

\bibitem{Falkowski:2015jaa}
A.~Falkowski, M.~Gonzalez-Alonso, A.~Greljo, and D.~Marzocca, {\it {Global
  constraints on anomalous triple gauge couplings in effective field theory
  approach}},  {\em Phys. Rev. Lett.} {\bf 116} (2016), no.~1 011801,
  [\href{http://arxiv.org/abs/1508.00581}{{\tt arXiv:1508.00581}}].

\bibitem{Lewin:1995rx}
J.~D. Lewin and P.~F. Smith, {\it {Review of mathematics, numerical factors,
  and corrections for dark matter experiments based on elastic nuclear
  recoil}},  {\em Astropart. Phys.} {\bf 6} (1996) 87--112.

\bibitem{ANGELI201369}
I.~Angeli and K.~Marinova, {\it Table of experimental nuclear ground state
  charge radii: An update},  {\em Atomic Data and Nuclear Data Tables} {\bf 99}
  (2013), no.~1 69--95.

\bibitem{FRICKE1995177}
G.~Fricke, C.~Bernhardt, K.~Heilig, L.~Schaller, L.~Schellenberg, E.~Shera, and
  C.~Dejager, {\it Nuclear ground state charge radii from electromagnetic
  interactions},  {\em Atomic Data and Nuclear Data Tables} {\bf 60} (1995),
  no.~2 177--285.

\bibitem{Picciau:2022xzi}
E.~Picciau, {\em {Low-energy signatures in DarkSide-50 experiment and neutrino
  scattering processes}}.
\newblock PhD thesis, Cagliari U., 2022.

\bibitem{creus2013light}
W.~Creus, {\em Light yield in liquid argon for dark matter detection}.
\newblock PhD thesis, University of Zurich, 2013.

\bibitem{LHCHiggsCrossSectionWorkingGroup:2016ypw}
{\bf LHC Higgs Cross Section Working Group} Collaboration, D.~de~Florian
  et~al., {\it {Handbook of LHC Higgs Cross Sections: 4. Deciphering the Nature
  of the Higgs Sector}},  \href{http://arxiv.org/abs/1610.07922}{{\tt
  arXiv:1610.07922}}.

\bibitem{Electroweak:2003ram}
{\bf LEP, ALEPH, DELPHI, L3, OPAL, LEP Electroweak Working Group, SLD
  Electroweak Group, SLD Heavy Flavor Group} Collaboration, D.~Abbaneo et~al.,
  {\it {A Combination of preliminary electroweak measurements and constraints
  on the standard model}},  \href{http://arxiv.org/abs/hep-ex/0312023}{{\tt
  hep-ex/0312023}}.

\bibitem{ALEPH:2013dgf}
{\bf ALEPH, DELPHI, L3, OPAL, LEP Electroweak} Collaboration, S.~Schael et~al.,
  {\it {Electroweak Measurements in Electron-Positron Collisions at
  W-Boson-Pair Energies at LEP}},  {\em Phys. Rept.} {\bf 532} (2013) 119--244,
  [\href{http://arxiv.org/abs/1302.3415}{{\tt arXiv:1302.3415}}].

\bibitem{ALEPH:2005ab}
{\bf ALEPH, DELPHI, L3, OPAL, SLD, LEP Electroweak Working Group, SLD
  Electroweak Group, SLD Heavy Flavour Group} Collaboration, S.~Schael et~al.,
  {\it {Precision electroweak measurements on the $Z$ resonance}},  {\em Phys.
  Rept.} {\bf 427} (2006) 257--454,
  [\href{http://arxiv.org/abs/hep-ex/0509008}{{\tt hep-ex/0509008}}].

\bibitem{SLD:2000jop}
{\bf SLD} Collaboration, K.~Abe et~al., {\it {First direct measurement of the
  parity violating coupling of the Z0 to the s quark}},  {\em Phys. Rev. Lett.}
  {\bf 85} (2000) 5059--5063, [\href{http://arxiv.org/abs/hep-ex/0006019}{{\tt
  hep-ex/0006019}}].

\bibitem{VENUS:1993pob}
{\bf VENUS} Collaboration, K.~Abe et~al., {\it {A Study of the charm and bottom
  quark production in e+ e- annihilation at s**(1/2) = 58-GeV using prompt
  electrons}},  {\em Phys. Lett. B} {\bf 313} (1993) 288--298.

\bibitem{VENUS:1997cjg}
{\bf VENUS} Collaboration, H.~Hanai et~al., {\it {Measurement of tau
  polarization in e+ e- annihilation at s**(1/2) = 58-GeV}},  {\em Phys. Lett.
  B} {\bf 403} (1997) 155--162,
  [\href{http://arxiv.org/abs/hep-ex/9703003}{{\tt hep-ex/9703003}}].

\bibitem{TOPAZ:2000evx}
{\bf TOPAZ} Collaboration, Y.~Inoue et~al., {\it {Measurement of the
  cross-section and forward - backward charge asymmetry for the b and c quark
  in e+ e- annihilation with inclusive muons at s**(1/2) = 58-GeV}},  {\em Eur.
  Phys. J. C} {\bf 18} (2000) 273--282,
  [\href{http://arxiv.org/abs/hep-ex/0012033}{{\tt hep-ex/0012033}}].

\bibitem{Breso-Pla:2021qoe}
V.~Bres\'o-Pla, A.~Falkowski, and M.~Gonz\'alez-Alonso, {\it {A$_{FB}$ in the
  SMEFT: precision Z physics at the LHC}},  {\em JHEP} {\bf 08} (2021) 021,
  [\href{http://arxiv.org/abs/2103.12074}{{\tt arXiv:2103.12074}}].

\bibitem{Janot:2019oyi}
P.~Janot and S.~Jadach, {\it {Improved Bhabha cross section at LEP and the
  number of light neutrino species}},  {\em Phys. Lett. B} {\bf 803} (2020)
  135319, [\href{http://arxiv.org/abs/1912.02067}{{\tt arXiv:1912.02067}}].

\bibitem{dEnterria:2020cgt}
D.~d'Enterria and C.~Yan, {\it {Revised QCD effects on the Z $\to b\bar{b}$
  forward-backward asymmetry}},  \href{http://arxiv.org/abs/2011.00530}{{\tt
  arXiv:2011.00530}}.

\bibitem{D0:1999bqi}
{\bf D0} Collaboration, B.~Abbott et~al., {\it {A measurement of the $W \to
  \tau \nu$ production cross section in $p\bar{p}$ collisions at $\sqrt{s} =
  1.8$ TeV}},  {\em Phys. Rev. Lett.} {\bf 84} (2000) 5710--5715,
  [\href{http://arxiv.org/abs/hep-ex/9912065}{{\tt hep-ex/9912065}}].

\bibitem{LHCb:2016zpq}
{\bf LHCb} Collaboration, R.~Aaij et~al., {\it {Measurement of forward $W\to
  e\nu$ production in $pp$ collisions at $\sqrt{s}=8\,$TeV}},  {\em JHEP} {\bf
  10} (2016) 030, [\href{http://arxiv.org/abs/1608.01484}{{\tt
  arXiv:1608.01484}}].

\bibitem{ATLAS:2016nqi}
{\bf ATLAS} Collaboration, M.~Aaboud et~al., {\it {Precision measurement and
  interpretation of inclusive $W^+$ , $W^-$ and $Z/\gamma ^*$ production cross
  sections with the ATLAS detector}},  {\em Eur. Phys. J. C} {\bf 77} (2017),
  no.~6 367, [\href{http://arxiv.org/abs/1612.03016}{{\tt arXiv:1612.03016}}].

\bibitem{ATLAS:2020xea}
{\bf ATLAS} Collaboration, G.~Aad et~al., {\it {Test of the universality of
  $\tau$ and $\mu$ lepton couplings in $W$-boson decays with the ATLAS
  detector}},  {\em Nature Phys.} {\bf 17} (2021), no.~7 813--818,
  [\href{http://arxiv.org/abs/2007.14040}{{\tt arXiv:2007.14040}}].

\bibitem{LHCb:2021bjt}
{\bf LHCb} Collaboration, R.~Aaij et~al., {\it {Measurement of the W boson
  mass}},  {\em JHEP} {\bf 01} (2022) 036,
  [\href{http://arxiv.org/abs/2109.01113}{{\tt arXiv:2109.01113}}].

\bibitem{ATLAS:2017rzl}
{\bf ATLAS} Collaboration, M.~Aaboud et~al., {\it {Measurement of the $W$-boson
  mass in pp collisions at $\sqrt{s}=7$ TeV with the ATLAS detector}},  {\em
  Eur. Phys. J. C} {\bf 78} (2018), no.~2 110,
  [\href{http://arxiv.org/abs/1701.07240}{{\tt arXiv:1701.07240}}]. [Erratum:
  Eur.Phys.J.C 78, 898 (2018)].

\bibitem{CDF:2012gpf}
{\bf CDF} Collaboration, T.~Aaltonen et~al., {\it {Precise measurement of the
  $W$-boson mass with the CDF II detector}},  {\em Phys. Rev. Lett.} {\bf 108}
  (2012) 151803, [\href{http://arxiv.org/abs/1203.0275}{{\tt
  arXiv:1203.0275}}].

\bibitem{D0:2012kms}
{\bf D0} Collaboration, V.~M. Abazov et~al., {\it {Measurement of the W Boson
  Mass with the D0 Detector}},  {\em Phys. Rev. Lett.} {\bf 108} (2012) 151804,
  [\href{http://arxiv.org/abs/1203.0293}{{\tt arXiv:1203.0293}}].

\bibitem{CDF:2022hxs}
{\bf CDF} Collaboration, T.~Aaltonen et~al., {\it {High-precision measurement
  of the $W$ boson mass with the CDF II detector}},  {\em Science} {\bf 376}
  (2022), no.~6589 170--176.

\bibitem{CMS:2014mgj}
{\bf CMS} Collaboration, V.~Khachatryan et~al., {\it {Measurement of the
  t-channel single-top-quark production cross section and of the $\mid V_{tb}
  \mid$ CKM matrix element in pp collisions at $\sqrt{s}$= 8 TeV}},  {\em JHEP}
  {\bf 06} (2014) 090, [\href{http://arxiv.org/abs/1403.7366}{{\tt
  arXiv:1403.7366}}].

\bibitem{ATLAS:2018gqq}
{\bf ATLAS} Collaboration, {\it {Measurement of the effective leptonic weak
  mixing angle using electron and muon pairs from $Z$-boson decay in the ATLAS
  experiment at $\sqrt s = 8$ TeV}},  {\em ATLAS-CONF-2018-037} (7, 2018).

\bibitem{D0:2011baz}
{\bf D0} Collaboration, V.~M. Abazov et~al., {\it {Measurement of
  $\sin^2\theta_{\rm eff}^{\ell}$ and $Z$-light quark couplings using the
  forward-backward charge asymmetry in $p\bar{p} \to Z/\gamma^{*} \to
  e^{+}e^{-}$ events with ${\cal L}=5.0$ fb$^{-1}$ at $\sqrt{s}=1.96$ TeV}},
  {\em Phys. Rev. D} {\bf 84} (2011) 012007,
  [\href{http://arxiv.org/abs/1104.4590}{{\tt arXiv:1104.4590}}].

\bibitem{CHARM:1986vuz}
{\bf CHARM} Collaboration, J.~Dorenbosch et~al., {\it {Experimental
  Verification of the Universality of $\nu_e$ and $\nu_\mu$ Coupling to the
  Neutral Weak Current}},  {\em Phys. Lett. B} {\bf 180} (1986) 303--307.

\bibitem{CHARM:1987pwr}
{\bf CHARM} Collaboration, J.~V. Allaby et~al., {\it {A Precise Determination
  of the Electroweak Mixing Angle from Semileptonic Neutrino Scattering}},
  {\em Z. Phys. C} {\bf 36} (1987) 611.

\bibitem{Blondel:1989ev}
A.~Blondel et~al., {\it {Electroweak Parameters From a High Statistics Neutrino
  Nucleon Scattering Experiment}},  {\em Z. Phys. C} {\bf 45} (1990) 361--379.

\bibitem{CCFR:1997zzq}
{\bf CCFR, E744, E770} Collaboration, K.~S. McFarland et~al., {\it {A Precision
  measurement of electroweak parameters in neutrino - nucleon scattering}},
  {\em Eur. Phys. J. C} {\bf 1} (1998) 509--513,
  [\href{http://arxiv.org/abs/hep-ex/9701010}{{\tt hep-ex/9701010}}].

\bibitem{Wood:1997zq}
C.~S. Wood, S.~C. Bennett, D.~Cho, B.~P. Masterson, J.~L. Roberts, C.~E.
  Tanner, and C.~E. Wieman, {\it {Measurement of parity nonconservation and an
  anapole moment in cesium}},  {\em Science} {\bf 275} (1997) 1759--1763.

\bibitem{Edwards:1995zz}
N.~H. Edwards, S.~J. Phipp, P.~E.~G. Baird, and S.~Nakayama, {\it {Precise
  Measurement of Parity Nonconserving Optical Rotation in Atomic Thallium}},
  {\em Phys. Rev. Lett.} {\bf 74} (1995) 2654--2657.

\bibitem{Vetter:1995vf}
P.~A. Vetter, D.~M. Meekhof, P.~K. Majumder, S.~K. Lamoreaux, and E.~N.
  Fortson, {\it {Precise test of electroweak theory from a new measurement of
  parity nonconservation in atomic thallium}},  {\em Phys. Rev. Lett.} {\bf 74}
  (1995) 2658--2661.

\bibitem{Qweak:2018tjf}
{\bf Qweak} Collaboration, D.~Androi\'c et~al., {\it {Precision measurement of
  the weak charge of the proton}},  {\em Nature} {\bf 557} (2018), no.~7704
  207--211, [\href{http://arxiv.org/abs/1905.08283}{{\tt arXiv:1905.08283}}].

\bibitem{PVDIS:2014cmd}
{\bf PVDIS} Collaboration, D.~Wang et~al., {\it {Measurement of parity
  violation in electron\textendash{}quark scattering}},  {\em Nature} {\bf 506}
  (2014), no.~7486 67--70.

\bibitem{Beise:2004py}
E.~J. Beise, M.~L. Pitt, and D.~T. Spayde, {\it {The SAMPLE experiment and weak
  nucleon structure}},  {\em Prog. Part. Nucl. Phys.} {\bf 54} (2005) 289--350,
  [\href{http://arxiv.org/abs/nucl-ex/0412054}{{\tt nucl-ex/0412054}}].

\bibitem{Argento:1982tq}
A.~Argento et~al., {\it {Electroweak Asymmetry in Deep Inelastic Muon - Nucleon
  Scattering}},  {\em Phys. Lett. B} {\bf 120} (1983) 245.

\bibitem{Gonzalez-Alonso:2016etj}
M.~Gonz\'alez-Alonso and J.~Martin~Camalich, {\it {Global
  Effective-Field-Theory analysis of New-Physics effects in (semi)leptonic kaon
  decays}},  {\em JHEP} {\bf 12} (2016) 052,
  [\href{http://arxiv.org/abs/1605.07114}{{\tt arXiv:1605.07114}}].

\bibitem{Cirigliano:2021yto}
V.~Cirigliano, D.~D\'\i{}az-Calder\'on, A.~Falkowski, M.~Gonz\'alez-Alonso, and
  A.~Rodr\'\i{}guez-S\'anchez, {\it {Semileptonic tau decays beyond the
  Standard Model}},  {\em JHEP} {\bf 04} (2022) 152,
  [\href{http://arxiv.org/abs/2112.02087}{{\tt arXiv:2112.02087}}].

\bibitem{UCNt:2021pcg}
{\bf UCN\ensuremath{\tau}} Collaboration, F.~M. Gonzalez et~al., {\it {Improved
  neutron lifetime measurement with UCN$\tau$}},  {\em Phys. Rev. Lett.} {\bf
  127} (2021), no.~16 162501, [\href{http://arxiv.org/abs/2106.10375}{{\tt
  arXiv:2106.10375}}].

\bibitem{UCNA:2017obv}
{\bf UCNA} Collaboration, M.~A.~P. Brown et~al., {\it {New result for the
  neutron $\beta$-asymmetry parameter $A_0$ from UCNA}},  {\em Phys. Rev. C}
  {\bf 97} (2018), no.~3 035505, [\href{http://arxiv.org/abs/1712.00884}{{\tt
  arXiv:1712.00884}}].

\bibitem{Markisch:2018ndu}
B.~M\"arkisch et~al., {\it {Measurement of the Weak Axial-Vector Coupling
  Constant in the Decay of Free Neutrons Using a Pulsed Cold Neutron Beam}},
  {\em Phys. Rev. Lett.} {\bf 122} (2019), no.~24 242501,
  [\href{http://arxiv.org/abs/1812.04666}{{\tt arXiv:1812.04666}}].

\bibitem{CHARM:1988tlj}
{\bf CHARM} Collaboration, J.~Dorenbosch et~al., {\it {EXPERIMENTAL RESULTS ON
  NEUTRINO - ELECTRON SCATTERING}},  {\em Z. Phys. C} {\bf 41} (1989) 567.
  [Erratum: Z.Phys.C 51, 142 (1991)].

\bibitem{Ahrens:1990fp}
L.~A. Ahrens et~al., {\it {Determination of electroweak parameters from the
  elastic scattering of muon-neutrinos and anti-neutrinos on electrons}},  {\em
  Phys. Rev. D} {\bf 41} (1990) 3297--3316.

\bibitem{CHARM-II:1994dzw}
{\bf CHARM-II} Collaboration, P.~Vilain et~al., {\it {Precision measurement of
  electroweak parameters from the scattering of muon-neutrinos on electrons}},
  {\em Phys. Lett. B} {\bf 335} (1994) 246--252.

\bibitem{SLACE158:2005uay}
{\bf SLAC E158} Collaboration, P.~L. Anthony et~al., {\it {Precision
  measurement of the weak mixing angle in Moller scattering}},  {\em Phys. Rev.
  Lett.} {\bf 95} (2005) 081601,
  [\href{http://arxiv.org/abs/hep-ex/0504049}{{\tt hep-ex/0504049}}].

\bibitem{CHARM-II:1990dvf}
{\bf CHARM-II} Collaboration, D.~Geiregat et~al., {\it {First observation of
  neutrino trident production}},  {\em Phys. Lett. B} {\bf 245} (1990)
  271--275.

\bibitem{CCFR:1991lpl}
{\bf CCFR} Collaboration, S.~R. Mishra et~al., {\it {Neutrino tridents and W Z
  interference}},  {\em Phys. Rev. Lett.} {\bf 66} (1991) 3117--3120.

\bibitem{ParticleDataGroup:2016lqr}
{\bf Particle Data Group} Collaboration, C.~Patrignani et~al., {\it {Review of
  Particle Physics}},  {\em Chin. Phys. C} {\bf 40} (2016), no.~10 100001.

\end{thebibliography}\endgroup

\end{document}